\documentclass[final,5p,times,twocolumn,authoryear]{elsarticle}

\usepackage{amssymb}
\usepackage{xcolor}
\usepackage{adjustbox}
\usepackage{longtable}
\usepackage{caption}
\usepackage{wasysym} 
\usepackage{ragged2e} 

\journal{Treatise on Geochemistry: vol. 3}

\begin{document}

\begin{frontmatter}



\title{Composition, Structure and Origin of the Moon}


\author[inst1]{Paolo A. Sossi}

\affiliation[inst1]{organization={Institute of Geochemistry and Petrology, ETH Zürich},
            addressline={Clausiusstrasse 25}, 
            city={Zürich},
            postcode={CH-8092}, 
            state={ZH},
            country={Switzerland}}

\author[inst2]{Miki Nakajima}

\affiliation[inst2]{organization={Department of Earth and Environmental Sciences, University of Rochester},
            addressline={227 Hutchison Hall}, 
            city={Rochester},
            postcode={14627}, 
            state={NY},
            country={USA}}

\author[inst1,inst3]{Amir Khan}
\affiliation[inst3]{organization={Institute of Geophysics, ETH Zürich},
            addressline={Sonneggstrasse 5}, 
            city={Zürich},
            postcode={CH-8092}, 
            state={ZH},
            country={Switzerland}}

\begin{abstract} 
Extensive sampling of the Moon, both directly and remotely, has afforded a unique view into its composition and structure. 
In the canonical model, the proto-Earth is struck by a Mars-sized body, where the Moon is derived predominantly from the impactor. However, the isotopic compositions of a variety of elements that show variability among planetary materials, initially O, but subsequently Cr and Ti, attest to a near-identical match between the composition of the Earth's mantle and the Moon. These observations, together with the low Fe content of the bulk Moon ($\sim$7.5 wt\%) compared to that of the Earth (33 wt\%) previously inferred from geophysical observations, have spurred the development of new physical models for its generation.
Here we critically examine the geophysical and geochemical properties of the Moon in order to identify the extent to which dynamical scenarios satisfy these observations. New joint inversions of existing lunar geophysical data (mean mass, moment of inertia, and tidal response) assuming a laterally- and vertically homogeneous lunar mantle show that, in all cases, a core with a radius of 300$\pm$20 km ($\sim$~0.8 to 1.5 \% the mass of the Moon) is required. However, an Earth-like Mg\# (0.89) in the lunar mantle results in core densities (7800$\pm$100 kg/m$^3$) consistent with that of Fe-Ni alloy, whereas FeO-rich compositions (Mg\# = 0.80--0.84) require lower densities (6100$\pm$800 kg/m$^3$). Consequently, an FeO-rich Moon implies the existence of a core with light- (e.g., S, C) and/or exotic (e.g. ilmenite) components, whereas Earth's mantle-like compositions typically result in an elastically more rigid planet in contradiction with the observed tidal response and may require a partially molten layer surrounding the core.
Geochemically, we use new data on mare basalts to reassess the bulk composition of the Moon for 70 elements, and show that the lunar core likely formed near 5 GPa, 2100 K and $\sim$1 log unit below the iron-wüstite buffer, based on the depletions of Ni, Co, Mo and W in the lunar mantle. Moreover, the Moon is depleted relative to the Earth's mantle in elements with volatilities higher than that of Li, with this volatile loss likely having occurred at low temperatures (1400$\pm$100 K). These temperatures are also able to reproduce the extent of mass-dependent stable isotope fractionation observed in moderately volatile elements (e.g., Zn, K, Rb), provided equilibrium was attained between the vapour and the condensed phase, which, owing to the low temperatures, was likely solid. On this basis, and given the low $^{87}$Sr/$^{86}$Sr of ferroan anorthosites (FANs) and high $^{238}$U/$^{204}$Pb in mare volcanics, additional volatile loss relative to the Earth must have been associated with Moon formation. The identical nucleosynthetic (O, Cr, Ti) and radiogenic (W) isotope compositions of the lunar and terrestrial mantles (following subtraction of the late veneer from the latter), strongly suggest the two bodies were made from the same material, rather than from an Earth-like impactor. The W isotope homogeneity among lunar rocks, coupled with only mild $^{142}$Nd variations, indicates that lunar mantle differentiation occurred at $\sim$4350 Ma, providing a minimum age for the formation of the Moon, whereas Rb-Sr in FANs and Lu-Hf and Pb-Pb zircon ages point to $\sim$4500 Ma. 
Taken together, there is no unambiguous geochemical or isotopic evidence for the role of an impactor in the formation of the Moon, implying perfect equilibration between the proto-Earth and Moon-forming material or alternative scenarios for its genesis. 


\end{abstract}

\begin{keyword}
Moon \sep Origin \sep Composition \sep Structure \sep Evolution \sep Mare \sep Highlands
\end{keyword}

\end{frontmatter}


\textbf{Key Points}
\begin{itemize}
\item The relationship between the Moon and the Earth is explored through geophysical and geochemical evidence in order to assess the plausibility of dynamical models
\item The Moon's density and moment of inertia suggest it is made largely from silicate material with only a small ($\sim$300 km radius) core
\item An Earth's mantle-like Mg\# in the lunar mantle requires a core made of Fe-Ni metal, whereas lower Mg\#s result in a core rich in light- and/or exotic components
\item Elemental abundances in the bulk silicate Moon are determined for 70 elements
\item Endogenous core formation caused the depletion of Ni, Co, Mo and W in the lunar mantle at $\sim$5 GPa, $\sim$2100 K and $\sim$$\Delta$IW-1
\item Volatile depletion on the Moon affected elements more volatile than Li and likely occurred near 1400 K at- or near equilibrium
\item The Moon is required to have formed from already volatile-depleted precursors in order to explain its present day Sr- and Pb isotope compositions, which give a lunar formation age close to 4500 Ma.
\item The similarity in nucleosynthetic- and radiogenic W isotope compositions as well as in elemental abundances between the Earth and Moon strongly suggest the latter was derived from the Earth's mantle after core formation
\item Canonical giant impacts are dynamically common but struggle to account for the isotopic homogeneity between the Earth and Moon, while the frequency of half-Earth impacts depends on the accretion model invoked yet should engender isotopic similarity
\item There is no one dynamical model that satisfies all geochemical and geophysical properties of the Moon

\end{itemize}


\section{Introduction}
\label{sec:intro}


As the only major satellite orbiting the inner, rocky planets, the existence of the Moon is something of an anomaly. Although of a typical size for a moon in absolute terms, its mass, 1.2\% that of the Earth's, means that it is the largest with respect to that of its host of any planet in the Solar System (excepting the Pluto-Charon system). Its constitution, too, is atypical with the 1737~km-radius Moon being comprised chiefly of silicate material and only a small metallic core with a radius in the range 300--400~km \citep{khan_etal2014,matsumoto_etal2015,garcia_etal2019}, comprising $\sim$1--4\% of its total mass. This is in contrast to other inner solar system planets, and even smaller bodies such as the asteroid 4 Vesta, whose core makes up at least 18 \% by mass of the body \citep{russell_etal2012}. The high total angular momentum of the present-day Earth-Moon system, 3.5$\times$10$^{36}$~kgm$^2$s$^{-1}$, exceeds that for the other inner planets, but lies on a correlation with mass raised to the power of 2/3 defined by the gas giants and asteroids \citep{fish1967angular,hartmann1967angular}. These peculiar traits imply that equally enigmatic processes played a role in the Moon's origin.\\

Fortunately, the 382 kg of rock samples collected during the Apollo missions between 1969 and 1972 provide concrete records by which to test the prevailing theories on the formation of the Moon \citep{heiken_etal1991}. These hand-collected samples are supplemented by 0.2 kg from the unmanned Luna missions of the 1970s, as well as samples from the more recent Chang'e missions, the most recent of which, Chang'e 5, collected 1.7 kg of material in late 2020. Moreover, the lunar meteorite collection, at $\sim$1026~kg \citep{korotev_irving2021} now far outweighs that returned by the Apollo missions, though the nature of the lithologies remain largely representative of those in the returned sample collection. \\

In-situ geophysical characterisation of the Moon has been instrumental in defining the nature of the lunar interior. Surface instrumentation in the form of seismometers, heat flow probes, magnetometers, and surface reflectors that were placed on the Moon by the Apollo astronauts have done much to enhance our understanding of its interior  \citep{dickey1994lunar,wieczorek2006,khan_etal2013}. More than 12000 moonquakes were recorded over a period of $\sim$8~yrs (1969--1977) \citep{nakamura1983seismic,nunn2020lunar} that have yielded information on the lunar radial seismic structure \citep{toksoz1974structure,nakamura1982apollo,khanmosegaard2002,lognonne_etal2003,matsumoto_etal2015,garcia_etal2019}, including its compositional, thermal, and mineralogical constitution \citep{khan2006earth,khan_etal2014,kronrod_kuskov2011,kronrod_etal2022}. \\

Remote missions were also instructive, notably the Lunar Reconnaissance Orbiter (2009 - present) charged with determining precise radius- and spectroscopic information using the LOLA (Lunar Orbiter Laser Altimeter) instrument \citep{smith2010LOLA}, and the GRAIL mission \citep{zuber_etal2013gravity,Wieczoreketal2013} that made precise measurements of the gravity field of the Moon. These have been supplemented in recent years by the Chandrayaan orbiters designed primarily to spectroscopically examine the surface of the Moon \citep{goswami2009chandrayaan}, the Korean Pathfinder Lunar Orbiter, as well as various payloads on the Artemis-I mission, launched in 2022. \\

Although the mass of the Moon had been determined to a precision of 0.5 \% by the end of the 19$^{th}$ century \citep[see][]{hughes2002}, it was not until much later that \cite{daly1946} noted the similarity in the mean density of the Moon (modern value = 3345.56$\pm$0.40~kgm$^{-3}$) with that of the Earth's upper mantle ($\sim$3350~kgm$^{-3}$ over Moon-relevant pressures, 0--5~GPa). Moreover, the Moon's mass is nearly homogeneously distributed in its interior; its mean normalised moment of inertia (I/MR$^2$) is 0.393112$\pm$0.000012 \citep{konopliv_etal2001,garcia_etal2019}, similar to, but marginally lower than that for a homogeneous sphere for which I/MR$^2$=0.4. To first-order, these observations permit that the Earth's mantle and Moon are made from similar material with little- to no Fe-rich metal in its core. However, such an interpretation is non-unique, owing to the possible combinations of mineralogies that lead to similar results \citep[e.g.,][]{khan2006earth}. Nevertheless, combining multiple geophysical datasets, produce lunar models with compositions akin to that of Earth's mantle, but almost invariably with higher FeO contents \citep[8.1~wt\% for the Earth's mantle, compared with 9--13~wt\%;][and references therein]{khan2006earth,kronrod_kuskov2011,Dauphasetal2014} and point to the existence of a small liquid core (radius $<$350~km) of either pure Fe or an Fe-Ni-X alloy with S being the most likely candidate incorporated as a light element \citep{hood_etal1999Prospector,williams2001lunar,khan_etal2004Core,khan_etal2014,wieczorek2006,weber_etal2011seismic,garcia_etal2011VPREM,harada_etal2014,matsuyama_etal2016grail,morard2018liquid,zhao2023core}.\\

On the basis of the thick anorthositic crust that forms a carapace over $\sim$80\% of the lunar surface, the Moon was thought to be enriched in refractory lithophile elements (RLE) relative to the Earth's mantle \citep{taylorbence1975}. Downward revision of the crustal thickness in the vicinity of Apollo stations 12 and 14 from $\sim$60~km based on early interpretation of Apollo seismic data \citep{toksoz1974structure} to $\sim$35--40~km obtained from re-analysis of the seismic data \citep{khanmosegaard2002,lognonne_etal2003} and subsequently supported by the GRAIL mission \citep{Wieczoreketal2013}, implies a commensurate decrease in the abundances of the RLE, Ca and Al, which are concentrated in plagioclase-dominated anorthosites. This result, combined with the observations that the lunar nearside is anomalously rich in Th \citep[and by extension, other incompatible refractory elements;][]{Prettymanetal2006} and the crust more porous than previously supposed \citep{Wieczoreketal2013}, results in no significant difference, or possibly even a deficit \citep{siegler_etal2022} in the refractory lithophile element budget of the Moon relative to the Earth's mantle \citep{warren2005,Dauphasetal2014}.  \\

One of the key findings from the Apollo missions was the striking similarity in the chemistry of mare basalts to those of terrestrial basalts \citep{wanke_etal1977,ringwoodkesson1977}. This kinship extends not only to the refractory lithophile elements, but, surprisingly, also to some ratios among refractory siderophile and refractory lithophile elements - for example the W/La ratio \citep{wanke_etal1973,wanke_etal1977} - across both mare and highlands rocks. The curiosity of this ratio is that it is also lower, by a factor of $\sim$20, in terrestrial and lunar basalts compared to chondritic meteorites. While early models mooted that such a depletion in W may be caused by metal saturation upon eruption of mare basalts or during lunar core formation \citep{wanke_etal1973}, these hypotheses promptly fell out of favour owing to the consistency (within a factor of 2--3) of the depletion of other siderophile elements between terrestrial and low-Ti mare basalts \citep{ringwoodkesson1977,RammenseeWanke1977}, including P, V, Mn and Cr \citep{drake_etal1989}. Not only does this concordance imply the limited influence of lunar core formation on their abundances in the lunar mantle, but it also lends strong support to the idea that the Moon was derived from the Earth's mantle after its core formed. The abundances of Ni and Co, which are lower in lunar mare magmatic products relative to terrestrial oceanic basalts, are exceptions. Extraction into a lunar core at relatively oxidising (Ni-rich, 1 log$_{10}$-unit below the iron-wüstite buffer, $\Delta$IW-1) conditions at an assumed 1 bar pressure was proposed by \cite{delano1986} and later by \cite{oneill1991origin} to account for this depletion. More recent models show that lunar core formation must have occurred at pressures similar to those found at the centre of the Moon \citep[$\sim$4--5~GPa,][]{righter2002,RaiVanWestrenen2014} with lunar core masses roughly 2\% of its total mass. These models permit, though do not require, formation of an endogenous lunar core from material that initially resembled Earth's present-day mantle. \\

By virtue of high precision isotope measurements, the aforementioned elemental similarity was found to extend to the abundances of stable isotopes whose variations are caused only by nucleosynthetic processes. While the isotopic closeness between the two bodies had long been recognised for $\Delta^{17}$O \citep{ClaytonMayeda1975}, the $\sim$0.3 per mille (300~ppm) variability among lunar rocks permitted some differences between the Earth and Moon. More modern methods narrowed the isotopic difference between the Earth's mantle and Moon to within $\sim$5~ppm \citep{wiechert2001,young2016oxygen,greenwood2018} though recent estimates indicate as much as a 10 ppm difference \citep{Canoetal2020}. This contrast remains, in all cases, insignificant relative to the $\sim$7~$\permil$ range observed among planetary materials \citep{mittlefehldt2008oxygen}. Confirmation that the two bodies are indistinguishable among elements with a different geochemical character came from Cr \citep{qin_etal2010,mougel2018} and Ti \citep{trinquier_etal2009,zhang_ti2012} isotopes, that agree to within $\pm$4 ppm, a factor $\sim$150 smaller than that observed among planetary materials. These observations prompted models in which Moon-forming material equilibrated with that derived from Earth's mantle in the context of a giant impact \citep{PahlevanStevenson2007,Pahlevanetal2011,Locketal2018}. \\

Equally, however, these scenarios must also be able to explain the severe depletion in (moderately) volatile elements observed in mare basalts relative to their terrestrial counterparts \citep{ganapathy1970trace,tera1970comparative}. The volatile-poor nature of mare basalts was initially ascribed to both degassing upon eruption \citep{ohara1970mascons} and to an intrinsic feature of the lunar interior \citep{ringwood1970special}. Support for the latter hypothesis came via the uniformly low K/Th of the lunar surface ($\sim360$, \citealt{Prettymanetal2006}) compared to those of other terrestrial objects (between 2000 - 7000, \citealt{Peplowskietal2011}) as deduced from gamma-ray measurements.
While the alkali metals are depleted by a factor of 5--6 \citep[e.g.,][]{kreutzberger_etal1986}, more volatile elements (such as Zn, Ag, Cd, and In) are 100--500 times more impoverished in mare basalts than in terrestrial basalts \citep{wolfanders1980}. \cite{oneill1991origin} proposed that the abundances of these elements in the lunar mantle may result from the addition of a small (4\%) H-chondrite component added post-core formation, given their near-chondritic relative abundances. However, the heavy mass-dependent stable isotope composition in one such volatile element, Zn \citep[$\delta ^{66/64}$Zn = 1.2$\pm$0.3~$\permil$ in a range of mare basalts compared to the terrestrial mantle;][]{panielloetal2012} provides evidence that the low abundances of these elements reflect partial evaporation rather than complete loss and then overprinting. This is because evaporation is expected to result in the preferential loss of the lighter isotope of a given element \citep[e.g.,][]{langmuir1916evaporation}. These conclusions were subsequently confirmed by the similarly heavy isotope composition of another moderately volatile element, K \citep[$\delta ^{41/39}$K = 0.4$\pm$0.1~$\permil$;][]{wangjacobsen2016}. The degree of isotope fractionation of a given volatile element scales with the extent of its depletion relative to the Earth \citep{niedauphas2019}, which has been interpreted as resulting from a Rayleigh evaporation process, and is also observed in Cl \citep{boyceetal2018,garganoetal2020} and H \citep{Greenwoodetal2011}. The locus and timing of this loss remains an open question \citep{DayMoynier2014}, with scenarios during spreading of the protolunar disk \citep{canup2015, niedauphas2019} and from a magma ocean during lunar assembly at the Earth's Roche limit \citep{tangyoung2020,charnoz2021tidal} having been investigated. \\

Additional constraints on the timing of volatile loss of the Moon are afforded by the Rb-Sr and U-Pb systems. In the former, the parent, Rb, is a volatile element with 50 \% nebular condensation temperature, T$_c^{50}$, of 752 K \citep{woodetal2019condensation} while the daughter, Sr, is a refractory element (T$_c^{50}$ = 1548 K). The roles are reversed for U-Pb; the T$_c^{50}$ for U is 1609 K while that of Pb is 495 K. Consequently, $^{87}$Sr/$^{86}$Sr and Pb isotope ratios reflect time-integrated volatile depletion. The initial $^{87}$Sr/$^{86}$Sr defined by ferroan anorthosite and Mg-Suite rocks that have chondritic $^{143}$Nd/$^{144}$Nd and give concordant ages is 0.69905$\pm$0.00005 at their crystallisation ages of $\sim$ 4.35 Ga \citep{halliday_porcelli2001,borg_etal2022}. This is marginally higher than the Solar System initial value given by calcium-aluminium-rich inclusions (CAIs), 0.698975$\pm$0.000008 \citep{hans_etal2013rb}. Extensive Pb loss is also documented in mare basalts, which must have been formed from high-$\mu$ ($\mu$ = $^{238}$U/$^{204}$Pb) sources, between $\sim$300 and 4600, for low-Ti- and KREEP-basalts, respectively \citep{snapeetal2019,connellyetal2022}. This reflects a factor $\sim$35--500 lower Pb contents in the Moon relative to the Earth's mantle. Together, these observations imply that the source material that formed the Moon had already experienced volatile depletion in the first few Myr of Solar System evolution. 
\\

While initially thought to be direct cumulates from a lunar magma ocean \citep{Woodetal1970}, the ferroan anorthosites define ages that overlap with those of lithologies that are clearly metamorphic in origin \citep[the Mg-Suite; see][for a recent review]{borg_carlson2023}. The view that multiple magmas (rather than a single magma ocean) may have contributed to the generation of FAN, first suggested by \cite{Haskin_etal1982}, has been reinforced by the observed variability in their trace element patterns \citep{floss1998lunar,pernetfisher2019,ji_dygert2023}. Because FAN crystallised at 4.36 Ga, after $^{182}$Hf was extinct \citep[$t_{1/2}$ = 8.9 Myr;][]{Vockenhuber_etal2004}, lunar rocks do not contain any significant W isotope anomalies that are not spallogenic \citep{Toubouletal2015,Kruijeretal2015} while variations in $\varepsilon ^{142}$Nd are muted in comparison to those found in Martian shergottites, defining a whole-Moon isochron of 4.33 Ga \citep{borg_etal2019}, in agreement with internal isochron ages. However, the Moon exhibits a $\sim$ 25 ppm $^{182}$W excess with respect to the present-day Earth's mantle, which has been variously interpreted as reflecting a proportionally lower fraction of late veneer on the Moon relative to Earth \citep{kruijer2017tungsten,kruijer_etal2021NatGeo} or partial sequestration of W in to the lunar core while $^{182}$Hf was extant at  $\sim$75 Myr after \textit{t}$_0$ \citep{thiemens2019early,Thiemensetal2021}. The former scenario permits a young Moon formation age (possibly as young as 190 Myr after CAI) whereas the latter requires an older age (around 75 Myr after CAI). Therefore, it is unclear whether the ages given by highland rocks directly record that of the crystallisation of the magma ocean, or whether the time lag required for complete crystallisation \citep{Mauriceetal2020} and remelting to have occurred \citep[as is necessary for the generation of the Mg Suite magmas;][]{shearer2015MgSuite} permits older Moon formation ages. \\



\cite{darwin1880}, son of Charles, was the first to account for tidal evolution of the Earth-Moon system, demonstrating that the two would have initially existed as a single body with a rotation period of $\sim$4 hours. Modern mechanisms that account for both geochemical and geophysical constraints are typically rationalised in the context of a giant impact scenario for the formation of the Moon \citep{HartmannDavis1975,CameronWard1976}. Although these early models advocated for a Mars-sized impactor, a result later also favoured by \cite{CanupAsphaug2001} owing to its success in reproducing the angular momentum of the Earth-Moon system, such `canonical' impact events all had the commonality that most of the material that later accreted to form the Moon was derived from the impactor \citep{Canup2004}. Owing to the growing body of evidence that supports a close genetic link between the Earth and Moon, subsequent studies have been focused on finding solutions that satisfy both the angular momentum and the chemical / isotopic composition of the Moon. These efforts have spawned a variety of models \citep[see][for recent reviews]{Halliday2022, Canupetal2021}, including an impact on a fast-spinning Earth \citep{CukStewart2012}, a hit-and-run collision \citep{Reuferetal2012, Asphaugetal2021} a half-Earth impact \citep{Canup2012}, multiple impacts \citep{Rufuetal2017} or impact on Earth with a pre-existing magma ocean \citep{Hosonoetal2019}. Perhaps the most significant advance has come from the realisation that many high-energy impacts result in the formation of a supercritical fluid throughout much of the protolunar disk, giving rise to a structure, dubbed a synestia, that promotes mixing and equilibration \citep{Locketal2018}. \\

In order to understand what mechanism(s) gave rise to the unique chemical and physical properties of the Moon, we embark upon a focused review of observations that permit constraints to be placed on the origin of the Moon. To do so, in section \ref{sec:geophys} we use the latest geophysical data to assess the major element composition of the bulk Moon, together with the nature of its mantle and core. This approach minimises the ambiguity introduced by the use of highlands rocks or mare basalts in inferring the abundances of the major elements in the Moon. The likelihood that the Moon's bulk composition differs substantially from that of the Earth's mantle is discussed. In section \ref{sec:geochem}, we supplement these geophysical approaches with new estimations of the trace element content of the lunar mantle using legacy- and more recent measurements of their abundances in products of lunar magmatism. In section \ref{sec:physchem}, we summarise the recent advances made in the stable isotope geochemistry of lunar lithologies to better characterise the source material from which the Moon formed, and the pressure / temperature path it took to reach its present state. In section \ref{sec:timing}, isotopic constraints on the timing of accretion of the Moon and its relationship with the Earth are investigated, and, finally, in section \ref{sec:dynamics}, the foregoing information is combined with physicochemical constraints in order to infer likely scenarios for its origin.



\section{Geophysical constraints on the composition and interior structure of the Moon}
\label{sec:geophys}

\subsection{State of the art}
\label{sec:geophys_soa}

The wide lithological diversity recorded in lunar rocks, not only between the plagioclase-bearing highlands rocks and the basaltic maria, but also among low- and high-Ti mare basalts far exceeds that observed in the ocean basins on Earth. Models for their petrogenesis cite the presence of varying degrees of cumulate minerals in the source regions; high-Ti basalt sources contain greater fractions of ilmenite as compared to low-Ti basalts \citep{longhi1992,NealTaylor1992}. A similar trend is mirrored by the lunar pyroclastic glasses \citep{delano1986,Guenther_etal2022}. Other evidence for gross mineralogical changes in the lunar mantle comes from the highlands Mg-Suite, which are cumulates formed from magmas with higher Mg\#s than recorded by mare magmatism \citep{shearer2015MgSuite}. It follows that the chemical heterogeneity observed in the lunar crust reflects that in the lunar mantle. 
Therefore, and unlike the Earth, 
the interior of the Moon likely preserves large-scale mineralogical heterogeneity \citep[e.g.,][]{ElkinsTanton2011}. Indeed, there are no lunar equivalents to xenolith- or massif peridotites that, by examination of chemical trends therein, permit determination of a primitive mantle composition \citep{McDonoughSun1995,palmeoneill2014}. Though a single lunar dunite (72,415) occurs in the Apollo collection, and dunitic and peridotitic clasts have been identified in Apollo 14 breccias \cite{lindstrom1984,shervaismcgee1999,treimansemprich2023}, the degree to which they represent the lunar mantle is uncertain. The corollary is that estimation of the bulk Moon's major element composition using only lithologies present in the lunar crust is fraught with uncertainty \citep[cf.][]{Jones_Delano1989}. \\

In order to break the degeneracy introduced by a poorly-mixed lunar mantle in determining its bulk composition by geochemical means, a variety of geophysical methods have been employed that mainly rely on data obtained from surface instrumentation installed during the Apollo missions. This included seismometers \citep{nakamura1982apollo,lognonnejohnson2007,khan_etal2013}, heat flow probes \citep{langseth_etal1976,rasmussen_warren1985}, laser retroreflectors \citep{dickey1994lunar,williams_etal2006LLR}, and magnetometers \citep{sonett1982EM,hood_etal1982,khan2006emsounding,Mittelholzetal2021,grimm2024}. The virtue of each of these methods is that they are able to provide a global, mean estimate of density, seismic velocity, and electrical and thermal conductivity, germane to constraining the bulk properties of the Moon. \\

\subsubsection{Mantle}
\label{sec:geophys_mantle}
Determining the chemical composition of the interior of a planetary body from geophysical observations can be achieved by making a number of simplifying assumptions. Chief among these are the model chemical system invoked, the condition of chemical equilibrium, and compositional homogeneity. Firstly, geophysically relevant chemical elements of the silicate mantle are Ca, Fe, Mg, Al, Si, and O that make up more than 98.5\% of the Earth's silicate mantle (Palme and O'Neill, 2014). All elements are bound to oxygen and Fe is taken to exist solely as FeO, giving rise to a 5-component chemical system; CaO-FeO-MgO-Al$_2$O$_3$-SiO$_2$ (CFMAS). Secondly, through thermodynamic equilibrium, a bulk composition can be converted to a mineral assemblage by Gibbs Free Energy minimisation \citep[e.g.,][]{kuskov_kronrod1998,stixrudelithgow2005,connollykhan2016}\footnote{Initial studies attempting to infer the bulk composition of the Moon \citep[e.g.,][]{buck_etal1980,hoodjones1987,mueller_etal1988} relied on equations-of-state predictions for a restricted suite of mantle mineral assemblages and comparison to Apollo-era derived seismic velocity models.}. Lastly, the mantle is considered to be compositionally homogeneous on account of the resolving power of the geophysical data \citep[e.g.,][]{khan2006earth}. Other parameters that are disregarded include variations in mantle water content and oxidation state. Although there are indications for the presence of water in the mantle of the Moon \citep{Greenwoodetal2011}, its abundance is too low \citep[likely no more than 100s of ppm;][]{saal2008volatile,Haurietal2015,Haurietal2017,mccubbinetal2021endogenous} to measurably affect geophysical properties. The reduced nature of the lunar interior \citep{sato1976} means that Fe$_2$O$_3$ is also disregarded as a chemical component. \\ 

Initial studies \citep[e.g,][]{kuskov_kronrod1998}, relied on comparison with the ``final" Apollo-era seismic velocity model of \cite{nakamura1983seismic}, and returned bulk lunar mantle compositions with refractory-element enrichment over Earth's mantle \citep{kuskov_kronrod1998}, similar to that posited by \cite{Taylor1982book} on petrologic arguments, with lower MgO ($\sim$30~wt\%) and higher FeO ($\sim$11~wt\%) than for Earth's mantle ($\sim$37~wt\% and 8~wt\%, respectively). Subsequent reprocessing and inversion of the Apollo lunar seismic data using techniques not available during the Apollo era \citep{khan_etal2000,khanmosegaard2002,lognonne_etal2003,Gagnepain-Beyneix_etal2006} together with selenodetic constraints on bulk mean density and moment of inertia, resulted in Earth's mantle-like RLE abundances in the bulk Moon (Khan et al. 2006a,b). FeO contents, however, remained high at $\sim$10--13~wt\%. Subsequent extensions of these model inversions \citep[e.g.,][]{khan_etal2014,kuskov_etal2014,kronrod_etal2022} included more chemical elements, namely Na and Ti, the latter of which is abundant in the source regions of mare basalts, and confirmed earlier conclusions that the bulk lunar mantle is enriched in FeO (11--13~wt\%) relative to Earth's upper mantle, whereas there is less consensus about Al$_2$O$_3$ enrichment. 

\subsubsection{Core}
\label{sec:geophys_core}
As evidenced in the similarity of the Moon's normalised mean moment of inertia (MoI) with that of a homogeneous sphere, the lunar core is at the limit of detectability: mean lunar mass and MoI are compatible with both a small ($<$300~km in radius) Fe-Ni-bearing and a larger ($\sim$400~km) light element- or ilmenite-bearing core. Analyses of the electromagnetic sounding data acquired during the Apollo era are sensitive only to mantle structure \citep{hood_etal1982,khan2006emsounding,khan_etal2014,Mittelholzetal2021,grimm2024}, whereas the observation of the Moon's induced magnetic field are only able to place upper bounds (400~km in radius) on an (assumed) perfectly conducting central sphere \citep{russell_etal1982,hood_etal1999Prospector,shimizu_etal2013}. \\

Although more than 12000 seismic events were recorded by the Apollo lunar surface experiment package \citep[ALSEP;][]{nakamura_etal1981alsep}, only about 60 of these could be employed to infer the structure of crust, mantle, and core, because of the presence of the highly scattering regolith and generally low signal-to-noise ratio of the detected moonquakes and meteorite impacts \citep{toksoz1974structure,nunn2020lunar}. Moreover, because of the geographical distribution of the seismic sources and the stations, little-to-no seismic energy penetrated the central parts of the Moon \citep[see][for a summary]{khan_etal2013,khan_etal2014}. While there are claims that core-reflected seismic phases have been observed in the coda of the Apollo seismic records \citep{weber_etal2011seismic,garcia_etal2011VPREM,garcia_etal2012}, the issue is far from resolved, as it is yet to be demonstrated that consistent results across different studies are obtained. For example, \cite{weber_etal2011seismic} detected P-wave reflections from a putative molten layer, core-mantle-boundary (CMB), and inner-core-boundary, whereas \cite{garcia_etal2011VPREM} only identified S-wave reflections from the CMB, highlighting the difficulty of extracting seismic phases in the scattering-dominated lunar seismograms. Moreover, the discrepancies in their results illustrate the difficulties associated with the type of data processing undertaken by both teams: where \cite{garcia_etal2011VPREM} favour a core with a radius of 380$\pm$40~km with an outer liquid part, \cite{weber_etal2011seismic} find a 150~km thick partially molten mantle layer overlying a 330~km radius core, consisting of a 90~km thick outer liquid and 240~km solid inner core, respectively. \\

Other geophysical evidence for a liquid core comes from analysis of $\sim$40~yrs of lunar laser ranging (LLR) data that measure the travel time between an observatory on the Earth and retroreflectors on the lunar surface installed as part of the Apollo and Soviet Lunokhod missions \citep[e.g.,][]{dickey1994lunar}. Through detection of the displacement of the Moon’s pole of rotation, which indicates that dissipation is acting on the rotation, \cite{williams2001lunar} noted that only a combination of dissipation due to monthly solid-body tides raised by the Earth and Sun and a fluid core (350--380~km in radius) with a rotation distinct from that of the solid body are able to account for the LLR data. Subsequent analyses of the amplitude and/or phase of the lunar tidal response are all compatible with a region of reduced rigidity in the deep lunar interior, which, following the initial suggestion by \cite{nakamura_etal1973} from seismic observations, has been interpreted as a partially molten layer sitting atop the core \citep[e.g.,][]{khan_etal2004Core,khan_etal2014,harada_etal2014,matsumoto_etal2015,matsuyama_etal2016grail,tan_harada2021,xiao_etal2022}. \\

While a consensus appears to have been achieved with regard to lunar core size with radii in the range 300--400~km and possibly a narrower range of 300--350~km, there is less agreement on core properties (P-wave velocity and density) as a consequence of the lack of direct sensitivity. This is reflected in the current range of core densities, spanning from 3900 to 7000~kg/m$^3$, corresponding to core compositions covering those expected for dense silicates/oxides over the Fe-FeS eutectic to pure Fe \citep[see][for a recent summary]{garcia_etal2019}.

\subsection{Uncertainties in geophysical constraints on the lunar interior}
\label{sec:geophys_unc} 
The pressure at the center of the Moon, $\sim$55~kbar, is sufficiently low so as to facilitate comparison of the physical properties of common mantle minerals, such as olivine, orthopyroxene and clinopyroxene, with those recovered from geophysical estimates, 
as they overlap with the accessible range in large-volume press apparatuses (piston cylinder, multi-anvil) and diamond anvil cells germane to spectroscopic measurements \citep[e.g.][]{hazen1976effects,angel_hughjones1994}. Thermodynamic databases \citep{robiehemingway1995,stixrudelithgow2005,hollandpowell2011} are employed in order to convert a bulk chemical composition into a mineral assemblage, and assign to it physical properties that can be used in a forward sense \citep[e.g.,][]{kronrod_kuskov2011} or inverted jointly with geophysical data \citep[e.g.,][]{khan2006earth,khan2006emsounding}. While thermodynamic equilibrium remains a convenient assumption, petrologic evidence indicates that disequilibrium, or igneous layering may exist in the lunar interior (section \ref{sec:geophys_soa}). Inversions of the composition of Earth's upper mantle assuming it can be described by mechanical mixtures of lithologies, highlight differences to equilibrium models \citep{munch_etal2020}, however, such approaches, as yet, remain untested for the Moon.    \\

\begin{figure}
  \begin{center}
\noindent\includegraphics[width=0.5\textwidth,angle=0]{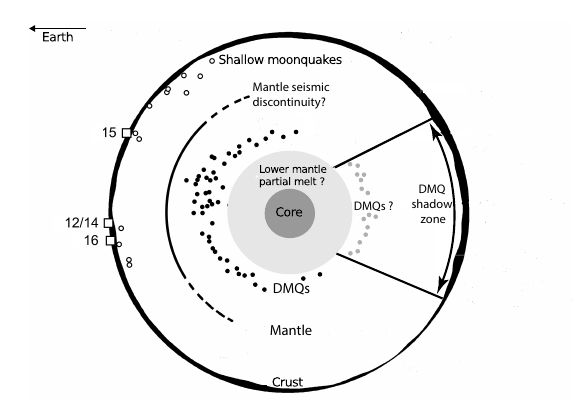}
  \end{center}
\caption{Schematic diagram of the internal structure of the Moon as viewed from geophysics. The Moon is differentiated into crust, mantle, and core. A mid-mantle seismic discontinuity -- a potential relic of magma ocean crystallisation in the form of a compositional boundary -- is not required by data. Geophysical evidence suggests the presence of a deep-seated partially molten layer, although whether the layer is located in the deep mantle or belongs the core is unclear.
The core is small and likely consists of a liquid metallic Fe-Ni alloy, plausibly together with a light element component (e.g., S or C, see section \ref{sec:geophys_inv_core}). The locations of the Apollo lunar seismic stations 12, 14, 15, and 16 on the lunar nearside are indicated by numbers. Shallow and deep (DMQ) moonquakes occur in the depth ranges 50--200~km and 800--1100~km, respectively. Farside DMQs have yet to be confirmed. The selenographical location of the indicated moonquakes is not accurate. Variations in crustal thickness are real but not to scale. Modified from \cite{wieczorek2006} and \cite{khan_etal2014}.}
\label{fig:moondiagram}
\end{figure}

Existing models devised within the aforementioned framework that satisfy geophysical constraints point to an FeO-enriched Moon relative to Earth's upper mantle \citep[e.g.,][]{khan2006earth,kronrod_etal2022}. Despite the richness of selenodetic data, there are, nevertheless, significant uncertainties in interpreting geophysical observations that prevent an unequivocal assessment of the composition of the lunar mantle. Notably, the quality of the Apollo-era seismic data, owing partly to the poorly-consolidated lunar regolith and partly to the distribution of moonquakes, are only well-resolved to depths of $\sim$1100~km \citep{nakamura1983seismic,khanmosegaard2002}. Related hereto is the question of layering in the lunar mantle. There are no seismological requirements for vertical layering to exist\footnote{The ``final" Apollo-era lunar seismic velocity model of \cite{nakamura1983seismic} contained a mid-mantle seismic discontinuity, but was only included because of computational convenience. The Apollo lunar seismic data are entirely compatible with a homogeneous mantle without large-scale radial velocity variations \citep{khan2006earth}.}, despite petrological observations from mare basalts for at least TiO$_2$ variations \citep[e.g.,][]{NealTaylor1992}.
Consequently, seismic wave speeds in the lunar interior are essentially constant as a function of depth \citep[e.g.,][]{khan2006earth} and the lunar mantle can be considered mineralogically homogeneous with respect to the level of accuracy and precision afforded by the Apollo Lunar seismic data, 
with the caveat that the seismic waves appear to be attenuated below $\sim$1100~km depth in the Moon (e.g., Nakamura et al., 1973). Indeed, a low-viscosity and 
highly dissipative zone in the lowermost lunar mantle has been mooted on the basis of LLR data \citep{harada_etal2014,khan_etal2004Core,khan_etal2014,matsumoto_etal2015}, but there are no essential requirements for differences in composition. The presence or absence of a liquid layer is largely dictated by the nature of the lunar viscoleastic response \citep{nimmo2012dissipation,efroimsky2012tidal,karato2013,harada_etal2014,walterová2023semimolten}, which cannot, as yet, be uniquely determined. For completeness, we note that a recent study relying on selenodetic constraints has indicated the existence of a lunar solid inner core with a radius of around 260~km \citep{Briaudetal2023}, yet considerable ambiguity remains as to the nature of the supporting evidence. A summary view of our current understanding of the interior of the Moon based on the geophysical evidence is shown in Figure~\ref{fig:moondiagram}. 

\subsection{Effect of mantle composition on interior structure}
\label{sec:geophys_inv} 
For the purpose of determining how mantle composition affects interior structure, we examine the case of a homogeneous lunar mantle composition, in which equilibrium is assumed in order to permit Gibbs Free Energy minimisation to sample the stable mineralogical assemblage, given the prevailing composition, pressure and temperature. Mantle compositions are modelled using the model chemical system CaO-FeO-MgO-Al$_2$O$_3$-SiO$_2$-TiO$_2$ after \cite{khan_etal2014}, assuming Earth's mantle-like refractory element contents and ratios; Al$_2$O$_3$ = 4.49~wt\% and CaO = 3.65~wt\% and TiO$_2$ = 0.21~wt\% with MgO, FeO and SiO$_2$ abundances that depend on the chosen model (Table~\ref{tab:inv_com}). In addition to the compositions compiled from the literature, we also considered a variable mantle compositional model, where we inverted for the major element components (including refractory oxides) following \cite{khan_etal2014}.
As for the selenotherm, we consider a range of variable mantle temperatures that span from sub- to super-adiabatic conditions. The core, owing to the scarcity of geophysical data sensitive to this region, is modelled as a homogeneous sphere of constant density. The inversions are designed to match the mean mass, mean moment of inertia, and tidal response in the form of the elastic degree-2 Love number\footnote{The tidal response of the Moon is formally quantified by Love numbers that depend upon the spherical harmonic degree and order of the tide-raising potential, where the ratio of the induced potential to the tidal potential is denoted by the Love number $k$. Here, we shall restrict ourselves to the amplitude of the degree-2 tide, i.e., $k_2$. The degree-2 Love number considered here ($k_2$=0.0232$\pm$0.00021) is that of \cite{khan_etal2014}, which represents the elastically-corrected value of \cite{williams_etal2014} obtained from Doppler tracking of GRAIL. The Love numbers are in the general case complex, consisting of a real part, which corresponds to the elastic response, and an imaginary part that describes the phase lag, which is proportional to dissipation within the Moon. To avoid having to introduce model-dependent visco-elastic parameters, only the elastic part of $k_2$ is considered.}, which help constrain the density and rigidity structure of the mantle, in addition to core density. Geophysical predictions are calculated according to the methods outlined in \cite{khan_etal2014} and the observations are reported in Table~\ref{tab:inv_com}. 
Uncertainties in density and elastic moduli computed using Gibbs free-energy minimisation are $<$0.5\% and $<$1\%, respectively \citep{connollykhan2016}. Pressure is computed by integrating the load from the surface assuming hydrostatic equilibrium. Finally, to avoid complexities associated with the lunar seismic data, we do not consider the latter as part of the inversion. Instead, we verified that the models were in general agreement with the seismic arrival time data set of \cite{nunn2020lunar} within 2--3 standard deviations about the mean value.\\

\begin{table*} \scriptsize
\centering
\caption{Radius and density of the lunar core, and the potential temperature of the mantle constrained by geophysical inversions for fixed bulk compositions and a model in which the mantle has a variable bulk composition. Uncertainties are 1 standard deviation about the mean. The geophysical observations considered include the degree-2 Love number ($k_2$), mean mass ($M$), and mean normalised moment of inertia ($I/MR^2$) after \cite{garcia_etal2019}.}
\begin{tabular}{l c c c c c c c}
 Model & R$_{\rm{core}}$ [km] & $\rho_{\rm{core}}$ [kg/m$^3$] & $f_{\rm{core}}$ & T$_{\rm{pot}}$ [K] & FeO [wt\%] & MgO [wt\%] & SiO$_2$ [wt\%]\\
\hline
Dauphas et al. (2014) & 317$\pm$7 & 7004$\pm$222 & 0.0127$\pm$0.0006 & 1008$\pm$68 & 10.6 & 34.3 & 46.8\\
Khan et al. (2006) & 315$\pm$8 & 6679$\pm$238 & 0.0119$\pm$0.0007 & 1051$\pm$60 & 12 & 34.3 & 45.5\\
O'Neill (1991) & 307$\pm$14 & 6305$\pm$525 & 0.0104$\pm$0.0008 & 1011$\pm$67 & 12.1 & 32.8 & 46.8\\
Taylor (1999) & 284$\pm$24 & 6036$\pm$728 & 0.0078$\pm$0.0011 & 976$\pm$57 & 13.1 & 31.7 & 46.8\\
Warren (2005) & 324$\pm$2 & 7301$\pm$58 & 0.0141$\pm$0.0003 & 1067$\pm$51 & 9.1 & 35.8 & 46.8\\
Variable composition* & 289$\pm$18 & 6292$\pm$812 & 0.0087$\pm$0.0017 & 1013$\pm$103 & 13.7$\pm$0.9 & 30.7$\pm$3.2 & 47.2$\pm$3.9\\
Earth & 322$\pm$2 & 7877$\pm$94 & 0.0149$\pm$0.0003 & 1117$\pm$39 & 8.1 & 36.8 & 44.9\\
\hline

\end{tabular}

$k_2$ = 0.0232$\pm$0.00021\\
$M$ = (7.34630$\pm$0.00088)$\times$10$^{22}$ kg\\
$I/MR^2$ = 0.393112$\pm$0.000012\\
$R$ = 1737.161~km \\
$^*$Al$_2$O$_3$ and CaO varied between 4.59$\pm$2.00 wt~\% and 3.57$\pm$1.03 wt~\%, respectively.
\label{tab:inv_com}
\end{table*}

 \begin{figure*}
     \centering
     \includegraphics[width=\textwidth]{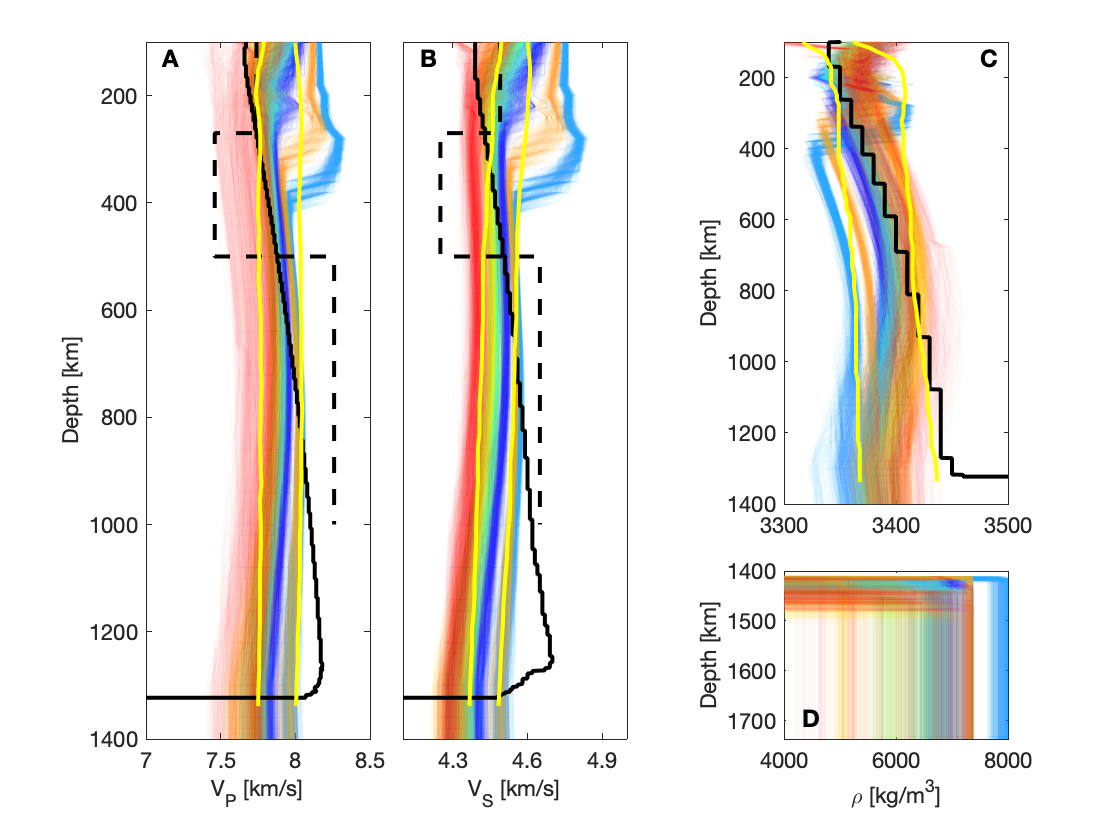}
     \caption{Inverted seismic P-wave (A) and S-wave (B) speed and density (C, D) profiles that fit the lunar tidal response and mean mass and moment of inertia for the range of compositional models listed in Table~\ref{tab:inv_com}. The inverted profiles correspond to \cite{Dauphasetal2014} (blue), Earth's mantle-like after \cite{palmeoneill2014} (light blue), \cite{khan2006earth} (cyan), \cite{oneill1991origin} (light green), \cite{Taylor1999} (light yellow), \cite{warren2005} (orange), and variable composition (red). Core P- and S-wave speeds are unconstrained and therefore not shown. 
     For comparison, models obtained from the Apollo lunar seismic data are shown as yellow \citep{khan2006earth} and black solid \citep[Model M1;][]{garcia_etal2019} and dashed lines \citep{nakamura1983seismic}, respectively. For the \cite{khan2006earth} model, uncertainties are included in the form of upper and lower bounds. Core density is not well constrained by these models and therefore omitted. See main text for details.}
     \label{fig:profiles}
 \end{figure*}

Seismic P- and S-wave speed and density profiles for all 7 mantle compositional models are shown in Figure~\ref{fig:profiles}. For comparison, the ``final" Apollo-era model of \cite{nakamura1983seismic} and the more recent models of \cite{khan2006earth} and \cite{garcia_etal2019} are also shown \citep[in the case of][we focus on their M2 model]{garcia_etal2019}. While the model of \cite{nakamura1983seismic} contains discontinuities, these are, as already discussed, not required by the seismic data and were only included for computational convenience. More generally, pressures inside the Moon are not high enough to cause mineral phase transformations and associated seismic discontinuities. In contrast, the models of \cite{khan2006earth} and \cite{garcia_etal2019} assume mantle homogeneity and are parameterised in terms of the thermo-chemical properties of the equilibrium mineral assemblage and the Adams-Williamson equation-of-state in combination with Birch's law, respectively, and are inherently smooth. In the model of \cite{khan2006earth}, as with the 7 compositional models used here, P- and S-wave speeds are seen to decrease toward the bottom of the mantle, which results from the decrease in seismic speed associated with increasing temperature with depth dominating over the opposing increase in wave speed associated with the increase in pressure with depth. The model of \cite{garcia_etal2019}, in contrast, increases from the top to the bottom of the mantle, which results from the particular modelling assumptions made (e.g., P- and S-wave speeds are obtained from density using constant, i.e., depth-independent, scaling factors and thermodynamic parameters). \\

Relative to \cite{nakamura1983seismic}, the compositional models are slightly faster in the upper (300--500~km depth) and slower in the lower part ($>$500~km) of the mantle, but are in good agreement with the models of \cite{khan2006earth}, and are, with the exception of the low mantle FeO, that is, the Earth- and \cite{warren2005} models, therefore also in general agreement with the Apollo lunar seismic data inasmuch as the models of \cite{khan2006earth} match the latter. 
In terms of density, there is good agreement with the models of \cite{khan2006earth} and \cite{garcia_etal2019}. 
In summary, the model variability displayed in Figure~\ref{fig:profiles} serves to underscore the lack of, and need for, improved constraints on the deep lunar interior, preferably through the acquisition of novel seismic data as envisaged with the return of seismology to the Moon with for instance, the upcoming Farside Seismic Suite mission \citep{panning_etal2022}. \\

\begin{figure}
     \centering
     \includegraphics[width=0.45\textwidth]{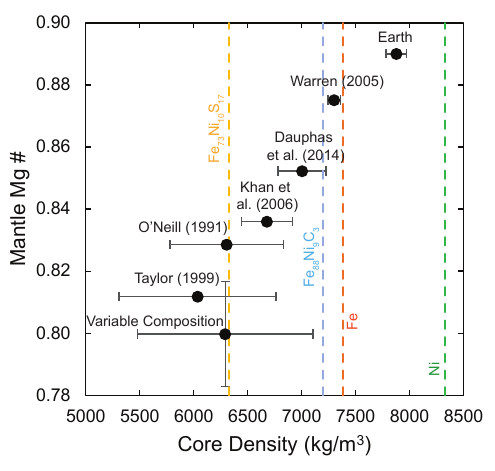}
     \caption{The relationship between Mg\# of the lunar mantle and the density of its core required to satisfy its bulk density and moment of inertia. Mantle compositions are fixed, except for the `Variable Composition' case (see Table~\ref{tab:inv_com}). Note that the Mg\#-core density distribution of the Variable Composition case is not Gaussian. Vertical lines correspond to densities of potential core-forming materials at 1900 K and 5.5 GPa, Yellow = Fe$_{73}$Ni$_{10}$S$_{17}$ (in at~\%) liquid, \cite{terasaki_etal2019}; Blue = Fe$_{88}$Ni$_{9}$C$_{3}$ (in at~\%) liquid, \cite{zhu2021density}; Orange = Fe liquid, \cite{andersonahrens1994}; Green = Ni liquid, \cite{naschmanghnani1998}.}
     \label{fig:mgno_density}
 \end{figure}

The density structure of the mantle is nearly indistinguishable among the 7 compositional models investigated here (Figure~\ref{fig:profiles}C--D), yielding values around 3400~kg/m$^{3}$, with higher Mg\# compositions yielding slightly lower average densities. Mantle potential temperatures derived from this region of the lunar mantle show a weak tendency for higher \textit{T$_{pot}$} in Mg-rich compositions (Table~\ref{tab:inv_com}). Although the core radii returned by the various mantle compositions all overlap within a small range (280--320~km), differences in core density arise, which are positively correlated with mantle Mg\# (Fig.~\ref{fig:mgno_density}). This dependence results from the requirement to match the mean mass and moment of inertia of the bulk Moon. Because the geophysical data together result in a relatively constant core radius, the greater mass deficit for marginally less dense, high-Mg\# mantle compositions must be compensated for by a relatively denser core 
to produce the observed mass of the Moon (Fig. \ref{fig:mgno_density}). \\

Not all models satisfy the elastic degree-2 tidal Love number of the Moon, with the more rigid, Mg-rich lithologies uniformly underpredicting \textit{k$_2$}, with values near 0.0210, compared to the observed value considered here (0.0232). In contrast, the FeO-rich compositions of Taylor (1999) and the variable-compositional model with $\ge$13~wt\% FeO (Table \ref{tab:inv_com}) are entirely consistent with the observed tidal response. Nevertheless, the degree-2 tidal Love number is not an unambiguous indicator of composition, owing to the i) weak differences in elasticity among silicate minerals as a function of their Mg\# \citep[e.g.,][]{speziale_etal2004}
and ii) the possible presence of a core-enveloping molten layer\footnote{The partially molten lower-mantle layer originated with the observation that strong shear wave arrivals from deep moonquakes were
only observed at the two closest stations of the four-station Apollo lunar seismic network, corresponding to bottoming depths of $\sim$1100~km, from what was then the only located farside deep moonquake. Waves having travelled deeper lacked the prominent shear wave arrivals and were inferred to have been attenuated as they transited a partially molten region \citep{nakamura_etal1973}. Moreover, the curious lack of farside moonquakes has long been considered to be an indication of an attenuating, possibly molten, region in the deep lunar interior \citep{nakamura2005}. Inspite hereof, the partially molten layer remains debated \citep[see][for recent summaries]{khan_etal2014,walterová2023semimolten}.} that can, without any chemical contrast with the surrounding solid mantle, produce an increase in $k_2$ sufficient to account for the observed value \citep[e.g.,][]{khan_etal2014,matsumoto_etal2015,garcia_etal2019}. Hence, although a solid lunar mantle with an Earth-like Mg\#  does not readily reproduce the $k_2$ of the bulk Moon, such compositions cannot be excluded based on \textit{k$_2$} alone. \\

In this context, two of the seven P- and S-wave speed models, Earth and \cite{warren2005}, contain negative velocity discontinuities in the upper mantle (at 300 and 400~km depth, respectively) that are an artefact resulting from a trade-off with temperature. Based on this observation, compositionally homogeneous models of the lunar mantle containing an Earth-like FeO budget are unlikely to be plausible, yet radially-heterogeneous models, where FeO content possibly varies with depth so as to be consistent with a bulk FeO content in the range 8--9~wt\%, for example, can not be excluded per se. The nature of the resultant seismic velocity profiles will, however, have to be tested further against the Apollo lunar seismic data, although the latter are unlikely to constrain compositional complexity beyond homogeneity. Thus, while geophysical evidence largely appears to favour a compositionally-uniform lunar mantle with an FeO content in excess of that of the Earth's, more complicated scenarios that act to produce an overall Earth-like composition cannot, at present, be ruled out.\\

Mare basalts, whose Mg\#s are in the range 0.45--0.55 \citep{NealTaylor1992}, require mantle sources that are more FeO-rich (Mg \# $\sim$0.71--0.79) than any mean bulk Moon composition tested in Table \ref{tab:inv_com}, given an olivine-melt K$_D^{Fe-Mg}$ = 0.33$\pm$0.02 at low TiO$_2^{liq}$ \citep{longhietal1978}. The most magnesian melt compositions, the Apollo 15 green glass clods, have inferred source regions with Mg\# of 0.84--0.87 \citep{delano1986} and come from deeper regions in the lunar mantle, 2.0--2.5 GPa, than do mare basalts, 0.5--1.5 GPa \citep{longhi1992,elkins2003experimental,barr2013experimental}, pointing to significant mineralogical heterogeneity. Higher Mg\# with increasing depth in the Moon is consistent with layering resulting from a magma ocean crystallisation sequence \citep{snyder1992chemical}.
Were Mg-rich compositions to exist predominantly in the lower mantle (as required for bulk lunar Mg\#s $\sim$0.89), this would imply that overturn \citep{hessparmentier1995, Elkinsetal2002} and mixing was either inefficient, or did not operate to excavate Mg-rich regions of the lunar mantle. Indeed, spectroscopic data \citep{Melosh2017,yamamoto2023,sunlucey2024lunar} indicate that low-calcium pyroxene-rich lithologies with Mg\#s $\sim$0.85 are abundant in the South Pole-Aitken basin, while olivine is present at much lower modal abundances (typically $<$4 \%) than in peridotitic upper mantle ($\sim$60 \%). Because this region is thought to have been formed through an impact that excavated portions of the lunar mantle down to 100 km depth \citep{Melosh2017}, such lithological characteristics may indicate incomplete (or no) overturn, as the lowermost mantle is expected to be olivine-rich \citep{Elkinsetal2002}. 

\subsection{Implications for core composition}
\label{sec:geophys_inv_core} 
The densities of the inverted core layer (Figure~\ref{fig:profiles}D) range from $\sim$6000$\pm$800 kgm$^{-3}$ in the most iron-enriched mantle compositions, to $\sim$7800$\pm$100 kgm$^{-3}$ for a composition equivalent to that of Earth's mantle. This diversity has significant implications for the inferred chemical composition of a nominal lunar core. Chemically, the lunar core is expected to be an alloy of Fe-Ni-rich metal together with some fraction of light components (H, C, O, S, Si), by analogy with the compositions of cores of terrestrial planets \citep[e.g.,][]{badro_etal2014,khan2022geophysical,Khan_etal2023,Samuel_etal2023}, as well as of iron meteorites \citep[e.g.,][]{scottwasson1975}. \\

In this framework, alloys that provide matches to the core of bulk density $\sim$6250~kgm$^{-3}$ recovered by \cite{weber_etal2011seismic}, require substantial amounts of S (10~wt\%) or Si (18~wt\%) \citep{terasaki_etal2019}. 
Silicon-containing alloys 
are strongly disfavoured on the grounds that lunar core formation would have taken place under relatively oxidising conditions \citep[near $\Delta$IW-1, e.g.,][section \ref{sec:physchem_corefm}]{oneill1991origin}, precluding the incorporation of significant amounts of Si into the core. Oxygen, like Si, enters metal at very high temperatures \citep{oneill_etal1998,gendre_etal2022}, also making it an unlikely candidate. Another cosmochemically abundant light element that may be present in the lunar core is C. However, C is highly volatile during condensation of the solar nebula \citep[e.g.,][]{woodetal2019condensation} and has low solubility in silicate melts \citep[e.g.,][]{yoshioka_etal2019}, such that it is unlikely to be sufficiently abundant in the bulk Moon to represent a major constituent of its core. Indeed, a similar conclusion was reached by \cite{righter_etal2017}, who calculate 3750~ppm C in the lunar core based on geochemical data in mare basalts and pyroclastic glasses \citep{bombardieri_etal2005,wetzel_etal2013} that yield a mantle abundance of $\sim$5~ppm. Although \cite{steenstra_etal2017} propose that the lunar core could host 0.5--4.8~wt\%~C, values at the upper end of this range are unlikely given that the C/Nb ratios in melt inclusions in pyroclastic glass 74220 are depleted by a factor $\sim$200 with respect to mid-ocean ridge basalts \citep{mccubbinetal2021endogenous}. The most plausible candidate, therefore, appears to be S, which is present in the lunar mantle at levels of 88--202~ppm \citep{dingetal2018,mccubbinetal2021endogenous,garganoetal2022}. \\

\begin{figure}
     \centering
     \includegraphics[width=0.5\textwidth]{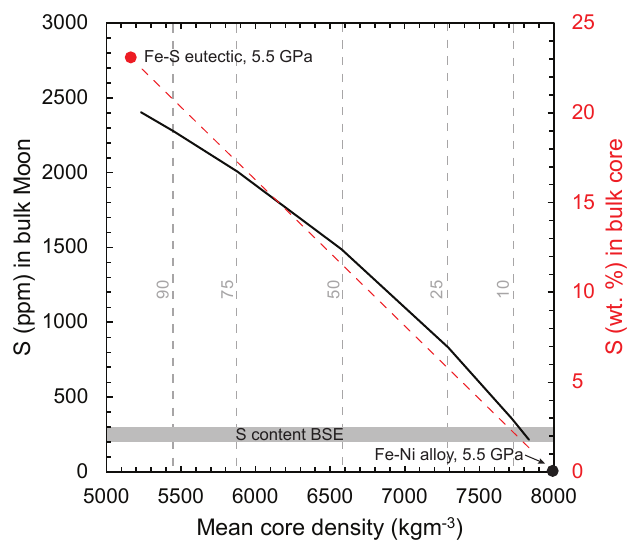}
     \caption{Calculated evolution of density of the lunar core (at 5.5 GPa) assuming it is produced from an ideal (i.e., mechanical) mixture of Fe-Ni alloy of density 8000 kgm$^{-3}$ (black point) and Fe-S eutectic liquid (77 wt. \% Fe, 23 wt. \% S) of density 5171 kgm$^{-3}$ (red point) and a constant total radius of 330 km (see Table \ref{tab:inv_com}). The total core mass fraction thus varies from 1.74 \% (pure Fe-Ni) to 0.46 \% (pure Fe-S). The solid black curve refers to the S content of the bulk Moon (in ppm) while the red dashed line denotes the S content of the bulk core (in wt. \%). Dashed vertical lines represent the modelled percentages of Fe-S eutectic composition that constitute the core. The S content of the bulk silicate Earth \citep[BSE,][]{palmeoneill2014} is shown as a horizontal bar. }
     \label{fig:S_core}
 \end{figure}

Taking these broad expectations into account, Figure~\ref{fig:mgno_density} overlays the densities of typical core-forming compounds (liquid Fe, Ni, Fe-Ni-C and Fe-Ni-S) on the densities constrained by the inversions (section~\ref{sec:geophys_inv}). The iron-rich mantle compositions of \cite{Taylor1999} or \cite{oneill1991origin} produce cores that overlap with the S-rich compositions shown on Figure~\ref{fig:mgno_density}. Such mantle compositions would imply a lunar core made almost entirely ($\sim$75~\% with the remainder being Fe-Ni alloy) of a eutectic Fe-S liquid, which, at 5.5~GPa, contains $\sim$23~wt\% S \citep{BuonoWalker2011}. Indeed, most of the two-layer geophysical models assume a mixture of a solid, Fe-Ni inner core together with a lower density, liquid outer core with the composition of the Fe-S eutectic \citep{weber_etal2011seismic,garcia_etal2011VPREM,williams_etal2014}. 
This amount of S, given the corresponding mass of the core, yields a bulk Moon S content of $\sim$2500 ppm (Figure~\ref{fig:S_core}). This compares with 250 ppm S in the BSE \citep{palmeoneill2014}, and the $\sim$7800 ppm in the bulk Earth, given a maximum of 2.4 wt\% S in the Earth's core \citep{dreibuspalme1996,badro_etal2014,hirose_etal2021}. \\

Any of the bulk Moon compositions are significantly poorer in Fe (6.3--10.6~wt\%; Table \ref{tab:inv_com}) than the bulk Earth (33~wt\%), indicating that the Moon is akin to a silica-rich fragment of a differentiated body. That is, it does not have a `chondritic' abundance of iron \citep[18--30~wt\% Fe,][]{wassonkallemeyn1988}. Therefore, although the Moon is often described as being an 'oxidised' body owing to its high FeO/Fe ratio, this description fails to capture the fact that it is depleted in iron with respect to the other terrestrial planets and chondrites, and cannot alone be formed by oxidising the majority of the iron present in an otherwise chondritic precursor. Rather, the Moon has been stripped of a metallic component. Given the strongly sidero-/chalcophile behaviour of sulfur, it follows that it must also be missing an S-rich component. If we consider, as a minimum estimate, that S is as siderophile as Fe, and given that the Moon has lost 3/4 of its Fe compared to an Earth-like precursor, then it should have no more than 2000 ppm S, which would allow for the Fe-rich mantle compositions predicted by the models of \cite{oneill1991origin} or \cite{Taylor1999}. However, this S content is likely to be a maximum estimate for two reasons, i) S is far more depleted in the BSE than is Fe, with some 96\% of its total budget in the Earth residing in the core ($\sim$87 \% for Fe) and ii) S, as a moderately volatile element, is likely to be as depleted in the Moon as elements with comparable condensation temperatures, such as Zn (factor $\sim$50 depletion relative to the BSE). While the lunar mare basalts themselves are only poorer in S by a factor of $\sim$4 relative to terrestrial basalts \citep[e.g.,][section \ref{sec:geochem_compmoon}]{dingetal2018}, this would nevertheless imply a maximum of 500 ppm S in the bulk Moon \citep[see also][]{righter_etal2017,steenstra_etal2017}. Indeed, were the Moon to have the bulk composition of the BSE (250 ppm), then Fig. \ref{fig:S_core} shows its core should have no more than $\sim$2~wt\% S (for a core mass fraction of 0.0125). If true, this would cast doubt on a ferroan lunar mantle composition, as this, in turn, would imply a low density core ($\sim$6000$\pm$800 kgm$^{-3}$, Fig. \ref{fig:mgno_density}), for which an appropriate light ingredient is lacking. \\

The presence of sulfur in the lunar core would also have left a discernible trace in the abundances of chalcophile metals in the complementary mantle \citep[e.g.,][]{mann_etal2009,wood_etal2014sulfur}. Given experimentally determined metal-silicate partition coefficients, sulfur-rich cores ($\sim$6 wt~\%) appear more consistent with the lunar mantle abundances of non-volatile moderately siderophile elements \citep[namely, Mo, W, Ni and Co;][]{schmitt1989,RighterDrake1996,RaiVanWestrenen2014,steenstra_etal2016}, though S-free models also provide a satisfactory solution within the uncertainties of the experimental data.  
Hence, at present, the partitioning of siderophile elements is unable to provide conclusive constraints on the nature or abundance of any potential light elements in the lunar core.  \\

Alternative postulates consider the possibility that the lunar core does not comprise solely Fe-Ni-S, but other, more exotic components, such as ilmenite \citep[e.g.,][]{khanmosegaard2001,devries_etal2010}. As ilmenite is among the densest phases (4700 kgm$^{-3}$) in the lunar mantle, it has a tendency to sink \citep[e.g.,][]{hessparmentier1995,Elkinsetal2002}, and TiO$_2$ is sufficiently abundant ($\sim$ 0.2--0.3~wt\%) for ilmenite to comprise some fraction of a lunar core. Indeed, geophysical studies \citep[e.g.,][]{khan_etal2014,matsumoto_etal2015} show that a TiO$_2$-rich layer in the lower mantle of the Moon is consistent with a range of geophysical data, and propose that the ilmenite-rich layer could be the seat of partial melt that had been proposed on the basis of geophysical observations \citep[e.g.,][]{nakamura_etal1973,nakamura_etal1974, williams2001lunar}. As ilmenite-bearing cumulates could have incorporated heat-producing elements (U, Th, and K), they could be responsible for having initiated melting at depth \citep{shearerpapike1993} and producing dynamically stable TiO$_2$-rich melts in the deep lunar interior \citep[e.g.,][]{delano1990,sakamaki_etal2010,vankanparker_etal2012}.
This model may obviate some of the difficulties in producing a low-density core, required in the FeO-enriched compositions, in the absence of volatiles. Nevertheless, some Fe-rich core is thought to be required on the basis of the remnant magnetism observed \citep[10--100 $\mu T$;][]{cisowski_etal1983,lawrence_etal2008,weisstikoo2014} in highland samples, declining to nil by 1920- to 800 Ma, interpreted to reflect the crystallisation of the lunar core \citep{mighani_etal2020}. \\

On the other hand, Mg-rich mantle compositions require near-pure Fe-Ni cores, a result that is readily reconcilable with the volatile-deprived nature of the Moon. As Fe and Ni mix almost ideally, their alloys have similar melting points with respect to the pure end-members \citep[$\sim$1980 K at 5.5 GPa;][]{urakawaetal1987}. 
Temperatures deduced from the inversions using different mantle compositions invariably yield temperatures at the lunar CMB of $\sim$1623--2073 K, consistent with independent estimates of the temperature of the lunar core \citep{antonangeli_etal2015}. At these temperatures, fusion of a predominantly Fe-Ni core may be expected, particularly in the presence of small quantities ($<$1~wt\%) of light elements. Forming a partially molten layer on the mantle side of an Earth's mantle-like composition is difficult 
as the solidus temperature of mantle peridotite (Mg \# = 0.89) is equivalent to that of Fe-Ni alloy, $\sim$1973 K at 5.5 GPa \citep{hirschmann2000}. The lack of readily fusible sources within an Earth's mantle-like bulk Moon renders fitting its \textit{k$_2$} and the LLR data more problematic, yet not impossible. \cite{nimmo2012dissipation} and \cite{karato2013} demonstrated the \textit{k$_2$} Love number of the Moon could be reproduced in the absence of a liquid layer, but showed difficulties in fitting other observables. \\

FeO-enriched and -poor lunar mantle compositions have significantly different, yet testable, implications for the bulk density, and hence composition of the lunar core. Both end-members result in considerable difficulties in reconciling certain aspects of the Moon. Clear detection and characterisation of the lunar core, particularly in terms of its bulk density and V$_P$, will therefore be key to testing hypotheses for the composition of the lunar mantle, and thus, its chemical relationship to the bulk silicate Earth.

\section{Geochemical constraints on the composition and interior structure of the Moon}
\label{sec:geochem}

\subsection{State of the art}

\subsubsection{Mantle}
\label{sec:geochem_mantle}

Most pre-eminent models for the bulk composition of the Moon, informed largely by geochemical data, proposed enrichments in refractory elements relative to Earth \citep[cf.][]{morgan_etal1978,Taylor1982book}. This presumption was founded on the perceived thickness of the anorthositic crust, 60--70~km \citep[e.g.,][]{nakamura1983seismic}, which was demonstrably shown to be an overestimate through reprocessing of the Apollo seismic data \citep{khanmosegaard2002,lognonne_etal2003}. Their more accurate determinations revealed a thickness of 38$\pm$3~km and 30$\pm$5~km, respectively, entailing a commensurate downward revision of RLE content of the bulk Moon ($\sim$6 wt.\% Al$_2$O$_3$) to more Earth's mantle-like values \citep[3.7 wt.\% Al$_2$O$_3$,][]{khan2006earth}, supported by inferences of a more porous crust from the GRAIL mission \citep{Wieczoreketal2013}. \cite{ryderwood1977}, on evidence of highland material excavated by impacts, had already asserted that the lower crust underlying FAN is likely to be noritic in composition, arguing against higher-than-terrestrial Al$_2$O$_3$ contents in the bulk Moon. Indeed, other contemporary geochemical models had also begun to recover Earth's mantle-like Al$_2$O$_3$ contents \citep[cf.][]{ringwood1977basaltic,wankedreibus1982,Ringwoodetal1987,Jones_Delano1989}.  These models came from the common observation of clearly defined chemical trends among lunar highlands rocks, whose compositions could be produced by mixing a pure anorthite end-member and ``primary matter", which was assumed to have chondritic Ca/Al or Al/Sc ratios \citep{wanke_etal1977}. The significance of this correlation is tied to the assumption that the lunar highlands represent material directly crystallised from the magma ocean \citep[cf.][]{ringwood1979book}. Noticing, however, that the `primary matter' component contained a markedly subchondritic Mg/Si ratio, 0.77 (Mg/Si$_\mathrm{{CI}}$ = 0.89), and that such a component resembled terrestrial komatiites, \citep{Ringwoodetal1987} refined the hypothesis by adding forsteritic olivine (Fo$_{88}$) to primary matter in order to recover either CI chondritic (Mg/Si = 0.89; 22 \% olivine) or terrestrial (Mg/Si = 1.04; 44\% olivine) ratios. This model is therefore based on an empirical resemblance of a fictional component in the lunar crust to Archean magmatic rocks. \\

A later postulate, that of \cite{oneill1991origin} is model-dependent in the framework of a giant impact, wherein some fraction (14 \%) of the Moon is assumed to have derived from a CI-like impactor and the remainder (86 \%) from the proto-Earth's mantle in order to fit the geophysically constrained Mg\# (0.83 at the time, though see section \ref{sec:geophys_inv}), arriving at similar bulk compositions. However, such a scenario no longer holds up to scrutiny when considering the identical isotopic composition of the two bodies, which implies that the impactor:proto-Earth ratio of the present-day Earth and Moon must be the same (see section \ref{sec:physchem_terrestrial}). Owing to the low relative volume of the lunar mantle sampled by mare magmatism, together with its mineralogical heterogeneity, \cite{warren2005} inferred that regions of high Mg\# exist in the lunar mantle from the observation that the most magnesian pyroclastic glasses were in equilibrium with a source with Mg \# of up to 0.87. This view is supported by trends among highland samples, which, when combined with estimates for the mass of the lunar crust, give Al$_2$O$_3$ content of 3.8 wt. \% in the bulk Moon, and yield Mg\# of 0.88--0.89 \citep{warren2005}.  Assuming that the lunar mantle can be modelled as simple binary mixtures of olivine and pyroxene, \cite{Dauphasetal2014} concluded, given recent estimates for the bulk density of the Moon and that those of its constituent minerals are functions of temperature, pressure and composition, that the Moon's Mg\# is 0.85, though Earth's mantle-like values cannot be excluded (see also section \ref{sec:geophys_inv}). \\

Studies providing comprehensive estimates of the trace element composition of the bulk Moon are few and far between. \cite{oneill1991origin} constructed a model for the composition of the lunar mantle, first deriving the abundances of volatile elements from mare basalt source data originally presented in \cite{ringwoodkesson1977} and \cite{wolfanders1980}, normalised to a refractory element of similar incompatibility. Second, owing to the correlation observed between the abundances of volatile elements between the Moon and H chondrites (with a slope = 0.04), he proposed that these volatiles were delivered by addition of 4 wt.~\%  H chondrites to the lunar mantle. As discussed in section \ref{sec:physchem_volatile}, new data renders this scenario unlikely. \cite{taylorwieczorek2014} employed correlations between volatile elements and refractory lithophile elements in both mare- and KREEP-rich basalts to refine the abundances of K, Rb, Cs, Sb, Bi, Zn, Cd, Br and Tl in the lunar mantle, assuming the RLE abundance is equal to that in Earth's mantle. The \cite{oneill1991origin} model tends to overestimate the abundances of the highly volatile elements compared to \cite{taylorwieczorek2014}, due to the imperfect assumption of the 4 wt. \% H chondrite contribution \cite[see Fig. 4 of][]{oneill1991origin}. Nevertheless, as these studies are ultimately drawn upon comparable primary data, and the key results that the Moon is depleted by a factor $\sim$ 5--6 in the alkali metals, and 100--500 $\times$ in highly volatile elements are sound.    \\

\cite{saal2008volatile} showed that the pyroclastic glass beads had lost significant fractions of their volatile elements upon extrusion. Chief among them is H, though moderately volatile elements were also inferred to have been vaporised, due to the presence of fine-grained coatings that often carapace individual glass beads \citep{meyer1975source,wasson1976volatile,liu2022direct}. The model for the composition of the bulk silicate Moon derived by \cite{Haurietal2015} is predicated upon reconstituting the bulk composition of the glass beads by addition of the volatiles hosted in their surface coatings, and typically yields higher volatile contents for the BSMoon than contemporary estimates \citep[e.g,][]{mccubbinetal2021endogenous}. However, recent analyses of diffusion profiles of moderately volatile elements in glass beads show that the pyroclastic liquids may have experienced several episodes of out- and ingassing \citep{su2023outgassing}, calling into question the assumption of closed-system behaviour. Indeed, \cite{Righter2018} point out that, were loss of an S-rich gas phase from glass beads to have transported moderately- and highly volatile elements (such as Cd, In and Bi), then their abundances should be inversely correlated with that of S, for which there is no evidence. \\

Recognising this ambiguity, \cite{ni2019melt} examined melt inclusions in mare basalts, and orange glass 74220, to determine pre-eruptive volatile contents. They demonstrated that orange glass 74220 has typical mare-like volatile/non-volatile element abundance ratios (e.g. S/Dy, Cl/K, Rb/Ba; though, crucially not H$_2$O/Ce), suggesting these magmas may not have been sourced from anomalously volatile-rich regions of the lunar interior. Nevertheless, there is evidence that mare basalt- and pyroclastic glasses were generated under different petrogenetic circumstances \citep{longhi1987}, with the latter 
showing evidence for KREEP in their sources, while their eruptive style \citep{shearerpapike1993} and lower $^{238}$U/$^{204}$Pb \citep[$<$50;][]{TeraWasserburg1976} attest to a higher volatile budget. This conclusion is supported by the $\Delta ^{33}$S of lunar glasses, which differs from those of mare basalts \citep{dottin2023}. As such, the exclusive reliance on pyroclastic glasses to determine the composition of the Moon with respect to volatile elements should be treated with caution. \\


\subsubsection{Core}
\label{sec:geochem_core}

There are no samples of the lunar core, hence its existence is inferred, on geochemical grounds, from its imprint on the composition of lunar rocks. Owing to the small mass ($\sim$0.8--1.5 \%) of the lunar core (Table \ref{tab:inv_com}), geophysical studies are equivocal as to its nature (i.e., solid or liquid) and composition (see section \ref{sec:geophys_core}), and they must therefore be complemented with geochemical analyses of the products of lunar magmatism. \\

\cite{ringwoodkesson1977} contended that the siderophile element pattern of the Moon is inherited from the Earth's mantle, largely based on the inferred abundances of Co, P, and W in the lunar mantle (themselves derived from low-Ti mare basalt data). That is, if one accepts that the depletions of these elements in the terrestrial mantle are likely \textit{only} to have been the result of processes endemic to core formation on the Earth \citep[i.e., at mean temperatures and pressures of the order $\sim$3000 K and 30 - 50 GPa;][]{wadewood2005}, then \cite{ringwoodkesson1977} argue that such conditions could not have prevailed during endogenous core formation on the Moon, the internal pressure of which does not exceed 5.5 GPa. Indeed, the small mass of the lunar core hinders the mantle depletion of all but the most siderophile elements. On this basis, \cite{RammenseeWanke1977} came to the same conclusion as \cite{ringwoodkesson1977}, and argued that the observed W/La ratio is too low (a factor $\sim$20) with respect to chondrites to have been imputed by lunar core formation alone. By contrast, on the evidence that eucrites have W/La similar to lunar basalts, \cite{NewsomDrake1982} suggested that the W/La similarity between lunar and terrestrial basalts was coincidental and could instead be explained by endogenous core formation on the Moon, were it to have occurred under sufficiently low \textit{f}O$_2$. Such an hypothesis encountered difficulties when considering that it would result in an overdepletion of other siderophile elements with respect to their observed abundances, notably P, Ni and Co, and would have required lunar core masses (5.5 wt. \%) and/or mantle FeO contents (3--6 wt. \%) \citep[see][for a discussion]{newsom1984depletion,ringwoodseifert1986} that are now known to be far outside those allowed geophysically (see Table \ref{tab:inv_com}).\\

\cite{oneill1991origin} supposed that the activity of the Ni$_2$SiO$_4$ component of olivine in the lunar mantle could be used to predict the Ni content of the lunar core, yielding, for an Fe$_2$SiO$_4$ activity of 0.17 (Mg\# = 0.83), segregation at $\Delta$IW-0.8 at an assumed temperature of 1573 K, resulting in 45 wt.~\% Ni in the metallic phase and \textit{f$_{core}$} = 0.007. However, as \cite{oneill1991origin} employed 1 bar thermodynamic data to solve for equilibrium, and subsequent metal-silicate partitioning data showed a marked decrease in Ni siderophility as a function of increasing pressure \citep{Walker_etal1993,ThibaultWalter1995}, the Ni mole fractions estimated by \cite{oneill1991origin} in the lunar core are too high by a factor $\sim$1.5. As a result, commensurately higher core masses ($\sim$1 wt.~\%) are required to account for the observed Ni depletion, had the lunar core formed at its present-day pressure ($\sim$5 GPa; see section \ref{sec:physchem_corefm}).  Nevertheless, the \cite{oneill1991origin} model provides a satisfactory fit for the abundances of some elements (e.g. Cu and P), but not others (Ga, Ge, Sn, Sb). With new experimental data in hand, \cite{RighterDrake1996} used the \textit{f}O$_2$ ($\Delta$IW-1) and lunar mantle abundances determined in the \cite{oneill1991origin} model (and the assumption that the \textit{bulk} Moon abundances are equal to those in Earth's mantle) to derive a core mass of $\sim$5~\% by mass, which formed at 2200 K and 3.5 GPa with 15 mol.~\% S. The higher pressures computed come from the need to explain the near-negligible Co depletion in the Moon with respect to Earth's mantle, which, at 1 bar, would require temperatures far in excess of the peridotite liquidus. \\

Most recent estimates have employed Monte-Carlo simulations that take into account uncertainties on parametric fits of metal-silicate partition coefficients to invert for the best-fit values of pressure, temperature and composition, \textit{P-T-X} \citep{righter2002,RaiVanWestrenen2014, sharp2015corefm,steenstra_etal2016}. The first study to converge upon lunar core properties that remain in line with modern-day estimates is that of \cite{righter2002}, with \textit{P} $\sim$ 5 GPa, \textit{T} $\sim$ 1873--2000 K, \textit{f}O$_2$ between $\Delta$IW-0.75 and -1.5, 0.20 $<$ \textit{X}Ni $<$ 0.31 and \textit{f$_{core}$} between 0.003 and 0.02. The range of values comes from the different forward models employed to estimate the abundances of Ni, Co, Mo, W, P, Re and Ga in the bulk Moon. \cite{RaiVanWestrenen2014} were able to narrow this range by considering the five non-volatile, moderately siderophile elements, Ni, Co, Mo, W, and P\footnote{although P is slightly volatile \citep[e.g.,][]{woodetal2019condensation} } to \textit{P} $\sim$ 4.5$\pm$0.5 GPa and \textit{T} = 2200 K at the somewhat lower \textit{f}O$_2$ of $\Delta$IW-2, yet required a core mass ($\sim$ 2.5 \%) higher than is now geophysically permitted (Table \ref{tab:inv_com}). Again, these authors relied upon pre-existing estimates for the abundances of these elements in the bulk Moon \citep[chiefly][]{oneill1991origin} which differ from those in the BSE, leading to considerable non-uniqueness in their \textit{P-T-X} estimates of core formation. \cite{steenstra_etal2016} extended the model of \cite{RaiVanWestrenen2014} by including chalcophile trace elements (most of which are moderately volatile) to assess the S content of the lunar core, concluding that it must have exceeded 6 wt. \% to account for the depletion of 15 elements in the lunar mantle. This conclusion, however, is predicated on the depletion of moderately volatile siderophile elements being the consequence of core formation alone, which is unlikely to be the case \citep[see][]{righter2019}. Indeed, \cite{steenstra2020} later argued that superliquidus temperatures (2600--3700 K) were needed to account for Ni and Co abundances in the lunar mantle, under the revised assumption of a BSE-like bulk Moon composition and an \textit{f}O$_2$ of $\Delta$IW-2 for \textit{f$_{core}$} = 0.015--0.02. However, increasing the \textit{f}O$_2$ to $\Delta$IW-1 and/or decreasing the core mass to 1 \% would also produce fits to the Ni and Co abundances of the lunar mantle without invoking superheating (see section \ref{sec:physchem_corefm}). These uncertainties highlight the degeneracy in determining the conditions of lunar core formation, even using the well-constrained elemental abundances of Ni and Co.

\subsection{Methods for constraining mantle and core composition}
\label{sec:geochem_methods}

As is evident from the foregoing discussion, each method for estimating the composition of the Moon is subject to considerable uncertainty. As such, geochemical estimates for the major element composition of the Moon were supplanted by geophysically constrained models (section \ref{sec:geophys_inv}). Therefore, here, we focus on the trace element composition of the Moon.\\

Based on cosmochemical expectations, the bulk lunar mantle should have \textit{i)} near-chondritic ratios of refractory lithophile elements (e.g., Al, Ti, Ca) and \textit{ii)} Mg\#s between 0.80--0.90 (cf. Table \ref{tab:inv_com}). The inferred sources of low-Ti mare basalts and green glass samples come closest to fulfilling these criteria, and have thus been taken as the most representative partial melting products of an hypothetical bulk silicate Moon \citep[e.g.,][]{ringwoodkesson1977, delano1986,HaskinWarren1991}. Despite fact that the maria, consisting of flows with total average thicknesses of $\sim$ 400 m \citep{dehon1979}, make up 17 \% of the surface by area, but only $\sim$ 1 \% by volume - $\sim$10$^{7}$ km$^3$ - of the lunar crust \citep{headwilson1992}, they have the advantage of representing partial melting products of the lunar mantle, for which experimental studies can be used to invert their chemical abundances to obtain those of their source regions \citep[e.g.,][see below]{longhi1992,browngrove2015}. To determine compositions of the elements in the basalt source regions, we exploit correlations between the element of interest and a refractory lithophile element with similar behaviour during partial melting, a commonly used approach \citep[e.g.,][]{McDonoughSun1995,yietal2000,taylorwieczorek2014}. In practice, this is achieved in two ways: \\

1) Direct use of the ratio X/RLE, where the identity of the RLE is selected such that the ratio X/RLE is independent of X \citep{wanke_etal1973}. The abundance of X in the lunar mantle is given by multiplying X/RLE by the abundance of the RLE supposed in the lunar mantle, which, here, is assumed to be equal to that in Earth's primitive mantle (i.e., 1.21 $\times$ CI, normalised to Mg). Typical ratios are In/Y \citep{yietal2000}, P/Nd \citep{wankedreibus1986}; K/U \citep{McDonoughSun1995}.\\

2) Based on the assumption that low Ti magmatism (both glasses and basalts) and MORB are formed by similar degrees of partial melting, the ratio (X/RLE)$_{\mathrm{low Ti}}$ / (X/RLE)$_{\mathrm{MORB}}$ is multiplied by the abundance of X in the terrestrial mantle \citep[e.g.,][]{palmeoneill2014} to yield X$_{Moon}$. That the behaviour of a given element remains constant during partial melting in both mantles is another caveat. This predominantly affects redox-sensitive elements whose valence state is liable to differ between the terrestrial- \citep[\textit{f}O$_2$ $\sim\Delta$IW+3.5;][]{frost2008} and lunar upper mantle \citep[\textit{f}O$_2$ $\sim\Delta$IW-1;][]{herd2008fO2}. \\

The considerable improvement in our understanding of the compositions of ocean floor basaltic magmas \citep{ArevaloMcDonough2010,jenneroneill2012}, and high precision analyses of mare magmatic products \citep[e.g.,][]{ni2019melt,garganoetal2020,gleissneretal2022} behoves a more detailed examination of the differences between terrestrial and lunar basalts.  In the following, we use the compositions of mare rocks, taking care to adhere to approaches (1) and (2) (above), together with geophysical constraints to provide estimates for the composition of the bulk silicate Moon for 70 elements (Table \ref{tab:lunar_abu}).

\subsection{Composition of the Moon}
\label{sec:geochem_compmoon}

\subsubsection{Refractory Lithophile Elements}
\label{sec:geochem_RLEs}

Owing to the near constancy ($\sim$10 \% relative) in chondritic meteorites \citep{wassonkallemeyn1988}, and their lack of affinity for the metallic phase during core formation \citep[e.g.,][]{ringwood1991solubilities}, the relative proportions of the RLEs are taken to be constant in the bulk Moon. Their absolute enrichment factors relative to CI chondritic reference were initially thought to be greater than that observed in the Earth's mantle \citep[2.83$\times$ CI,][]{palmeoneill2014} on the basis of the thickness of the Ca-, Al-rich anorthositic crust \citep[e.g.,][]{Taylor1999}. However, this view was overturned following revisions to the crustal structure from Apollo seismic data that led to a thinner and anorthositic crust \citep[][see section \ref{sec:geophys_mantle}]{khan_etal2000,lognonne_etal2003} with significant portions of gabbroic-noritic material \citep{khan_etal2000,Wieczoreketal2013} both leading to reduced contributions of the crust to the Al- and Ca budget of the bulk Moon. Moreover, the crust over which the Apollo seismic data were collected possesses a higher porosity (12 \%, on average, in the upper several km) than crystalline anorthosite resulting in a lower mean density \citep[2550 $\pm$ 250 kgm$^{-3}$;][]{Wieczoreketal2013}.\\

In independent support of Earth-like RLE abundances, remote sensing gamma-ray measurement of the concentrations of radioactive RLEs, namely U and Th, show distinct hemispherical asymmetry with most elevated values being detected on the nearside Procellarum KREEP Terrain (PKT) \citep{lawrence1998gammaray,Jolliffetal2000}. Calibration of the gamma-ray data with respect to the composition of lunar meteorites and Apollo samples \citep{warren2005,Prettymanetal2006,siegler_etal2022} yielded mean surface Th contents of $\sim$1.49 ppm, which, when extrapolated to a mantle concentration, resulted in Th between 0.045 and 0.08 ppm \citep[Earth's mantle = 0.08 ppm;][]{palmeoneill2014}. The major sources of uncertainty contributing to this range of estimates relate to the Th content of the present-day lunar mantle (and, given a constant U/Th ratio of 3.7$\pm$0.2, the U content) as well as its Urey ratio that are used to compare compositional models with lunar heat flux measurements \citep{siegler_etal2022}. Based on these observations, we assume the RLE enrichment in the Moon to be equivalent to that in Earth's mantle, i.e., 2.83$\times$ CI.

\subsubsection{Volatile Lithophile Elements}
\label{sec:geochem_VLEs}

\begin{figure}
     \centering
     \includegraphics[width=0.5\textwidth]{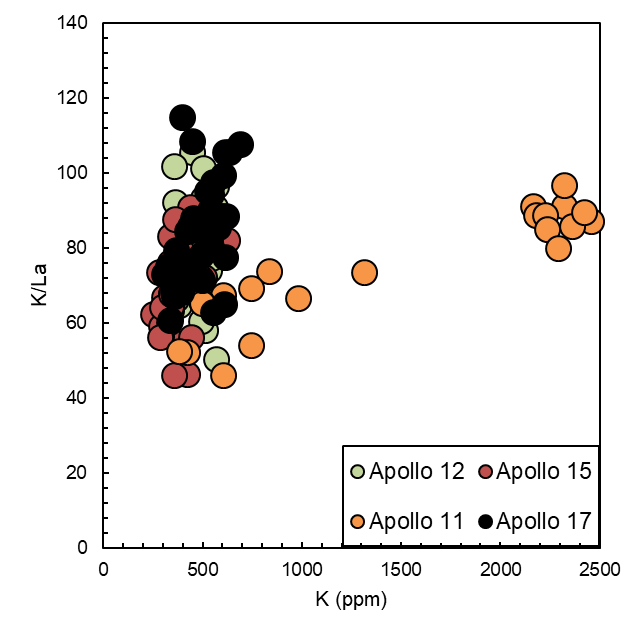}
     \caption{Data for K (ppm) vs. K/La compiled from mare basalts \citep[modified after][]{neal2007}, subdivided into those from the Apollo 11, 12, 15 and 17 missions.}
    \label{fig:K-KLa}
 \end{figure}

\textit{Alkali metals - Li, Na, K, Rb and Cs.} 
The incompatibility of alkali metals increases down the group, with Li being relatively compatible in olivine and Na in clinopyroxene and plagioclase, while K, Rb and Cs are highly incompatible, and are therefore often compared with other highly incompatible RLEs (La, Th, U, Ba) and one another \citep[e.g.,][]{oneill1991origin,taylorwieczorek2014}. The Li/Yb ratio is typically used for estimating lunar Li abundances \citep{oneill1991origin,normantaylor1992,ni2019melt}, however, this ratio differs among Apollo 12 and 15 Low-Ti mare basalts, whereas we find Li/Ce is constant among all mare basalts, producing an Li content of 1.08 $\pm$ 0.55 (\ref{tab:lunar_abu}) that is consistent with estimates for the Earth's primitive mantle \citep[1.39 $\pm$ 0.10;][]{marschall2017boron}. Sodium is referenced to Sr, owing to their preference for plagioclase over other minerals, for which all basaltic groups define similar Na/Sr ratios to give 403$\pm$81 ppm in the bulk Moon, though there are no clear trends within a mare basalt suite. Our result matches the estimate of \cite{ni2019melt} derived from melt inclusions, 398 ppm. Among mare basalts, K correlates better with La (\textit{r}$^2$ = 0.93; Fig. \ref{fig:K-KLa}) than it does with U (\textit{r}$^2$ = 0.84) due to its slightly higher compatibility, giving 52.6 $\pm$ 9.6 ppm, higher than the 36.9 $\pm$ 9.6 ppm reported by \cite{taylorwieczorek2014} using the same method, although their Fig. 6 implies a concentration of 46.2 ppm, indistinguishable from our estimate, that of \cite{Dauphasetal2022_MVE}, 44 ppm, \cite{ni2019melt}, 45.5 ppm, and of \cite{Haurietal2015}, 57 ppm. Rubidium is determined independently by the $^{87}$Sr/$^{86}$Rb ratio of lunar magmatic rocks \citep{halliday_porcelli2001,borg_etal2022} to yield an estimate (0.15 $\pm$ 0.01 ppm) in agreement with that derived from Rb/La ratios \citep[0.13$\pm$0.005 ppm;][]{taylorwieczorek2014}, the 0.1470 ppm of \cite{Haurietal2015} but higher than the 0.096 ppm of \cite{Dauphasetal2022_MVE}. Using their Rb/Sr ratio (0.0053$\pm$0.0006) and the Sr content of the BSE (22 ppm) yields a higher estimate, 0.117$\pm$0.013 ppm. The good correlation between Rb and Cs in all lunar samples \citep[\textit{r}$^2$ = 0.93;][]{HaskinWarren1991} is exploited to derive 6.2$\pm$0.5 ppb Cs in the bulk silicate Moon in agreement with \cite{Haurietal2015}, 6.7 ppb, though marginally higher than the 5$\pm$0.3 given in \cite{taylorwieczorek2014} on the basis of Cs/La ratios (\textit{r}$^2$ = 0.88) or \cite{ni2019melt}, 4.1 ppb.\\

\textit{Zn.} Precise estimates for the Zn content of the BSMoon are now afforded by high precision Zn isotope measurements of a variety of lunar lithologies \citep{moynier_etal2006,panielloetal2012,Katoetal2015,garganoetal2022}. These studies reveal remarkable homogeneity in the Zn contents of mare basalts, both High- and Low-Ti, lying within the range 1.5$\pm$0.8 ppm (excepting 8.1 ppm in the Apollo 11 High-Ti basalt, 10017, which has anomalously light $\delta^{66}$Zn, -8.0 $\pm$ 4.5 \permil). Considering the slight incompatibility of Zn in mantle minerals \citep{kohnschofield1994,lerouxetal2011,davisetal2013FRTE} and the concomitant enrichment of Zn in MORB relative to their mantle sources \citep{jenneroneill2012}, a BSMoon Zn abundance of 1$\pm$0.6 ppm is proposed. This is similar to the factor $\sim$200 depletion in Zn/Fe relative to MORB reported by \cite{albarede_etal2015}, but significantly lower than the 7.7 ppm given by \cite{Haurietal2015} using the reconstructed composition of orange glass 74220, and the 2.9 ppm reported by \cite{ni2019melt} using Zn/Fe ratios on the same sample, but higher than the 0.4 $\pm$ 0.15 ppm based on Zn/Sc ratios of \cite{taylorwieczorek2014}. Using Sc as a normalising element likely underestimates Zn abundances in the BSMoon, as Sc is $\sim$10$\times$ more compatible than Zn in pyroxenes whereas Zn has a preference for olivine \citep[e.g.,][]{mallmannoneill2009,davisetal2013FRTE}. However, Zn/Fe ratios are also likely misleading, as Zn is present in trace proportions such that its partitioning during partial melting follows Henry's Law, whereas that of Fe is dependent upon phase equilibria, meaning a single partition coefficient is not applicable. Indeed, the Zn/Fe ratio in lunar pyroclastic glasses ranges from 10$^{-6}$ to 10$^{-4}$ \citep{albarede_etal2015,mccubbinetal2021endogenous}, rendering Zn abundance estimates using Zn/Fe ratios uncertain. \\

\textit{B, F, Cl, Br and I.} Boron abundances, together with those of the halogens are more difficult to determine in the BSMoon, owing to both their volatility and the lack of samples for which their concentrations have been quantified. Boron abundances have been measured in only a handful of basalts, and our estimate comes from the constancy of B/Ce ratios \citep[as observed for terrestrial magmas;][]{marschall2017boron} in Apollo 12 mare basalts. Nevertheless, our value of 0.10$\pm$ 0.05ppm is similar to the 0.0743 ppm reported by \cite{Haurietal2015} and roughly half the BSE abundance \citep[0.19$\pm$ 0.02 ppm;][]{marschall2017boron}. Halogen abundances chiefly rely upon the measurements of \cite{mccubbin2015}, \cite{boyceetal2018}, \cite{ni2019melt} and \cite{garganoetal2020} in pyroclastic glasses, melt inclusions and lunar mare basalts. Chlorine, the most often-studied halogen, has abundances that correlate well with K, yielding Cl/K = 0.006 $\pm$ 0.001 \citep[see][]{mccubbinetal2021endogenous}, which, given K of 52.6 ppm, yields 0.31 $\pm$ 0.10 ppm Cl, similar to 0.142--0.205 ppm \citep{Haurietal2015} and 0.39 ppm \citep{ni2019melt}, but lower than the 1.0--4.1 ppm proposed by \cite{mccubbinetal2021endogenous}. Given that Cl is as incompatible as K during partial melting, and the Cl contents of mare basalts and glasses are typically $\sim$1--6 ppm \citep[e.g.,][]{ni2019melt,garganoetal2020}, the estimates of \cite{mccubbinetal2021endogenous} appear too high. The content of fluorine, the most abundant and least volatile of the halogens, is typically estimated by comparison to a LREE, La \citep{oneill1991origin} or Nd \citep{Haurietal2015}, however, \cite{ni2019melt} note that F/Nd ratios are variable among mare basalts, yet, in the absence of another proxy, use it to derive F contents of 5 ppm in the BSMoon, while \cite{mccubbinetal2021endogenous} propose 3.1 - 4.9 ppm, also based on F/Nd of melt inclusions and mare basalts. We utilise the constant F/Lu ratios in Low-Ti basalts \citep{garganoetal2020} to give a similar estimate, 2.5$\pm$0.9 ppm. Bromine, although showing behaviour akin to that of Cl in MORB \citep{kendrick2012halogen}, is not correlated with any other element in lunar mare basalts \citep{garganoetal2020}, and hence we ascribe a tentative BSMoon estimate of 0.001 ppm to Br, which is, in turn, similar to 0.00079 ppm suggested by \cite{taylorwieczorek2014}. Iodine correlates with Cl in Low-Ti mare basalts \citep{garganoetal2020}, defining a constant I/Cl ratio of 0.02 $\pm$ 0.003, significantly higher than that in MORB \citep[1.3 $\pm$ 1.0 $\times$ 10$^{-4}$; ][]{kendrick2012halogen}, yielding 6.2 $\pm$ 2.6 $\times$ 10$^{-4}$ ppm in the BSMoon. 

\subsubsection{Mn, Cr and V.} 
\label{sec:geochem_MnCrV}
These elements are afforded special status in lunar geochemistry on account of their peculiarity among terrestrial and lunar magmas relative to those from other telluric bodies in the inner Solar System. They have similar 50 \% nebular condensation temperatures at 10$^{-4}$ bar \citep[e.g.,][]{woodetal2019condensation} that increase in the order Mn ($\sim$1150 K) $<$ Cr ($\sim$1290 K) $<$ V ($\sim$1380 K). 

The philosophy behind the determination of their lunar abundances is contingent upon the empirical correlations between the three elements and Fe among lunar highlands rocks \citep{ringwoodkesson1977,seifertringwood1988,drake_etal1989}. This coherence stems from their existence as divalent cations at the \textit{f}O$_2$ of the formation of the lunar crust, although this is expected to break down should spinel be present. \cite{oneill1991origin} describes the MnO/FeO ratio as \textit{``the surest known of all items of lunar chemistry"}, with a constant value of 0.014$\pm$0.001 across all lunar rocks \citep{lauletal1974, HaskinWarren1991}. Previous estimates simply assumed FeO = 12.7 wt. \% to derive Mn = 1350 ppm, and thus an enrichment of $\sim$30 \% over Earth's mantle, but significantly lower than in the mantles of Vesta \citep[$\sim$4000 ppm;][]{mittlefehldt2015asteroid} or Mars \citep[$\sim$3000 ppm;][]{khan2022geophysical}. Our approach differs in that we are agnostic as to the lunar mantle FeO content (see section \ref{sec:geophys_inv}), which could vary between $\sim$8 and $\sim$14 wt~\% in the BSMoon (Table \ref{tab:inv_com}). Yet we exploit the observation that MnO is constant among all mare basalts (0.27$\pm$0.02 wt. \%), suggesting D(Mn)$_{min/melt}$ close to unity, in agreement with experimentally determined partition coefficients for Mn$^{2+}$ \citep{kohnschofield1994,longhietal1978}. 
The green glass average MnO, 0.23$\pm$0.03 \citep{delano1986} yields 1158$\pm$150 ppm in the BSMoon, and hence an FeO content of 10.6$\pm$1.4 wt.~\%, which is similar to the mean value of the geophysically derived FeO contents. Therefore, we adopt 1236$\pm$331 ppm to reflect this range, while noting that assuming Earth-like FeO contents would result in $\sim$900 ppm Mn.

Vanadium, although lumped in with the RLEs, is the least refractory element of the group, as evidenced by increasing Al/V ratios from CI (159) to CV (182) chondrites \citep{jonespalme2000}. Indeed, the Earth's mantle's Al/V ratio is 275, though much of the V depletion relative to Al and chondrites likely arises from core formation \citep{Siebertetal2013}. Here we assume that V is present in Earth's mantle-like abundances, given no reason to expect its depletion, neither by volatility (\textit{cf.} Li, which is more volatile) nor by core formation (\textit{cf.} W, which is more siderophile), hence V = 86 $\pm$ 5 ppm in the BSMoon. The Cr/V ratio is near-constant among all mare basalts and pyroclastic glasses (24.4$\pm$5.3, \textit{n} = 116), which, for 86 $\pm$ 5 ppm, yields 2098 $\pm$ 461 ppm, marginally depleted with respect to, but overlapping within uncertainty of the BSE (2520 $\pm$ 250 ppm).

To reproduce the V ($\sim$175 ppm) and Cr ($\sim$4000 ppm) contents of Low-Ti Apollo 12 mare basalts with the highest Mg \# (0.58) and of the lunar green glass \citep[V = 160$\pm$3 ppm, Cr = 3458$\pm$248 ppm;][]{meyer2005}, bulk D(V,Cr)$_{min/melt}$ of $\sim$0.5 are required. This is readily achievable for Cr, because, for all Cr as Cr$^{2+}$ at the conditions of partial melting at $\Delta$IW-1 \citep[e.g.][]{mallmannoneill2009,oneillberry2021},  D(Cr)$_{min/melt}$ is $\sim$0.5 in all mantle silicate minerals \citep{seifertringwood1988,hansonjones1998,mallmannoneill2009,sossioneill2016}. However, V becomes distinctly more compatible at lower \textit{f}O$_2$ as it converts from V$^{4+}$ under Earth's mantle conditions to V$^{3+, 2+}$ in the lunar mantle, with a bulk D(V)$_{min/melt}$ close to unity at $\Delta$IW-1 \citep{mallmannoneill2009}. However, their experiments were performed in the iron-free system, while \cite{kohnschofield1994} have shown D(M$^{2+})_{min/melt}$ is inversely correlated with the Non-Bridging Oxygen / Tetrahedral cation (NBO/T) proportion of the silicate liquid. As such, D(V)$_{min/melt}$ might be expected to be lower for high NBO/T ($\sim$2), FeO-rich mare basaltic liquids; a prediction supported by experiments performed in Fe-Ni capsules on the green glass composition, in which D(V)$_{ol/melt}$=0.3, D(V)$_{opx/melt}$=0.6 and D(V)$_{cpx/melt}$=0.9, yielding bulk D(V)$_{min/melt}$ $\sim$0.5 \citep{seifertringwood1988}

\onecolumn
\begin{small}
\begin{longtable}{p{0.13\linewidth}p{0.08\linewidth}p{0.08\linewidth}p{0.3\linewidth}p{0.25\linewidth}}

\caption{Abundances of elements in the bulk silicate Moon}\\
	\hline
        \textbf{Element} & \textbf{X} & \textbf{1s} & \textbf{Comments} & \textbf{References} \\ 
	\hline
        \textbf{H (\%)} & - & - & ~ & ~ \\ 
        \textbf{Li (ppm)} & 1.08 & 0.55 & Li/Ce = 0.42±0.11 in all mare basalts; n = 110 vs. Li/Ce = 0.54±0.24 in MORB & \cite{neal2007,magna2006,day2016lithium,marschall2017boron} \\ 
        \textbf{Be (ppm)} & 0.062 & 0.012 & RLE, PM-like & \cite{palmeoneill2014} \\ 
        \textbf{B (ppm)} & 0.10 & 0.05 & B/Ce = 0.056±0.006 in Apollo 12 basalts, n = 5. Uncertain. & \cite{neal2007} \\ 
        \textbf{C (\%)} & - & - & ~ & ~ \\ 
        \textbf{N (\%)} & - & - & ~ & ~ \\ 
        \textbf{F (ppm)} & 2.5 & 0.9 & F/Lu = 36±7 in Low Ti mare basalts & \cite{garganoetal2020} \\ 
        \textbf{Na (ppm)} & 403 & 81 & Na/Sr = 18.3±3.7 in all mare basalts; n = 136 & \cite{neal2007} \\ 
        \textbf{MgO (\%)} & - & - & See Table \ref{tab:inv_com} & ~ \\ 
        \textbf{Al$_2$O$_3$ (\%)} & 4.49 & 0.36 & RLE, PM-like & ~ \\ 
        \textbf{SiO$_2$ (\%)} & - & - & See Table \ref{tab:inv_com} & ~ \\ 
        \textbf{P (ppm)} & 18.7 & 7.5 & P/Nd in mare basalts converge to a ratio of \~10±5 in primitive examples, mean P/La = 35.1±9.4, n = 116 & \cite{neal2007,HaskinWarren1991} \\ 
        \textbf{S (ppm)} & 81 & 15 & S/Tb in mare basalts, melting model & \cite{dingetal2018,gleissneretal2022,garganoetal2022} \\ 
        \textbf{Cl (ppm)} & 0.31 & 0.10 & Cl/K = 0.006±0.001 in mare basalts and melt inclusions & \cite{Haurietal2011,mccubbin2015,mccubbinetal2021endogenous,boyceetal2018,ni2019melt,garganoetal2020} \\ 
        \textbf{K (ppm)} & 52.6 & 9.6 & K/La = 77±14 in all mare basalts; n = 136 & \cite{neal2007} \\ 
        \textbf{CaO (\%)} & 3.65 & 0.29 & RLE, PM-like & ~ \\ 
        \textbf{Sc (ppm)} & 16.4 & 1.6 & RLE, PM-like & \cite{palmeoneill2014} \\ 
        \textbf{Ti (ppm)} & 1265 & 127 & RLE, PM-like & \cite{palmeoneill2014} \\ 
        \textbf{V (ppm)} & 86 & 5 & RLE, PM-like & ~ \\ 
        \textbf{Cr (ppm)} & 2098 & 461 & Constant Cr/V in all mare basalts; 24.4±5.3  n = 116 & \cite{neal2007,sossietal2018} \\ 
        \textbf{Mn (ppm)} & 1236 & 331 & Green glass 0.23 $\pm$ 0.03 wt~\% MnO, D$_{\mathrm{Mn}}~\sim1$, MnO/FeO = 0.014$\pm$0.001, FeO between 8 and 14 wt~\%, See Table \ref{tab:inv_com} & \cite{lauletal1974,HaskinWarren1991,neal2007} \\ 
        \textbf{FeO (\%)} & - & - & See Table \ref{tab:inv_com} & ~ \\ 
        \textbf{Co (ppm)} & 90 & 5 & Green glass = Co/(FeO+MgO) = 1.95$\pm$0.05 & \cite{delano1986} \\ 
        \textbf{Ni (ppm)} & 470 & 50 & Green glass = 180±20 ppm & \cite{delano1986} \\ 
        \textbf{Cu (ppm)} & 2.3 & 1.1 & Primitive Low-Ti basalts = 9.4±2.7 ppm (n = 6) vs. MORB = 100±20 ppm, PM = 25±5 & \cite{gleissneretal2022,sunetal2020CuS} \\ 
        \textbf{Zn (ppm)} & 1.0 & 0.6 & Average, all mare basalts; n = 34, normalised to primitive MORB/PM & \cite{panielloetal2012,Katoetal2015,ni2019melt,garganoetal2022} \\ 
        \textbf{Ga (ppm)} & 0.94 & 0.47 & Correlation with Mg\# in mare basalts 3.2 vs. 15 ppm in MORB at Mg\# = 0.7 & \cite{neal2007,katomoynier2017gallium,borg_etal2019} \\ 
        \textbf{Ge (ppm)} & - & - & No high precision data & ~ \\ 
        \textbf{As (ppm)} & - & - & No high precision data & ~ \\ 
        \textbf{Se (ppm)} & 0.014 & 0.003 & S/Se = 5558±316 in Low Ti mare basalts (n = 7) & \cite{gleissneretal2022} \\ 
        \textbf{Br (ppm)} & 0.001? & - & Uncertain & \cite{wolfanders1980,garganoetal2020} \\ 
        \textbf{Rb (ppm)} & 0.15 & 0.01 & Rb/Sr = 0.00674±0.0005 (Sr isotopes) & \cite{halliday_porcelli2001,borg_etal2022} \\ 
        \textbf{Sr (ppm)} & 22.0 & 2.2 & RLE, PM-like & \cite{palmeoneill2014} \\ 
        \textbf{Y (ppm)} & 4.1 & 0.4 & RLE, PM-like & \cite{palmeoneill2014} \\ 
        \textbf{Zr (ppm)} & 10.3 & 1.0 & RLE, PM-like & \cite{palmeoneill2014} \\ 
        \textbf{Nb (ppm)} & 0.73 & 0.08 & Nb/Ta = 17±2, n = 37 & \cite{munker2010high,thiemens2019early} \\ 
        \textbf{Mo (ppm)} & 0.019 & 0.010 & Average of ID data of mare basalts (n = 4); batch melting F = 0.1 to 0.15. Uncertain. & \cite{burkhardt2014evidence,leitzke2017redox} \\ 
        \textbf{Ru (ppm)} & 1.6$\times 10^{-4}$ & 5.0$\times 10^{-5}$ & 0.00023±2 × CI & \cite{Dayetal2016} \\ 
        \textbf{Rh (ppm)} & 3.0$\times 10^{-5}$ & 9.1$\times 10^{-6}$ & 0.00023±2 × CI & \cite{Dayetal2016} \\ 
        \textbf{Pd (ppm)} & 1.5$\times 10^{-4}$ & 1.0$\times 10^{-4}$ & 0.00023±2 × CI & \cite{Dayetal2016} \\ 
        \textbf{Ag (ppm)} & 9.9$\times 10^{-5}$ & 9.0$\times 10^{-6}$ & Constant Ag/Sm in Low-Ti basalts; 2.3(0.2)$\times 10^{-4}$, n = 7 & \cite{gleissneretal2022} \\ 
        \textbf{Cd (ppm)} & 5.2$\times 10^{-5}$ & 1.1$\times 10^{-5}$ & Constant Cd/Dy in all mare basalts; 7.1(1.6)×10$^{-5}$, n = 12 & \cite{gleissneretal2022} \\ 
        \textbf{In (ppm)} & 1.1$\times 10^{-4}$ & 5.0$\times 10^{-5}$ & Constant In/Y in Low-Ti basalts; 2.5(1.2)×10$^{-5}$, n = 8 & \cite{gleissneretal2022} \\ 
        \textbf{Sn (ppm)} & 0.023 & 0.008 & Sn/Sm = 0.018 in mare basalts; n = 8 & \cite{wangetal2019tin} \\ 
        \textbf{Sb (ppm)} & - & - & No high precision data & ~ \\ 
        \textbf{Te (ppm)} & 2.1$\times$10$^{-4}$ & 0.3$\times$10$^{-4}$ & Se/Te = 66 $\pm$ 7 (\textit{n} = 7) in Low-Ti mare basalts & \cite{gleissneretal2022} \\ 
        \textbf{I (ppm)} & 6.2$\times 10^{-4}$ & 2.6$\times 10^{-4}$ & Cl/I = 0.02±0.003 in Low Ti basalts (n = 7) & \cite{garganoetal2020} \\ 
        \textbf{Cs (ppm)} & 0.0062 & 0.0005 & Rb/Cs =25.4±7.4, n = 69 all mare basalts; log(Cs) = 1.01log(Rb)+1.65 all lunar rocks & \cite{HaskinWarren1991,neal2007} \\ 
        \textbf{Ba (ppm)} & 6.85 & 1.03 & RLE, PM-like & \cite{palmeoneill2014} \\ 
        \textbf{La (ppm)} & 0.683 & 0.068 & RLE, PM-like & \cite{palmeoneill2014} \\ 
        \textbf{Ce (ppm)} & 1.753 & 0.175 & RLE, PM-like & \cite{palmeoneill2014} \\ 
        \textbf{Pr (ppm)} & 0.266 & 0.040 & RLE, PM-like & \cite{palmeoneill2014} \\ 
        \textbf{Nd (ppm)} & 1.341 & 0.134 & RLE, PM-like & \cite{palmeoneill2014} \\ 
        \textbf{Sm (ppm)} & 0.435 & 0.043 & RLE, PM-like & \cite{palmeoneill2014} \\ 
        \textbf{Eu (ppm)} & 0.167 & 0.017 & RLE, PM-like & \cite{palmeoneill2014} \\ 
        \textbf{Gd (ppm)} & 0.586 & 0.029 & RLE, PM-like & \cite{palmeoneill2014} \\ 
        \textbf{Tb (ppm)} & 0.108 & 0.016 & RLE, PM-like & \cite{palmeoneill2014} \\ 
        \textbf{Dy (ppm)} & 0.724 & 0.072 & RLE, PM-like & \cite{palmeoneill2014} \\ 
        \textbf{Ho (ppm)} & 0.160 & 0.024 & RLE, PM-like & \cite{palmeoneill2014} \\ 
        \textbf{Er (ppm)} & 0.468 & 0.047 & RLE, PM-like & \cite{palmeoneill2014} \\ 
        \textbf{Tm (ppm)} & 0.074 & 0.011 & RLE, PM-like & \cite{palmeoneill2014} \\ 
        \textbf{Yb (ppm)} & 0.477 & 0.048 & RLE, PM-like & \cite{palmeoneill2014} \\ 
        \textbf{Lu (ppm)} & 0.071 & 0.011 & RLE, PM-like & \cite{palmeoneill2014} \\ 
        \textbf{Hf (ppm)} & 0.301 & 0.030 & RLE, PM-like & \cite{palmeoneill2014} \\ 
        \textbf{Ta (ppm)} & 0.04 & 0.002 & RLE, PM-like & \cite{palmeoneill2014} \\ 
        \textbf{W (ppm)} & 1.1$\times 10^{-2}$ & 2.0$\times 10^{-3}$ & Constant W/U in all mare basalts; 0.48±0.07, n = 17 & \cite{munker2010high,kruijer2017tungsten,thiemens2019early} \\ 
        \textbf{Re (ppm)} & 9.0$\times 10^{-6}$ & 6.0$\times 10^{-6}$ & 0.00023±2 × CI & \cite{Dayetal2016} \\ 
        \textbf{Os (ppm)} & 9.7$\times 10^{-5}$ & 3.0$\times 10^{-5}$ & 0.00023±2 × CI & \cite{Dayetal2016} \\ 
        \textbf{Ir (ppm)} & 8.7$\times 10^{-5}$ & 4.0$\times 10^{-5}$ & 0.00023±2 × CI & \cite{Dayetal2016} \\ 
        \textbf{Pt (ppm)} & 2.1$\times 10^{-4}$ & 1.3$\times 10^{-4}$ & 0.00023±2 × CI & \cite{Dayetal2016} \\ 
        \textbf{Au (ppm)} & 3.4$\times 10^{-5}$ & 1.0$\times 10^{-5}$ & 0.00023±2 × CI & \cite{Dayetal2016} \\ 
        \textbf{Hg (ppm)} & - & - & No high precision data & ~ \\ 
        \textbf{Tl (ppm)} & 3.3$\times 10^{-5}$ & 1.1$\times 10^{-5}$ & Constant Tl/Te in Low-Ti basalts; 1.5(0.3)×$10^{-4}$, n = 7 & \cite{gleissneretal2022} \\ 
        \textbf{Pb (ppm)} & 0.0045 & $^{-0.0010} _{+0.0018}$ & $^{238}$U/$^{204}$Pb ($\mu$ ratio) in Low-Ti basalts = 350±100 & \cite{snapeetal2019,connellyetal2022} \\ 
        \textbf{Bi (ppm)} & - & - & No high precision data & - \\ 
        \textbf{Th (ppm)} & 0.0849 & 0.0127 & RLE, PM-like & \cite{palmeoneill2014} \\ 
        \textbf{U (ppm)} & 0.0229 & 0.0034 & RLE, PM-like & \cite{palmeoneill2014} \\ \hline
        
    \label{tab:lunar_abu}
\end{longtable}
\end{small}
\twocolumn

and in line with those determined for olivine-melt pairs for lunar basalt analogues \citep{jing2024}. Additional experiments examining the partitioning of Mn, Cr and V in lunar composition melts would be useful in better constraining their mantle abundances.

\subsubsection{Non-volatile Siderophile Elements}
\label{sec:geochem_RSEs}
\textit{Co and Ni.} Cobalt abundances were estimated 
based on correlations between CoO and FeO and MgO in lunar pyroclastic glasses, yielding \citep[90$\pm$5][]{delano1986} for the BSMoon. This figure highlights that the Co contents of the terrestrial \citep[102$\pm$5][]{palmeoneill2014} and lunar mantles are indistinguishable within uncertainty. The Ni content of the Moon is taken from \cite{delano1986} based on the Ni concentrations of green glass beads, together with its partition coefficient during melting of the lunar mantle, yielding 470$\pm$50 ppm. Should the Mg \# of the bulk lunar mantle exceed that in equilibrium with the green glass \citep[$\sim$ 0.84--0.87 for liquid Mg \# of 0.62--0.67, respectively;][]{elkins2003experimental}, the Ni content would be a minimum estimate owing to the compatibility of Ni in olivine and the K$_{D, olivine-melt}^{Fe-Mg}$ of 0.33$\pm$0.02 \citep{longhietal1978}. \\

\textit{Mo and W.} These elements bear special significance in constraining models for lunar origin, as they are the only two elements that are refractory \citep[though see][]{oneill1991origin, sossietal2019} as well as moderately siderophile. Molybdenum abundances in terrestrial and lunar rocks were the subject of investigations by Horton Newsom and colleagues throughout the 1980s \citep{newsom1984depletion,newsom1986constraints}. These authors based their estimates of the Mo content of the BSMoon on correlations with Nd, yielding an Mo/Nd ratio of 0.0016 in mare basalts, some 27 $\times$ lower than that in terrestrial basalts, 0.043 \citep{newsom1984depletion}, with the modern value = 0.060$\pm$0.018 \citep{jenneroneill2012}. However, 
the Mo/Nd ratio used by \cite{newsom1986constraints} is scattered among mare basalts, as noted by \cite{HaskinWarren1991}. Indeed, our compilation indicates log[Mo/Nd] = -2.20 $\pm$ 0.46, \textit{n} = 54. More recently, \cite{leitzke2017redox} performed experiments to assess the effect of oxygen fugacity on Mo partitioning between plausible lunar mantle minerals and silicate melt. They observed increasing D(Mo)$_{min/melt}$ to lower \textit{f}O$_2$, consistent with the stoichiometry of the equilibrium MoO$_2$ ($l$) + 1/2O$_2$ (\textsl{g}) = MoO$_3$ ($l$). Based on thermochemical data and metal-silicate partitioning experiments \citep{holzheid_etal1994,oneilleggins2002} and supplemented by X-Ray Absorption Near-Edge Structure measurements \citep[XANES;][]{righter2016valence}, \textit{X}MoO$_3$/\textit{X}MoO$_2$ = 1 near $\Delta$IW-1. Thus, at the \textit{f}O$_2$ of mare basalt source regions, Mo exists largely as Mo$^{4+}$ and the bulk D(Mo)$_{min/melt}$ ranges from $\sim$0.5 (in the absence of metal) to $\sim$5 (for a metal-bearing source) \citep{leitzke2017redox}. We use the average of Mo abundances determined in mare basalts \citep[0.033 $\pm$ 0.015, \textit{n} = 4;][]{burkhardt2014evidence} and apply the melting model of \cite{leitzke2017redox} in the absence of metal (see below) to estimate 0.019$\pm$0.010 Mo (in ppm) in the BSMoon. 
We stress that additional high-precision determinations of the Mo contents of mare basalts would be invaluable.  

\begin{figure}
     \centering
     \includegraphics[width=0.5\textwidth]{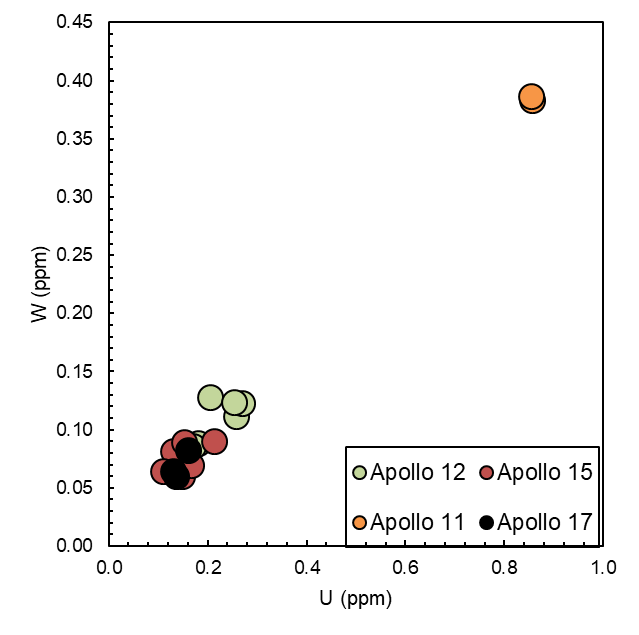}
     \caption{High precision ID-TIMS data for W (ppm) vs. U (ppm) abundances compiled from mare basalts \citep{munker2010high,kruijer2017tungsten,thiemens2019early}, subdivided into those from the Apollo 11, 12, 15 and 17 missions.}
    \label{fig:W-U}
 \end{figure}

Tungsten, as a highly incompatible element, has long been recognised to correlate near-perfectly with La \citep{wanke_etal1973,wanke_etal1977,ringwoodkesson1977} and with U \citep[][Fig. \ref{fig:W-U}]{palmerammensee1981,oneill1991origin}. Unlike Mo, W is mostly hexavalent, even at the reducing conditions of mare basalt petrogenesis \citep{oneill2008solubility}, and its D(W)$_{min/melt}$ remains $<$0.1 \citep{fonseca_etal2014} during partial melting, unless the \textit{f}O$_2$ descends below $\Delta$IW-2. Here we use W abundances in a selection of mare basalts determined by ID-MS \citep{munker2010high,kruijer2017tungsten,thiemens2019early} together with U abundances on the same samples to derive W/U = 0.47$\pm$0.07, \textit{n} = 17, and hence a W abundance of 0.011$\pm$0.002 ppm. This value is indistinguishable from that in the BSE \citep[0.013$\pm$0.002 ppm][]{kleinewalker2017tungsten}.  \\

\textit{Platinum-Group Elements (PGEs).} Platinum Group Element abundances of the BSMoon, on the basis of their concentrations in mare basalts, were deduced to be present in the lunar mantle at 0.00023 $\pm$ 0.00002 $\times$ CI abundances (Day et al. 2016). Translating PGE abundances in mare basalts to those of their mantle sources is predicated on the assumption that, upon partial melting, no PGE-bearing phases remained in the residue. 
Few - if any - primitive lunar magmas have S concentrations sufficiently high (typically between 600 - 2000 ppm) to saturate in a sulfide phase in the mantle \citep{dingdasgupta2017,steenstra2018evidence,brenan2019abundance}, nor during crustal-level differentiation \citep{steenstra2018evidence}. However, some lunar magmas could exceed the SCSS should the activity of the FeS component in the sulfide phase be significantly below unity (e.g. due to the presence of Fe-rich metal or Ni-rich sulfide in solid solution), as used to model U/W and Hf/W trends in mare basalts \cite{fonseca_etal2014}. \cite{brenan2019abundance}, propose, as a consequence, that mare basalts were indeed sulfide saturated, and that their PGE abundances do not reflect that in their mantle sources, but rather contamination with $\leq$ 1 \% (by mass) of regolith material. Metal and sulfide are both accessory phases in Apollo 12 \& 15 mare basalts \citep[e.g.,][]{brett_etal1971apollo,brett1976}, and most appear to have PGE element patterns consistent with fractional crystallisation rather than regolith contamination, supported by Pd/Ir ratios in whole rock mare basalts \citep{daypaquet2021} and other chalcophile metals \citep{day2018geochemical}. Nevertheless, contamination of mare basalt melts by regolith or of the lunar mantle by late-accreting chondritic material are two processes that are inherently difficult to disentangle. Therefore, we adopt the PGE abundances in the BSMoon determined by \cite{Dayetal2016}, noting that these may be maximum estimates.

\subsubsection{Volatile Chalcophile and Siderophile Elements}
\label{sec:geochem_VSEs}

\textit{Moderately volatile - S, Se, Te, Ga, Cu, P}. Sulfur, as the most abundant of the aforementioned elements, 
can be determined from its abundances in lunar mare basalts. Typically, S is ratioed to Dy \citep{saaletal2002,mccubbinetal2021endogenous}. Assuming the highest S/Dy ratios are representative, \cite{mccubbinetal2021endogenous} determine 88 - 202 ppm S in the lunar mantle. However, it should be noted that S concentration in a silicate melt in equilibrium with a sulfide phase, is constrained to the SCSS \citep[e.g.,][]{dingdasgupta2017}, whereas Dy, as a trace element, has no such restriction. Owing to the homogeneity in the S and chalcophile- and highly siderophile element contents of Low-Ti basalts \cite[][see section \ref{sec:geochem_RSEs}]{garganoetal2022,gleissneretal2022,daypaquet2021}, it is unlikely that they were sulfide saturated at their sources unless the activity of FeS (\textit{a}FeS) was low ($<$0.5). As such, their S contents are likely to be reflective of those in their sources.
While S is prone to degassing from lunar magmas \cite{saaletal2002,Rengglietal2017}, the uniformity in the $\delta ^{34}$S isotopic composition of mare basalts (= 0.58$\pm$0.05) implies they lost $\leq$10\% of their S budgets during emplacement \cite{wingfarquhar2015sulfur}. We use a combination of the S/Tb ratio and partial melting models \citep[after][]{gleissneretal2022} to determine the S content of the BSMoon to 81$\pm$15 ppm. The Se content follows from the constancy of S/Se ratios in Low-Ti basalts \citep[5558$\pm$315;][]{gleissneretal2022}, which is a factor $\sim$2 greater than that in Earth's mantle \citep[2690$\pm$700][]{wangbecker2013}. Tellurium abundance is given by the constant Se/Te ratio, 66 $\pm$ 7 (\textit{n} = 7) in the same samples, similar to the Se/Te in MORB, but an order of magnitude higher than in peridotites \citep[7.9$\pm$1.6, \textit{n} = 63][]{wangbecker2013}.

\begin{figure*}
     \centering
     \includegraphics[width=1\textwidth]{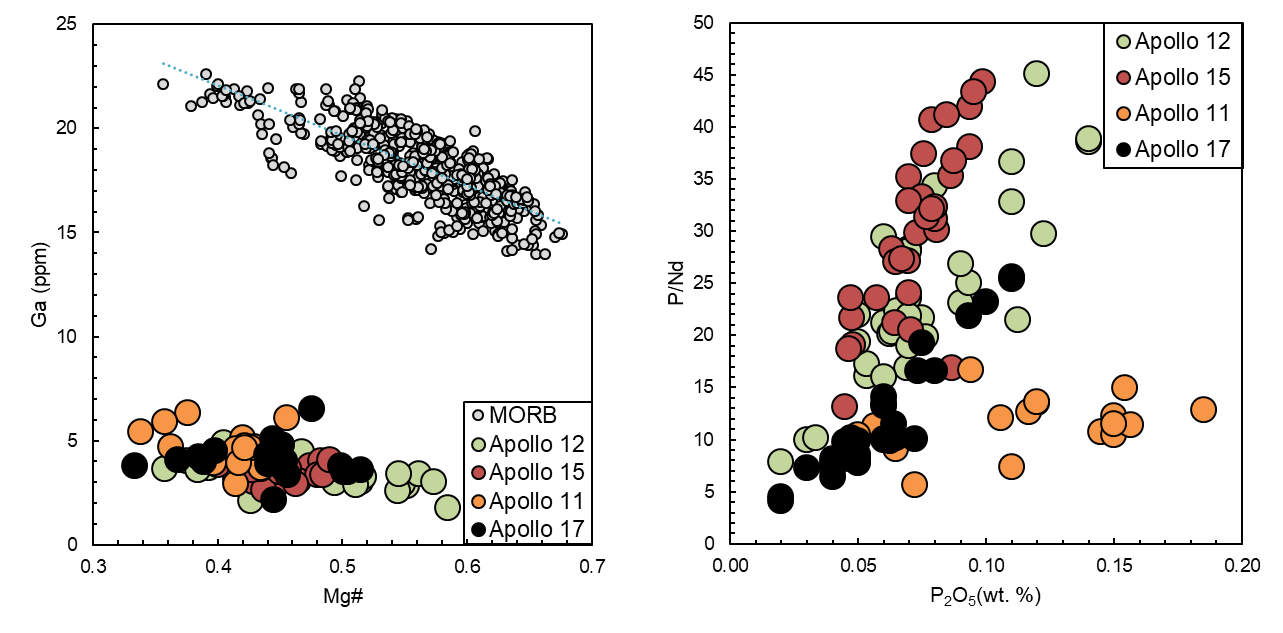}
     \caption{\textbf{a)} Distribution of Ga (ppm) as a function of Mg\# in mare basalts \citep[modified after][]{neal2007} and MORB \citep{jenneroneill2012} and \textbf{b)} arrays formed by mare basalts \citep[modified after][]{neal2007} as a function of P/Nd.}
    \label{fig:Ga-P}
 \end{figure*}

Copper contents are variable in mare basalts, ranging typically from 1--20 ppm \citep{herzogetal2009, gleissneretal2022}. 
However, amongst the highest Mg\# Low-Ti basalts, Cu abundances are homogeneous at 9.4$\pm$2.7 ppm. Given the $\sim$ 100 ppm in primitive MORB \citep{jenneroneill2012} and the 25 $\pm$ 5 ppm in Earth's mantle \citep{sunetal2020CuS}, we derive a BSMoon abundance of 2.3$\pm$1.1 ppm, indistinguishable from the 2.1 ppm proposed by \cite{oneill1991origin} and 3$\pm$1.5 ppm by \cite{righter2019} using a similar approach. Gallium contents are estimated by first noting that there is a weak inverse correlation of Ga with Mg\# in all mare basalts (\textit{r$^2$} = 0.21; Fig. \ref{fig:Ga-P}a), resulting in 3.2 ppm at the highest Mg\# (0.58) compared with 15 ppm in MORB at similar Mg\# \citep[0.62,][]{jenneroneill2012}. For 4.4 ppm Ga in the BSE, the mare basalt/MORB ratio implies 0.94$\pm$0.47 ppm in the BSMoon, in agreement with 1$\pm$0.5 found by \cite{righter2019}. However, this is a factor $\sim$2 lower than estimated by \cite{mccubbinetal2021endogenous}, who used the Ga/Al ratio given the concordance of Ga$^{3+}$ and Al$^{3+}$ in terrestrial magmas \citep{deargolloschilling1978}. However, Ga and Al show no such correlation among lunar lithologies \cite[][]{HaskinWarren1991}, neither should the two elements be compared given that Al is a major element and hence does not show Henrian behaviour. 

Although phosphorus often mirrors Nd during magmatic processes \citep{wankedreibus1986,oneill1991origin}, the P/Nd ratios of mare basalts is not constant as a function of P content, rather, they form linear arrays in P/Nd vs. P space that converge to a common value near P/Nd = 10$\pm$5 at low P contents (Fig. \ref{fig:Ga-P}b). Instead, P/La ratios are essentially uniform in all mare basalts, 35.1$\pm$9.4 \citep[as also noted by][]{righter2019} and, combining the two estimates yields 18.7$\pm$7.5 ppm P. This is in excellent agreement with the estimates of 20 ppm \citep{oneill1991origin} and 30$\pm$10 ppm \citep{righter2019}, confirming the $\sim$3- to 5-fold depletion of P relative to the BSE \citep[87$\pm$13 ppm;][]{palmeoneill2014}.  \\

\textit{Highly volatile - Pb, Sn, Ag, In, Cd, Tl}. Owing to the fact that these elements are both highly volatile and at least weakly siderophile, they are present in vanishingly small quantities in mare basalts, rendering their BSMoon abundances difficult to estimate. Of these, Pb is by far the most abundant, and can also be determined by the measured $\mu$ ($^{238}$U/$^{204}$Pb) ratios in mare basalts. Although variability exists among the High- and Low-Ti basalt types, there is consensus as to the homogeneity of the Low-Ti basalt source with $\mu$ = $\sim$ 300 - 400 \cite{snapeetal2019,connellyetal2022}. Using a mean value of 350$\pm$50, and a U abundance of 0.0229 ppm, we obtain 0.0045 $^{-0.0010} _{+0.0018}$ ppm Pb. This estimate is 2.5$\times$ higher than that of \cite{oneill1991origin} but a factor 5 lower than the value given by \cite{Haurietal2015}. 
Tin is rarely measured in lunar rocks, with an average of 45$\pm$20 ppb reported by \cite{wolfanders1980}. Modern values obtained in the context of Sn isotope measurements \citep{wangetal2019tin} are a factor $\sim$2 higher, 86$\pm$34 ppb (\textit{n} = 8) with no systematic difference between Low- and High-Ti basalts. Although these authors used the Sn/Sm ratio to estimate the lunar mantle abundances by analogy with the concordant behaviour of these two elements during terrestrial magmatic processes, this is not likely to hold during mare basalt petrogenesis, as Sn assumes its stannous state near the IW buffer \citep{roskosz_etal2020}, 
Nevertheless, we adopt the value of \cite{wangetal2019tin}, 23$\pm$8 ppb, noting that it is likely to be a minimum estimate, even though it is higher than the 6$\pm$4 ppb derived by \cite{righter2019}. 
Indium, as a trivalent cation under all plausible \textit{f}O$_2$ conditions, is known as a sister element to Y \citep[e.g.,][]{yietal2000}. In Low-Ti mare basalts, the In/Y ratio is (2.5$\pm$1.2)$\times$10$^{-5}$ \citep{gleissneretal2022}, resulting in 0.11$\pm$0.05 ppb In in the BSMoon, similar to that given by \cite{oneill1991origin}. The remaining elements, Cd, Ag and Tl, are moderately \citep[Cd, Ag;][]{witteickschen_etal2009,li2022partitioning} to highly incompatible \citep[Tl;][]{prytulak2017thallium} and show reasonable correlations with the REE in mare basalts \citep{gleissneretal2022}. \cite{yietal2000} provide evidence for constant Cd/Dy in oceanic basalts on Earth, and we find no variation of Cd/Dy, stable at (7.1$\pm$1.6)$\times$10$^{-5}$, with Cd content in Low-Ti mare basalts \citep{gleissneretal2022}. This compares with a Cd/Dy of 0.048 in mantle peridotites \citep{witteickschen_etal2009}, meaning Cd is depleted by a factor 680 in the Moon (vs. BSE, 35$\pm$7 ppb), at 0.052$\pm$0.011 ppb. This is a factor $\sim$2 lower than the 0.13 ppb of \cite{oneill1991origin} and 6 $\times$ lower than given by \cite{righter2019}, both using the older \cite{wolfanders1980} determinations with a mean Cd content in Low-Ti mare basalts of 1.99 $\pm$ 0.35 ppb, compared to 0.45$\pm$0.12 ppb in \cite{gleissneretal2022}. In Earth's mantle, Ag is largely hosted in sulfide, a product of its strong chalcophile affinity \citep[e.g.,][]{kiseevawood2015}. However, sulfide is unlikely to be a residual phase during the petrogenesis of lunar basalts, and under these conditions Ag behaves as a moderately- to highly incompatible lithophile element \citep{li2022partitioning}. In support of this assertion, mare basalts define constant Ag/Sm (2.3$\pm$0.2 $\times$ 10$^{-4}$) that results in an Ag abundance of 0.10 $\pm$ 0.01 ppb in the the BSMoon, similar to the 0.18 ppb proposed by \cite{Haurietal2015} and 3--4 $\times$ lower than given by \cite{righter2019} and \cite{oneill1991origin}. Thallium is a highly incompatible element during partial melting on Earth, with D(Tl)$_{min/melt}$ similar to Cs, Ba or U \citep{heinrichsetal1980,prytulak2017thallium}. In Low-Ti mare basalts, Tl correlates adequately with La and with Te. The corresponding ratios of Tl/La, (4.8$\pm$1.6) $\times$ 10$^{-5}$ and Tl/Te (1.5$\pm$0.3) $\times$ 10$^{-4}$, imply Tl contents of 0.033 $\pm$ 0.011 ppb, consistent with other estimates; 0.042 ppb \citep{oneill1991origin} and 0.03$\pm$0.01 ppb \citep{righter2019}, but significantly lower than the 1.5 ppb of \cite{Haurietal2015}. 

\section{Physicochemical evolution of the Moon}
\label{sec:physchem}

\subsection{Volatile Depletion}
\label{sec:physchem_volatile}
\begin{figure*}[!ht]
     \centering
     \includegraphics[width=1\textwidth]{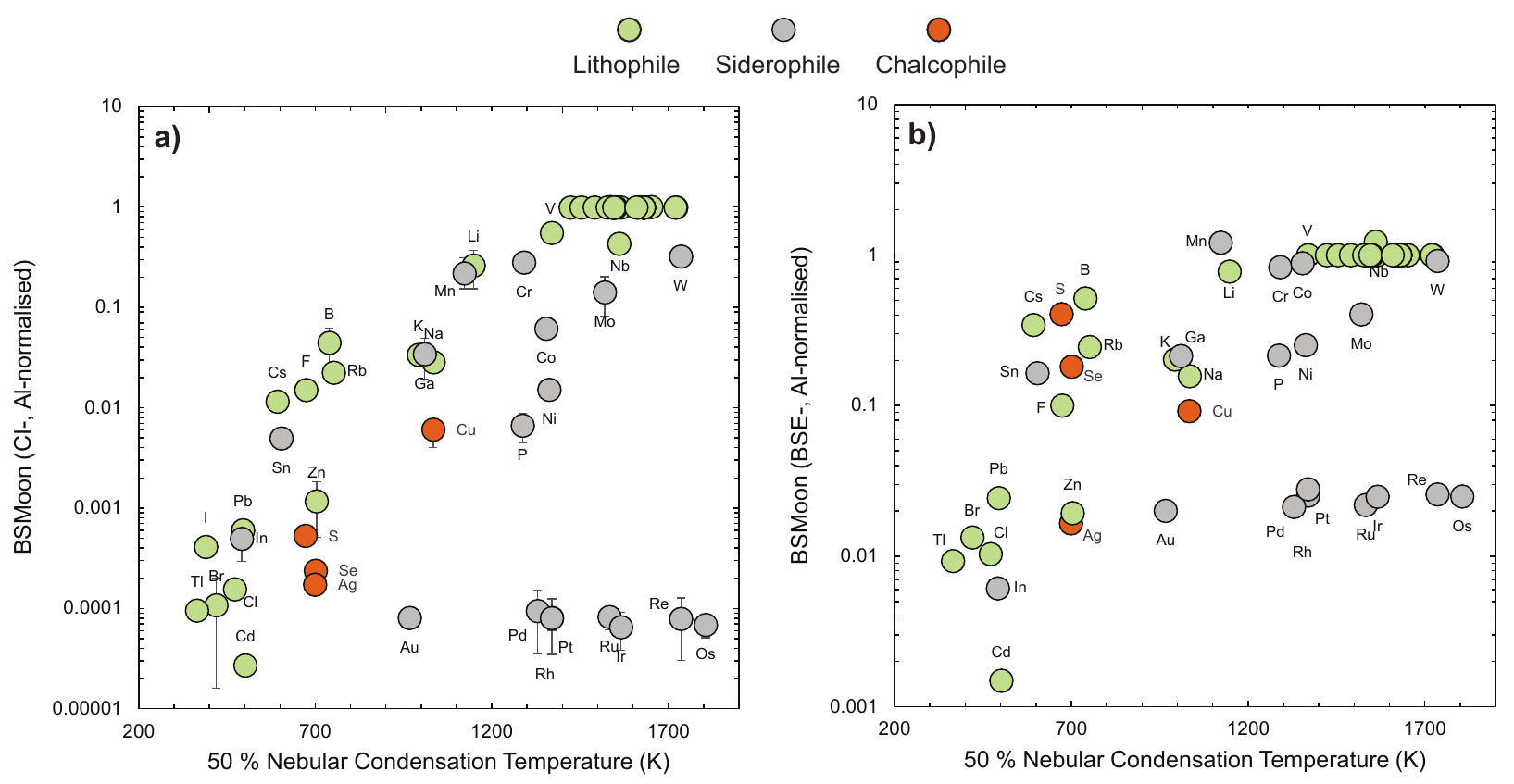}
     \caption{The abundances of elements in the bulk silicate Moon (BSMoon; Table \ref{tab:lunar_abu}) normalised to Al- and \textbf{a)} CI chondrites and \textbf{b)} the bulk silicate Earth (BSE), plotted as a function of their 50 \% nebular condensation temperatures \citep{woodetal2019condensation}. Elements are colour-coded by their nominal affinity for silicate (lithopile; green), metal (siderophile, grey) and sulfide (chalcophile, orange) during planetary accretion and differentiation.}
     \label{fig:Moon_Tc}
 \end{figure*}

One of the characteristic features of the Moon that emerges from the composition of its mantle (Table \ref{tab:lunar_abu}) is the extensive depletion in volatile elements compared to both CI chondrites (Fig. \ref{fig:Moon_Tc}a) and the bulk silicate Earth (Fig. \ref{fig:Moon_Tc}b). Although the volatile-poor nature of the Moon has long been recognised (section \ref{sec:geochem_compmoon}), new determinations of the abundances of trace volatile elements in lunar mare basalts and pyroclastic glasses \citep[e.g.,][]{ni2019melt,garganoetal2020,gleissneretal2022} and their associated isotopic compositions \citep[][]{panielloetal2012, wangjacobsen2016, niedauphas2019} have afforded better-constrained conclusions to be drawn from their distributions. \\

\cite{kreutzberger_etal1986}, through experimentation in which lunar analogue silicate melts were heated under controlled oxygen fugacity, demonstrated that the alkali elements evaporate in the order Na $>$ K $>$ Rb $>$ Cs, \textit{i.e.,} down the group. \cite{oneill1991origin} assessed the likelihood that lunar volatile depletion was caused by evaporation from silicate liquids (or analogues thereof) on the basis of thermodynamic data for their presumed vaporisation reactions. The recognition that the Li content of the lunar mantle is equivalent, within uncertainty, to that of the Earth's while Na is depleted by a factor of $\sim$5 was used to argue that volatile loss must have occurred at low temperatures, near 1400 K \citep[][see also Table \ref{tab:lunar_abu}]{oneill1991origin}; too low for silicate melts to have been stable. That the isotopic composition of Li (expressed as $\delta ^{7/6}$Li) of lunar mare basalts and their terrestrial counterparts are equivalent within measurement uncertainty substantiates the notion that Li behaved conservatively during the formation of the Moon \citep{magna2006}. The absence of volatile loss of any element more refractory than Li is supported by the similarity in the stable isotope compositions of these elements in the Earth and Moon, including Fe \citep{sossimoynier2017, poitrassonetal2019}, Si \citep{armytage2012silicon}, Mg \citep{sedaghatpouretal2013}, Ti \citep{millet2016titanium} and Ca \citep{simondepaolo2010}, even when chondrites and basaltic achondrites exhibit a significant range. \\

In detail, the abundances of the alkalis are correlated with those of refractory lithophile elements among \textit{all} lunar lithologies \citep[including the ferroan anorthosites,][]{ringwoodkesson1977, HaskinWarren1991}, while the stable isotope compositions of Zn and K compared to Earth's mantle \citep{panielloetal2012,wangjacobsen2016} are uniformly heavy. Both observations, together with the uniformly low K/Th ratio of the lunar surface \citep[$\sim$360,][]{Prettymanetal2006} attest to the depletion in volatiles being a characteristic feature of the whole Moon, as originally suggested by \cite{ringwood1970special}, rather than pertaining to localised volatile loss during eruption. Temperature constraints on volatile depletion are therefore instrumental in distinguishing between plausible dynamical scenarios for lunar formation (Section \ref{sec:dynamics}). \\

To first order, there is evidence that elements taken to be refractory under solar nebular conditions, such as the rare-earth elements (REEs), behaved as such during Moon formation, on the grounds that volatility-driven abundance patterns, as observed in some CAIs \citep{boynton1975,hu2021REE} are not detected in lunar rocks. This is not unexpected, in that the Moon almost certainly formed long ($>$50 Myr, see section \ref{sec:timing}) after the nebular gas had dispersed \citep[$\sim$ 5 Myr;][]{weiss2021nebula}. An important corollary thereof is that the H$_2$-rich nebular gas would have no longer imposed an \textit{f}O$_2$ on the material from which the Moon was forming. Instead, as pointed out by \cite{oneill1991origin} and developed more recently by \cite{VisscherFegley2013_MVE,canup2015,Locketal2018,Tartese2021,ivanov2022} and \cite{fegley2023chemical}, any vapour phase would have been generated from proto-lunar material itself. In order for such a vapour to impart volatile depletion on the residual, Moon-forming matter, it must have been subsequently removed from the system. Considering that the present-day Moon is overwhelmingly comprised of silicates, they would have also likely been a major constituent at the locus of vapour production. Consequently, vaporisation reactions of the type;

\begin{equation}
M^{x+n}O_{(x+n)/2}(l,s) = M^{x}O_{n/2}(g) + \frac{n}{4}O_2(g),
\label{eq:vap_general}
\end{equation}
would have prevailed during lunar volatile loss, where $M^{x+n}O_{(x+n)/2}$ pertains to the oxide component of the element, \textit{M} dissolved in silicate (liquid or solid), $M^{x}O_{n/2}$ the equilibrium gaseous (oxide) species and \textit{n} the number of electrons exchanged. In this scenario, \textit{f}O$_2$ is set by the dissociative vaporisation of silicate material \citep[e.g.,][]{VisscherFegley2013_MVE,sossifegley2018}.
However, even small quantities of metallic iron could have buffered the \textit{f}O$_2$ of the system to values $\sim$ 1 log unit below the Iron-Wüstite buffer ($\Delta$FMQ $-4.5$). Assuming equilibrium, the resultant partial pressure (\textit{p}) of M$^{x}O_{n/2}$ is given by:

\begin{equation}
\ln p(M^{x}O_{n/2}) = \frac{-\Delta G^o_{\ref{eq:vap_general}}}{RT}+\ln a(M^{x+n}O_{[x+n]/2})- \ln\frac{n}{4}{f(O_2)}
\label{eq:partialpressure}
\end{equation}

where \textit{a}($M^{x+n}$O$_{[x+n]/2}$) refers to the activity of $M^{x+n}$O$_{(x+n)/2}$ in the condensed phase(s) and $\Delta G^o_{\ref{eq:vap_general}}$ is the Gibbs free energy of reaction, taking the reference pressure (1 bar) and a convenient standard state for $M^{x+n}$O$_{n/2}$, such as the pure liquid/solid component at the temperature of interest and 1 bar. Without knowledge of the mass of vapour relative to that of condensed material, the absolute quantity of the element, \textit{M}, vaporised is not resolved. A more tractable computation is to define the \textit{relative} fraction of an element, \textit{M$_1$}, evaporated with respect to a comparator element, \textit{M$_2$}, as the total mass of the system cancels from the operation \citep{oneill1991origin}. In silicate/metal systems, a single stable gaseous species prevails for most moderately volatile elements \citep{sossietal2019}, such that the mass fraction of an element in the vapour, $f^{vap}$, assuming ideal gas behaviour, is proportional to the partial pressure of the dominant species:

\begin{equation}
f^{vap} = \frac{X({M^{x}O_{n/2})}}{X_0} = \frac{p(M^{x}O_{n/2})}{X_0 P_T}
\label{eq:f_vap}
\end{equation}

where $X_0$ is the mole fraction of the element initially in the system and $P_T$ is the total pressure. By definition, the fraction remaining in the condensed phase is then ($1 - f_{vap}$), which is itself equivalent to:

\begin{equation}
(1 - f^{vap}) = \frac{X({M^{x+n}O_{[x+n]/2})}}{X_0} = \frac{a(M^{x+n}O_{x+n/2})}{X_0\gamma(M^{x+n}O_{[x+n]/2})}
\label{eq:1-f_vap}
\end{equation}

When combining eqs. (\ref{eq:f_vap} and \ref{eq:1-f_vap}) with eq. (\ref{eq:partialpressure}) for two elements, \textit{M$_1$} and \textit{M$_2$}, the total pressure cancels, as does the initial amount of the element in the system, while \textit{f}O$_2$ also cancels provided the vaporisation reactions of the two elements have the same \textit{n}, resulting in:

\begin{equation}
\frac{f^{vap}_{M1}}{(1-f^{vap}_{M1})} = \frac{f^{vap}_{M2}}{(1-f^{vap}_{M2})} \left(\frac{\gamma _{M1}}{\gamma _{M2}}\right) f(O_2)^\frac{\Delta{n_{M2-M1}}}{4} e^{\frac{(\Delta G^o_{M2}-\Delta G^o_{M1})}{RT}}
\label{eq:relative_vap}
\end{equation}

where ($\Delta G^o_{M2}-\Delta G^o_{M1}$) relates to the free energy of the exchange reaction between vapour and condensed phase(s), such as:

\begin{equation}
Li(g) + NaO_{0.5}(s,l) = Na(g) + LiO_{0.5}(s,l)
\label{eq:Li-Na}
\end{equation}

For a given composition, the relative vaporisation of elements in reactions of the type of eq. (\ref{eq:Li-Na}) are sensitive only to temperature, provided their respective activity coefficients in the condensed phase(s) of interest are known. The uncertainty in activity coefficients is also mitigated given that the operative variable is their \textit{ratio}. Elements of a particular valence can be arranged into components of the same moiety in silicate liquids; if these components bear resemblance to crystalline phases, then they are expected to mix near-ideally \citep[e.g.,][]{ghiorsocarmichael1980}. Enthalpies of fusion will also cancel for two crystalline phases, provided they have similar melting points. This approximation is particularly useful for putative alkali oxide components (e.g., LiAlSi$_3$O$_8$, KAlSi$_3$O$_8$, NaAlSi$_3$O$_8$). For divalent cations, experimental data for the activities of their oxide species (e.g., FeO, NiO, CoO) in silicate melts indicate that they mix near-ideally relative to the pure liquid metal oxide \citep{oneilleggins2002, woodwade2013}, while SiO$_2$, as a major component of silicate melts, has activities close to unity. \\

\begin{figure}[!ht]
     \centering
     \includegraphics[width=0.5\textwidth]{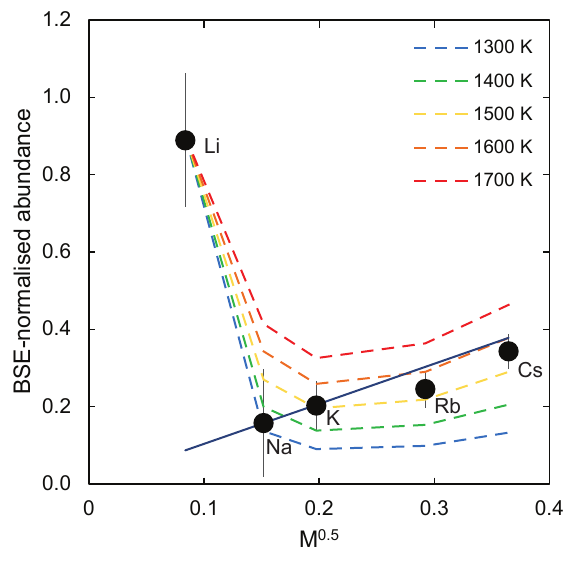}
     \caption{The bulk silicate Earth (BSE)-normalised abundances of the alkali metals in the bulk silicate Moon as a function of the square root of their molar masses. The coloured, dashed lines correspond to the modelled abundances of Na, K, Rb and Cs, using eq. \ref{eq:relative_vap}, assuming ($1-f_{Li}^{vap}$) = 0.9, at different temperatures; 1300 K (blue), 1400 K (green), 1500 K (yellow), 1600 K (orange), 1700 K (red). The solid black line is calculated according to the equation ($1-f_{A}^{vap}$) = ($1-f_{Na}^{vap}$) $\times \sqrt{M_{A}/M_{Na}}$, where A = Li, Na, K, Rb and Cs.}
     \label{fig:alkali_volatility}
 \end{figure}

Elements for which thermodynamic properties of stable silicate phases (and activities in silicate liquids) are available are listed in Table \ref{tab:evap_table}. In constraining temperature, only the alkali elements are used, as their vaporisation stoichiometries eliminate any dependence on \textit{f}O$_2$ (eq. \ref{eq:Li-Na}). Moreover, the entropy change across alkali oxide exchange reactions are similar (Table \ref{tab:evap_table}), such that ($\Delta G^o_{M2}-\Delta G^o_{M1}$) is nearly independent of temperature, and relative volatilities are proportional to exp(1/$T$). The corollary is that low temperatures exaggerate the discrimination between elements that are readily vaporised (e.g., Rb) from those that are not (e.g., Li). The abundances of the alkali metals in the BSMoon relative to the BSE are shown in Fig. \ref{fig:alkali_volatility}. \\

\begin{table}[!ht]
    \centering
    \caption{Thermodynamic properties of evaporation reactions for selected moderately volatile elements of the form eq. \ref{eq:vap_general}. All data from \cite{sossietal2019} except for Ga and In from \cite{bischof2023}. }
    \begin{tabular}{ccccc}
    \hline
        \textbf{} & \textbf{$\Delta$H$^o$ (kJ/mol)} & \textbf{$\Delta$S$^o$ (kJ/mol.K)} & \textbf{$n$} & \textbf{$\gamma_0$$^{*}$} \\ \hline
        \textbf{Li} & 419.0 & 0.141 & 1 & 0.2  \\ 
        \textbf{Cs} & 217.0 & 0.113 & 1 & 3.0$\times 10^{-5}$ $^\dag$  \\ 
        \textbf{Rb} & 220.5 & 0.119 & 1 & 3.0$\times 10^{-5}$  \\ 
        \textbf{Ga} & 754.7 & 0.248 & 3 & 0.036  \\ 
        \textbf{K} & 236.1 & 0.122 & 1 & 7.3$\times 10^{-5}$  \\ 
        \textbf{Na} & 267.0 & 0.121 & 1 & 5.0$\times 10^{-4}$  \\ 
        \textbf{Zn} & 411.7 & 0.177 & 2 & 0.17  \\ 
        \textbf{Ag} & 279.5 & 0.141 & 1 & 0.22  \\ 
        \textbf{In} & 690.8 & 0.255 & 3 & 0.017  \\ 
        \textbf{Ge} & 471.2 & 0.230 & 2 & 0.03  \\ 
        \textbf{Cd} & 355.0 & 0.194 & 2 & 0.05  \\ \hline
        \multicolumn{5}{l}{\parbox{8.5cm}{$^{*}$\footnotesize{$\gamma_0$ calculated for the oxide in silicate liquid at a reference temperature, $T_0$, of 1673 K, which yields $\gamma$ at the desired temperature, $T$, using the equation ln$\gamma$ = $\frac{T_0}{T}$ln$\gamma_0$.\\
        $^{\dag}$\footnotesize{$\gamma_0$ for CsO$_{0.5}$ assumed to be equivalent to that for RbO$_{0.5}$.}}}}
    \end{tabular}
    \label{tab:evap_table}
\end{table}

\begin{figure}[!ht]
     \centering
     \includegraphics[width=0.5\textwidth]{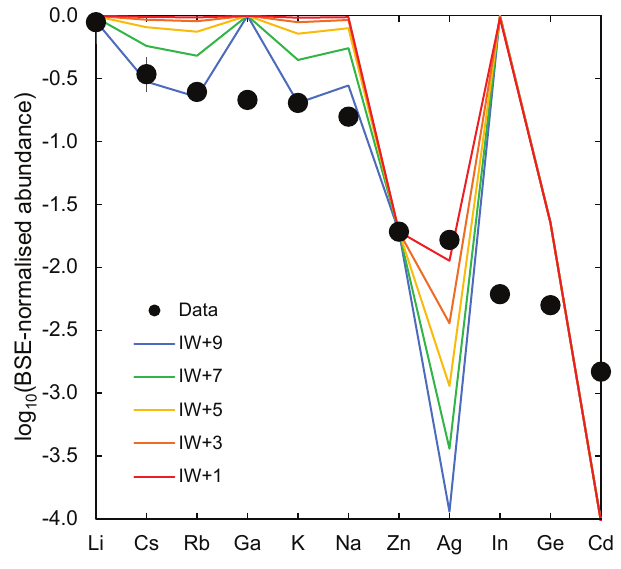}
     \caption{The log$_{10}$ bulk silicate Earth (BSE)-normalised abundances of selected volatile elements in the bulk silicate Moon ordered by their depletion factors ($1-f^{vap}$). The coloured lines correspond to the modelled abundances in the bulk silicate Moon, using eq. \ref{eq:relative_vap},  assuming ($1-f_{Zn}^{vap}$) = 0.02, at different oxygen fugacities relative to the iron-wüstite (IW) buffer; +9 (blue), +7 (green), +5 (yellow), +3 (orange), +1 (red) and at constant temperature (1500 K).}
     \label{fig:volatiles_all_Moon}
 \end{figure}

The computation illustrates the necessity for alkali vaporisation to have occurred at low temperatures, roughly 1500 K, in order to preserve 90 \% of the Moon's Li budget, whilst simultaneously vaporising 80 \% of that of Na, from a BSE-like initial composition. Should volatile loss have occurred at higher temperatures ($>$1800 K) then the alkali metals should each be depleted to a similar extent relative to one another (i.e., $f^{vap}$ for Li and Na would be similar), a feature that is not observed. Since these temperatures are too low to engender vapour loss from a fully-formed Moon via thermal escape \citep[see][section \ref{sec:lunarvolatileloss}]{tangyoung2020,charnoz2021tidal}, separation of vapour from small mm- to cm-sized precursor material to the Moon may have been responsible for its present-day volatile element budget. Such low temperatures are also a far cry from the liquidus temperature of mantle peridotite \citep{hirschmann2000}, suggesting that vapour loss likely occurred sub-solidus in equilibrium with silicates, plausibly olivine, pyroxene and feldspar, and possibly metallic solid/liquid iron-rich alloy.
In order to assess whether iron-rich alloys were present, exchange reactions that depend on \textit{f}O$_2$ should be examined. The same treatment applied to the alkali metals is extended to Ga, Zn, Ag, In, Ge and Cd (Fig. \ref{fig:volatiles_all_Moon}). \\

Assuming a constant temperature fixed by the relative abundances of the alkalis, what emerges is that very high \textit{f}O$_2$s ($>$IW+7) are required to simultaneously reproduce the abundances of the alkali metals and Zn (Fig. \ref{fig:volatiles_all_Moon}). However, such conditions would overestimate Ag loss, whose vaporisation stoichiometry mirrors those of the alkalis. Moreover, Ga and In, owing to their \textit{n} = 3 evaporation reactions, behave as refractory elements under oxidising conditions, whereas their lunar abundances are slightly (Ga) or markedly (In) lower than in the bulk silicate Earth. There is therefore a stark disconnect between the degree of depletion among the alkali metals (+Ga) and the highly volatile elements (Cd, Ag, Ge, In, Zn). The measured abundances of the latter group in lunar mare basalts was initially theorised to be the result of addition of a small amount ($\sim$ 4 \% mass of the Moon) of chondritic material \citep{wolfanders1980,oneill1991origin}. However, this scenario would have resulted in a chondritic Zn isotope composition of the Moon, which is at-odds with the highly fractionated Zn isotope composition of lunar mare basalts \citep[$\delta ^{66}$Zn $\sim$ +1.5 $\permil$][]{panielloetal2012,Katoetal2015} relative to terrestrial basalts ($\delta ^{66}$Zn $\sim$ +0.3 $\permil$). The prevailing interpretation, therefore, is that the lunar abundances of these elements are residual from vapour loss event(s). \\

In aid of determining the nature of such an event(s), \cite{jonespalme2000} noted that the apparent depletions of the alkali metals in the Moon are inversely proportional to their volatilities deduced by \cite{kreutzberger_etal1986}, leading the authors to suggest that the loss of heavier alkalis was stymied by a mass transport process. Indeed, the abundances of alkali metals in the Moon with respect to the Earth increase as a function of the square root of their molar mass (Fig. \ref{fig:alkali_volatility}). However, this correlation may be fortuitous, in that the abundances of the heavier alkalis are poorly defined in both the Earth and Moon \citep[cf.][]{mezger_etal2021}. Such a pronounced mass-dependent control on vapour loss efficacy would have also resulted in extensive kinetic stable isotope fractionation of these elements, which is not observed (see below). \\

Therefore, given that the observed depletion factors of the elements are likely to reflect their equilibrium volatilities, the combination of temperatures that are below the peridotite liquidus ($\sim$ 1400 K) and the recognition that there is no single oxygen fugacity that satisfactorily accounts for the volatile element composition of the Moon leads to the conclusion that vapour loss may have occurred under differing assumptions than those considered here. Plausible alternatives include vapour segregation \textit{i}) in stages rather than in a single event (and hence with an evolving, non-BSE-like composition), \textit{ii}) between gas and solid(s), rather than gas and liquid(s),  and \textit{iii}) under different thermodynamic conditions, for example, in the presence of an H, C, S, N, and/or Cl-rich atmosphere or with liquid Fe-Ni alloy stable. \\

The difficulty in reproducing the abundances of the alkali metals on the one hand, and highly volatile elements on the other, during a single event was noted by \cite{ivanov2022}, who proposed that the Moon was formed by mixing between a high temperature (4000 K) condensate with a lower temperature (2000 $-$ 2500 K) condensate in the presence of liquid metal. However, their model predicts strong ($\sim$ 50 \%) depletions in the main components, Mg, Si, and Fe, which are not observed.  Although \cite{Locketal2018} presented an ostensible match between the composition of liquid condensates at 3000 $-$ 3500 K and $\sim$ 10 bar and that of the bulk silicate Moon, these computations considered only a subset of volatile elements (i.e., Li, Rb, In, Cd and Cs were absent). \cite{fegleylodders2017} also showed that such temperatures would lead to the development of pronounced negative Ce anomalies in the Moon, which are not observed\footnote{Though LREE depletions are detected for the Moon based on Ce isotopes \citep{hasenstab2023cerium}, these are inferred to have been inherited from the Earth rather than from vaporisation.}. These temperatures are, however, significantly higher than what is often experimentally achievable \cite[e.g.,][]{sossifegley2018}, behooving caution when extrapolating these data to potential Moon-forming conditions.  \\

\begin{figure}[!ht]
     \centering
     \includegraphics[width=0.5\textwidth]{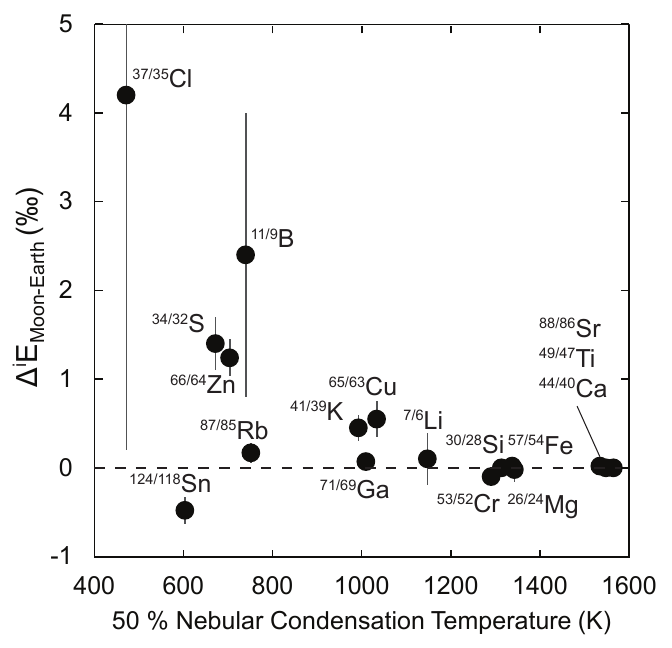}
     \caption{The mass-dependent stable isotope difference between the Moon and the Earth expressed in delta notation ($\permil$), $\Delta ^i$E$_{Moon-Earth}$ as a function of the 50 \% nebular condensation temperature of the element \citep{woodetal2019condensation}. Estimates for the composition of the Moon are inferred from measurements of mare basalts, whereas terrestrial data largely come from peridotites. See Table \ref{tab:stables} for data sources.}
     \label{fig:Stables_Tc}
 \end{figure}

The equilibrium approach detailed above provides an additional testable hypothesis; because equilibrium mass-dependent isotopic fractionation scales with 1/$T^{2}$ \citep{urey1947}, observed Moon-Earth stable isotope differences ($\Delta ^i$E$_{Moon-Earth}$, where E = the element of interest) should match those determined between the relevant vapour and condensed species at the equivalent temperatures. To determine the isotopic composition of the Moon, mare basalts are widely used, owing to the lack of shallow-level degassing they experienced \citep[e.g.,][]{garganoetal2022}, in contrast to pyroclastic glasses \citep{saalhauri2021}. A compilation of  $\Delta ^i$E$_{Moon-Earth}$ inferred from the compositions of mare basalts (Table \ref{tab:stables}), cast against the 50 \% nebular condensation temperature of the element substantiates the notion that there exists a causative relationship between the degree of isotopic fractionation between the Moon and the Earth and the volatility of the element (Fig. \ref{fig:Stables_Tc}). Only elements more volatile than Li exhibit significant departures from the 0 value (\textit{i.e.,} from Moon-Earth equivalence). The homogeneity of mare basalts with respect to these stable isotope ratios has been used as evidence to suggest that there was a lunar scale process that lead to this isotope fractionation, invariably equated with a giant impact event \citep[e.g.,][section \ref{sec:dynamics}]{panielloetal2012,wangjacobsen2016}.  \\

\begin{table*}[!ht]
    \centering
    \caption{Summary of BSE-normalised depletion factors (\textit{f$_{Moon}$} = X$_{BSMoon}$/X$_{BSE}$), mass dependent stable isotopic data ($\Delta ^i$E$_{Moon-Earth}$) and their associated uncertainties (standard deviation for abundances, 2 sigma for isotope ratios) for various elements between the Earth and Moon.}
    \begin{footnotesize}
    \begin{tabular}
    {p{0.06\linewidth}p{0.06\linewidth}p{0.04\linewidth}p{0.03\linewidth}p{0.06\linewidth}p{0.1\linewidth}p{0.03\linewidth}p{0.43\linewidth}}
    \hline
         {\textbf{Element}} & {\textbf{Ratio}} & {\textbf{$f_{Moon}$}} & \Centering{\textbf{$\pm$}} & {\textbf{T$_c$ (K)$^*$}} & {\textbf{$\Delta ^i$E$_{Moon-Earth}$}} & \Centering{\textbf{$\pm$}} & \Centering{\textbf{Reference }} \\ \hline
        \textbf{Cl} & 37/35 & 0.01 & 0.00 & 472 & 4.20 & 4.00 & \cite{boyceetal2018,garganoetal2022} \\ 
        \textbf{Sn} & 124/116 & 0.16 & 0.01 & 604 & -0.48 & 0.15 & Wang et al. (2019)  \\ 
        \textbf{S} & 34/32 & 0.32 & 0.07 & 672 & 1.40 & 0.30 & \cite{wingfarquhar2015sulfur,saalhauri2021,garganoetal2022,dottin2023}  \\ 
        \textbf{Zn} & 66/64 & 0.02 & 0.01 & 704 & 1.24 & 0.21 & \cite{panielloetal2012,Katoetal2015,garganoetal2022}  \\ 
        \textbf{B} & 11/9 & 0.53 & 0.21 & 740 & 2.40 & 1.60 & \cite{zhai1996boron,marschall2017boron}  \\ 
        \textbf{Rb} & 87/85 & 0.25 & 0.01 & 752 & 0.17 & 0.11 & \cite{pringle2017rubidium, niedauphas2019}  \\ 
        \textbf{K} & 41/39 & 0.21 & 0.01 & 993 & 0.45 & 0.15 & \cite{wangjacobsen2016}  \\ 
        \textbf{Ga} & 71/69 & 0.21 & 0.10 & 1010 & 0.07 & 0.10 & \cite{katomoynier2017gallium, wimpenny2022gallium, render2023gallium} \\ 
        \textbf{Cu} & 65/63 & 0.09 & 0.03 & 1034 & 0.55 & 0.20 & \cite{dayetal2019rr}; Paquet et al. (submitted)  \\ 
        \textbf{Li} & 7/6 & 0.89 & 0.22 & 1148 & 0.10 & 0.30 & \cite{magna2006}  \\ 
        \textbf{Cr} & 53/52 & 0.83 & 0.10 & 1291 & -0.10 & 0.04 & \cite{bonnand2016Cr,sossietal2018} \\ 
        \textbf{Si} & 30/28 & 1.00 & 0.25 & 1314 & 0.00 & 0.03 & \cite{savage2010silicon,armytage2012silicon,zambardi2013silicon} \\ 
        \textbf{Fe} & 57/54 & 1.00 & 0.25 & 1338 & 0.02 & 0.03 & \cite{sossimoynier2017,poitrassonetal2019} \\ 
        \textbf{Mg} & 26/24 & 1.00 & 0.25 & 1343 & -0.02 & 0.14 & \cite{sedaghatpouretal2013}  \\ 
        \textbf{Ca} & 44/40 & 1.00 & 0.25 & 1535 & 0.02 & 0.08 & \cite{simondepaolo2010,valdes2014Ca}  \\ 
        \textbf{Sr} & 88/86 & 1.00 & 0.25 & 1548 & 0.00 & 0.05 & \cite{moynier2010sr}  \\ 
        \textbf{Ti} & 49/47 & 1.00 & 0.25 & 1565 & 0.00 & 0.02 & \cite{millet2016titanium,greber2017titanium}  \\ \hline
        \multicolumn{8}{l}{$^{*}$\footnotesize{50 \% nebular condensation temperatures from \cite{woodetal2019condensation}.}}
    \end{tabular}
    \end{footnotesize}
    \label{tab:stables}
\end{table*}

\begin{figure*}[!ht]
     \centering
     \includegraphics[width=1\textwidth]{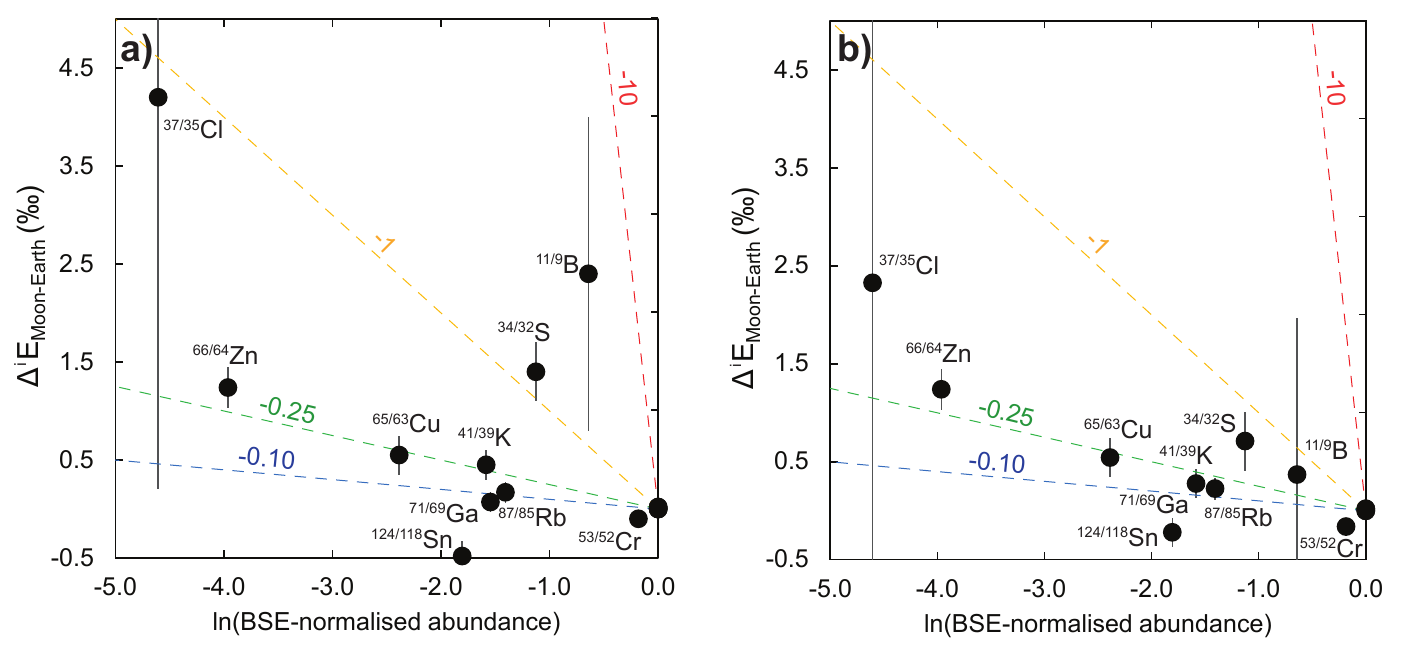}
     \caption{The mass-dependent stable isotope difference between the Moon and the Earth expressed in delta notation ($\permil$), $\Delta ^n$E$_{Moon-Earth}$ as a function of the natural logarithm of their BSE-normalised abundance in the bulk silicate Moon (equivalent to ln\textit{f} in eq. \ref{eq:rayleigh_isotope}, where \textit{f} = the fraction of the element remaining in the residue). The slope of the lines gives the fractionation factor between the Moon and the Earth, $\Delta ^i$E$_{Moon-Earth}$, and are shown in colour for values of -10 (red), -1 (yellow), -0.25 (green) and -0.1 (blue). Part \textbf{a)} shows the raw data and \textbf{b)} scales the isotopic data by the mass difference, $\Delta E^{i/j}$ and $\beta$$\Delta E^{k/j}$  scale with $\beta$ = ((1/$m_j - $1/$m_i$)/(1/$m_j - $1/$m_k$)), relative to that of $^{66/64}$Zn. Estimates for the composition of the Moon are inferred from measurements of mare basalts, whereas terrestrial data largely come from peridotites. See Table \ref{tab:stables} for data sources.}
     \label{fig:Rayleigh_Stables}
 \end{figure*}

The extent of stable isotope fractionation in a given volatile element is correlated with the extent to which it is depleted in comparison to a bulk silicate Earth-like precursor (Fig. \ref{fig:Rayleigh_Stables}a), which is particularly evident when the isotope fractionation is scaled by relative atomic mass unit (Fig. \ref{fig:Rayleigh_Stables}b). On the basis of this correlation, all models for the depletion of volatile elements from the Moon invoke Rayleigh fractionation \citep[e.g.,][]{panielloetal2012}, irrespective of the mechanism by which it occurs \citep[e.g.,][]{niedauphas2019};

\begin{equation}
\delta E^i_{s,l} - \delta E^i_{0} \simeq [\Delta E^{i,eq}_{g-s,l}+(1-S_i)\Delta E^{i,kin}_{g-s,l}] \ln f \simeq \Delta E^i_{Moon-Earth}
\label{eq:rayleigh_isotope}
\end{equation}

in which a finite reservoir of an isotope of mass \textit{i}, of a given element, $E$, with an initial isotopic composition $\delta^i_{0}$ (here, equated with the BSE) evolves to a new isotopic composition ($\delta E^i_{s,l}$; here the Moon) proportional to the natural logarithm of the fraction of the element remaining in the condensed phase(s) ($f$), and the fractionation factor, $\Delta$, between gas, \textit{g}, and condensed phase(s), \textit{s,l}. Depending on the degree to which equilibrium is attained at the gas-condensed phase interface, (\textit{p$_i$}/\textit{p$_{i,sat}$}), $\Delta E^{i,eq}_{g-s,l}$ is scaled between pure equilibrium evaporation (\textit{p$_i$}/\textit{p$_{i,sat}$} $\rightarrow$ 1), in which the fractionation factor is determined only by differences in bond stiffness between the two phases, and pure kinetic evaporation (\textit{p$_i$}/\textit{p$_{i,sat}$} $\rightarrow$ 0), in which $\Delta E^{i,kin}_{g-s,l}$, for a gas with a Maxwellian velocity distribution, is given by 10$^3$ln$\sqrt{m_j/m_i}$, where \textit{i} $>$ \textit{j}. \\

With few exceptions, isotopic differences between the Moon and the Earth are explicable with fractionation factors, $\Delta E^{i}_{Moon-Earth}$ $\sim$ $-$0.25 $\permil$ (Fig. \ref{fig:Rayleigh_Stables}. As mass-dependent stable isotope fractionation factors between two isotope ratios, $\Delta E^{i/j}$ and $\beta$$\Delta E^{k/j}$  scale with $\beta$ = ((1/$m_j - $1/$m_i$)/(1/$m_j - $1/$m_k$)) \citep{young2002laws}, the scatter in Fig. \ref{fig:Rayleigh_Stables}a is largely removed by weighting the $\Delta E^{i}_{Moon-Earth}$ factor by the relative mass differences of \textit{i} and \textit{j}. The observation that the $\Delta E^{i}_{Moon-Earth}$ of both alkali (K and Rb) and highly volatile elements (Zn, Cl) fall along a curve of constant slope (within uncertainty) supports the hypothesis that their loss from the Moon was related to a single event. Nevertheless, deviations from the line (in particular for the isotopes of S, Sn and Cr) indicate factors other than the extent of element loss are at play (see below), while the uncertainties on Cl and B isotopic ratios span a range of plausible fractionation factors. In either case, that these fractionation factors are much smaller (i.e., closer to 0) than those predicted for pure kinetic isotope fractionation has been taken as evidence of near-equilibrium conditions of vapour loss during formation of the Moon \citep{panielloetal2012,wangjacobsen2016}.  \\

Because equilibrium mass dependent stable isotope fractionation factors ($\Delta E^{i,eq}_{g-s,l}$) decrease proportional to 1/$T^2$ \citep{urey1947}, should equilibrium at the vapour-solid/liquid segregation have been established, the isotopic difference between the Moon and the Earth can be used as a thermometer. In an equilibrium scenario in which gas-liquid segregation occurs at $\sim$3500 K, as envisaged for a synestia \citep[][section \ref{sec:dynamics}]{Locketal2018}, isotope fractionation factors are essentially nil. Therefore, under these conditions, the synestia model lacks a mechanism for engendering isotopic fractionation in the condensed phase. Should the equilibrium condition be relaxed, \cite{niedauphas2019} and \cite{Dauphasetal2022_MVE} demonstrated that the observed isotopic fractionation between Moon and Earth could be achieved, even at 3500 K, if the saturation of the gas (\textit{p$_i$}/\textit{p$_{i,sat}$}) in the gas phase at the evaporating surface was equal to 0.99. 
However, such an interpretation is non-unique, as no attempt has yet been made to combine the modelled isotopic fractionation factors with the extent of volatile loss of each element. For example, at 1600 K, the value of \textit{p$_i$}/\textit{p$_{i,sat}$} required increases to $\sim$ 0.995, while it reaches unity at $\sim$ 1300 K \citep[see also][]{Tartese2021}. Therefore, although the isotopic composition of volatile elements in the Moon alone cannot discriminate between these scenarios, it should be highlighted that, in each case, a very high degree of equilibrium (\textit{p$_i$}/\textit{p$_{i,sat}$} $>$ 0.99) is required to account for the observations. Physically, this likely pertains to a high density (i.e., low mean free path) and/or stationary (i.e., non-advecting) gas phase at the locus of volatile depletion, such that mass transport rates of the evaporating species were slow with respect to vaporisation rates \citep[see also][section \ref{sec:dynamics}]{wangjacobsen2016,niedauphas2019,sossietal2020,tangyoung2020,charnoz2021tidal}. \\

An additional uncertainty arises from determining the degree to which the observed Moon-Earth fractionation of some elements reflects volatile loss. Notably, the $\delta ^{71}$Ga of ferroan anorthosites correlates with the An content (molar Ca/[Ca+Na] ratio) of the constituent plagioclase \citep{wimpenny2022gallium, render2023gallium}, with inferred bulk Moon values of $\delta ^{71}$Ga being indistinguishable from that of the BSE. Gallium is also a moderately siderophile/chalcophile element, whose depletion in the BSMoon may also result, in part, from core formation \citep[e.g.,][]{righter2019}. The same is true of Cu, Sn, S and Fe, which compromises their utility for constraining the conditions of volatile loss on the Moon. Consequently, the isotopes of elements with a purely lithophile affinity, such as the alkalis and Zn, are more faithful witnesses to volatile depletion of the Moon. The isotopic compositions of Cr and Sn show lighter values than in the BSE \citep{sossietal2018,wangetal2019tin}, where $\Delta$Cr = $\sim$ -0.10 $\pm$0.04 $\permil$/amu and $\Delta$Sn = -0.06 $\pm$0.02 $\permil$/amu. In the event that Cr and Sn isotope variations are the direct consequence of vapour loss processes, then equilibrium (\textit{p$_i$}/\textit{p$_{i,sat}$} = 1) is required, as any degree of kinetic loss would produce heavy isotope compositions in the residue (i.e., the BSMoon; eq. \ref{eq:rayleigh_isotope}). \\

Should equilibrium between gas- and condensed phase(s) have prevailed, then condensation/evaporation could have only occurred at low temperatures ($\sim$1400 K), a scenario consistent with the relative abundances of the alkali metals (Fig. \ref{fig:alkali_volatility}). A corollary of such low temperatures is that condensation must have occurred predominantly into solid silicate and metal grains, rather than into silicate or metallic liquid. The oxides of alkali metals (AO$_{0.5}$) have low activities when dissolved in silicate liquids \citep[10$^{-4}$ and below,][]{charles1967,mathieu2011,sossietal2019,zhang2021loss}, causing a reduction in the partial pressure of their gaseous species relative to the pure oxide. By contrast, alkalis are hosted in plagioclase feldspar\footnote{Li is an exception and condenses into orthopyroxene.} during nebular condensation \citep{woodetal2019condensation}, in which deviations from ideality are essentially non-existent \citep[$\gamma$NaAlSi$_3$O$_8$ $\sim$ 1 $-$ 2;][]{orville1972plagioclase,hollandpowell1992} or modest \citep[$\gamma$KAlSi$_3$O$_8$ $\sim$ 10;][]{ghiorso1984fsp}. However, the partial pressure of Na, \textit{p}Na, above pure albite at 1673 K and above an hypothetical silicate melt at the same \textit{f}O$_2$ (6.1 $\times$ 10$^{-7}$ bar) and \textit{X}NaO$_{0.5}$ (0.115) in which the activity coefficient is 10$^{-4}$ \citep{sossietal2019,wolf2023vaporock} is essentially the same, 2.5 $\times$ 10$^{-6}$ bar and 4.8 $\times$ 10$^{-6}$ bar, respectively. Therefore, condensation of the alkalis into solids, rather than into silicate liquids, will not appreciably alter their volatilities relative to the highly volatile elements.\\

Gallium and In are two elements whose oxide species also have low activity coefficients in silicate liquids \citep{woodwade2013}, and appear to behave in an anomalously refractory manner with respect to other volatile elements (Fig. \ref{fig:volatiles_all_Moon}). The Moon is depleted in Ga and In by a factor $\sim$ 5 and $\sim$ 150, respectively (Table 
\ref{tab:lunar_abu}), yet their volatilities from silicate melts suggest minimal depletion, unless the \textit{f}O$_2$ is below $\Delta$IW+1 (Fig. \ref{fig:volatiles_all_Moon}). Other than the higher relative $f$O$_2$, plausible reasons for their relatively refractory behaviour in silicate melts include their condensation into metal in the solar nebula, and the absence of complexing anions relative to those present in the solar nebula. This latter scenario was examined by \cite{waiwasson1979}, who attributed the volatility of Ga during nebular condensation to the formation of GaCl(g) and GaOH(g) species. \\

Whether these species (and their analogues for In and the  alkalis) were extant in the Moon-forming system depends upon the nature of the source material. Presuming it was largely derived from the proto-Earth's mantle, then chlorine alone would not have been abundant enough ($\sim$ 30 ppm) relative to the abundances of the alkali metals (Na alone is present at $\sim$100 $\times$ Cl) to stabilise significant proportions of chloride species. This conclusion holds even assuming chondritic Na/Cl ratios, which are invariably near 10 for a variety of chondrite groups \citep{wassonkallemeyn1988}.  \\

Hydrogen, on the other hand, is present at $\sim$ 1000 ppm levels (when expressed as H$_2$O) in the bulk silicate Earth \citep{Marty2012}, comparable to Na abundances. The presence of H stabilises hydrides (\textit{e.g.,} LiH) or hydroxides (\textit{e.g.,} LiOH) from monatomic gases \citep[e.g.,][]{Fegleyetal2016}, the formation constants of which are related to the fugacities of H$_2$ or H$_2$O in the gas phase via the homogeneous equilibria: \\

\begin{equation}
M^0(g) + H_2(g) = M^{n+}H_{n}(g)
\label{eq:hydride_fm}
\end{equation}

\begin{equation}
M^0(g) + H_2O(g) =  M^{n+}[OH]_{n}(g)
\label{eq:hydroxide_fm}
\end{equation}

where M stands for the metal of interest. For an ideal gas at a given temperature and pressure, the \textit{f}H$_2$/\textit{f}H$_2$O ratio is fixed, provided the \textit{f}O$_2$ is known. For plausible Moon-forming environments, oxygen fugacities are expected to lie between $\Delta$IW-2 and $\Delta$IW+4 \citep{oneill1991origin,VisscherFegley2013_MVE,Tartese2021}, resulting in \textit{f}H$_2$/\textit{f}H$_2$O of $\sim$ 10 or $\sim$ 0.01, respectively (the dependence on temperature is small). However, the Gibbs Free Energies of formation of alkali hydrides are such that \textit{f}H$_2$/\textit{f}H$_2$O $>$ 100 are typically required before the fugacities of alkali hydrides and -hydroxides become subequal at 1500 K. Therefore, while reaction \ref{eq:hydride_fm} can be ignored for alkali metals in lunar environments, GaH and InH are predicted to be stable at \textit{f}O$_2$s near IW and below. Equilibrium constants for equation \ref{eq:hydroxide_fm} differ significantly depending on the identity of $M$. In particular, the \textit{f}LiOH/\textit{f}Li ratio is $\sim$600 at the IW buffer at 1500 K and 1 bar, while the corresponding ratios for Na- and K-bearing gas species are $\sim$1 and $\sim$4, respectively, at the same conditions \citep[see also][]{ivanov2022}. The result is that the volatility of Li would be expected to increase considerably with respect to its higher-\textit{Z} counterparts as \textit{f}H$_2$O increases. The observation that Li is not resolvably depleted in the Moon with respect to the Earth (in contrast to Na or K) therefore suggests that the environment in which vapour loss occurred was probably nearly anhydrous (\textit{i.e.,} H$_2$O-poor). This conclusion would be reinforced were the pressure to have exceed 1 bar, as higher pressures favour the formation of the associated species \citep[eq. \ref{eq:hydroxide_fm},][]{sossifegley2018}. \\

What emerges from both the thermodynamics and stable isotope data is that there is no unequivocal evidence that volatile loss from the Moon occurred under high temperatures, although such a scenario cannot be excluded. Rather, it appears more likely that volatile loss from the Moon occurred at low temperatures ($\sim$1400 K) and hence did so predominantly in the solid-vapour system. A plausible environment for vapour to have separated from residual solids may have been a proto-lunar disk (see Section \ref{sec:dynamics}). However, the total amount of vapour lost must have been negligible, as the five oxides that together comprise $\sim$ 98.5 \% of the Earth's mantle, SiO$_2$, Al$_2$O$_3$, MgO, FeO and CaO were not vaporised (and lost) to any appreciable extent from the Moon or its precursor material (relative to refractory elements). Even supposing that Moon-forming material lost much of its budget of H$_2$O during this process (1000 ppm in the BSE), the total mass loss it underwent was likely $<$ 1 \% of its present-day mass.





\subsection{Core Formation}
\label{sec:physchem_corefm}

The lunar core, as constrained by geophysical inversions, has a bulk density between $\sim$6000 and $\sim$8000 kgm$^{-3}$ (Table \ref{tab:inv_com}, section \ref{sec:geophys_core}). As these densities far exceed those of common silicate minerals, to first order, the lunar core must be predominantly composed of the cosmochemically abundant metals, Fe, Ni, Co with a smaller, unknown fraction of light elements. Geophysical inversions (section \ref{sec:geophys_inv}) indicate that, for BSE-like Mg\#s in the lunar mantle (0.89), the core is essentially devoid of light elements, whereas Mg\#s of $\sim$0.80 imply 5--15 wt~\% S in the core (Fig. \ref{fig:S_core}). Mole fractions, \textit{X}, of the metal, \textit{M} (where \textit{M} = Fe, Ni and Co) can be predicted self-consistently based on their free energies of formation at 1 bar, corrected for their non-idealities in multicomponent liquid silicate- and metal alloys, according to reactions of the type;

\begin{equation}
M^{n+}O_{n/2}(l) = M^{0}(l) + \frac{n}{4}O_2(g),
\label{eq:met-sil}
\end{equation}

where \textit{n} is the oxidation state of the metal cation in the silicate. At equilibrium, the mole fraction, \textit{X(M$^0$)}, in the alloy is given by;

\begin{equation}
X(M^{0}) = \frac{X(M^{n+}O_{n/2}).\gamma(M^{n+}O_{n/2}).K}{f{O_2}^\frac{n}{4}.\gamma(M^{0})}
\label{eq:K-met-sil}
\end{equation}

The equilibrium constants, \textit{K}, are assessed using data compiled from experimental data of the free energies of formation for metals and metal oxides \citep{OneillPownceby1993}. Activity coefficients, $\gamma$ of \textit{M}$^{n+}$O$_{n/2}$ in silicate liquids are compiled from literature data \citep{oneilleggins2002,oneill2008solubility,woodwade2013} while those for $M^{0}$ in the alloy phase are calculated using \texttt{METALACT} \citep{wood2017}. Equilibrium constants are modified for pressure using the approach of \cite{wadewood2005}. Furthermore, if the composition of the metallic phase is subject to the constraint,

\begin{equation}
X(Fe) + X(Ni) + X(Co) = 1
\label{eq:met-comp}
\end{equation}

then the composition of the metal is uniquely known for a given pressure, temperature, \textit{f}O$_2$ and \textit{X}($M^{n+}O_{n/2}$) of each \textit{M}. The values of \textit{X}(NiO) and \textit{X}(CoO) in the lunar mantle are set to 6.0 $\times$ 10$^{-4}$ and 1.2 $\times$ 10$^{-5}$, respectively (see section \ref{tab:lunar_abu}). \\

\begin{figure*}[!ht]
     \centering
     \includegraphics[width=1\textwidth]{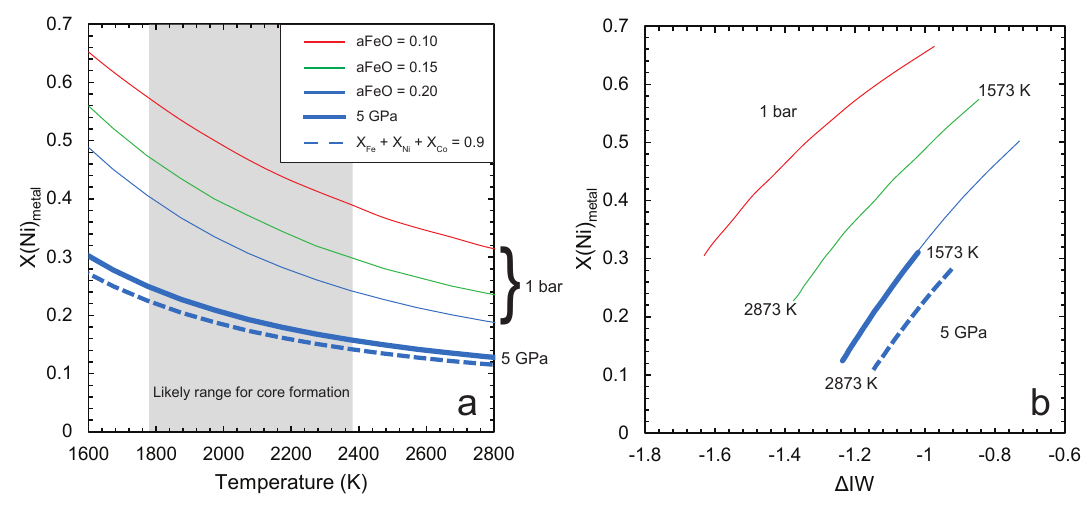}
     \caption{Modelled composition of lunar core-forming alloy phase in equilibrium with a mantle with FeO activity in the silicate, \textit{a}FeO, of 0.1 (red), 0.15 (green), 0.2 (blue) at 1 bar (thin lines) and 5 GPa (thick lines) assuming \textit{X}Fe + \textit{X}Ni + \textit{X}Co = 1 (solid lines) or \textit{X}Fe + \textit{X}Ni + \textit{X}Co = 0.9 (dashed line) both at \textit{a}FeO = 0.2. The resulting mole fraction of Ni in the alloy, \textit{X}Ni, is plotted as a function of \textbf{a)} Temperature (K) and \textbf{b)} oxygen fugacity relative to the iron-wüstite buffer ($\Delta$IW). The grey shaded region denotes the likely temperature range for core formation based on the liquidi of peridotite and Fe-Ni alloy.}
     \label{fig:Ni_fO2_core}
 \end{figure*}

Given that the mole fraction of FeO the lunar mantle, \textit{X}(FeO), is likely to lie in the range 0.06 to 0.12 (Table \ref{tab:inv_com}) and its activity coefficient in Mg-rich silicate liquids is $\sim$ 1.75 at 1673 K \citep{woodwade2013}, the composition of the core-forming metal on the Moon is calculated at \textit{a}(FeO) = 0.1, 0.15 and 0.2 at 1 bar. Calculations are also performed at 5 GPa for \textit{a}(FeO) = 0.2, as well as for a light-element enriched case, in which the right-hand side of equation \ref{eq:met-comp} is set equal to 0.9 (\textit{i.e.,} 10 mol \% of light elements). Resulting lunar core compositions for a range of likely temperatures, \textit{T} (K) $\in$ (1573, 2873), are shown in Figs. \ref{fig:Ni_fO2_core}a and b. \\

The model indicates that, at 1 bar, the lunar core should have a Ni-rich composition, with \textit{X}(Ni) between $\sim$ 0.3 and 0.7, where both increasing temperature and \textit{a}(FeO) cause \textit{X}(Ni) to decrease (Fig. \ref{fig:Ni_fO2_core}). \cite{oneill1991origin} performed a similar exercise (though in the olivine-orthopyroxene-metal system) and recovered equivalent results; that the Ni fraction of the metallic phase is $\sim$0.5 at 1573 K and at Fe$_2$SiO$_4$ activities relevant to the Moon (0.17). However, the calculations of \cite{oneill1991origin} were performed at 1 bar. Given that the volume change of eq. \ref{eq:met-sil} for Ni is positive at a given $T$ and $f$O$_2$, increasing pressure causes the logarithm of the partition coefficient of Ni, log[D(Ni)] = log[XNi$_{met}$/XNi$_{sil}$] to decrease proportional to $-P \Delta V / RT$ \citep{liagee2001}. Consequently, for a given temperature and \textit{a}FeO, the equilibrium \textit{X}Ni is lower (by a factor of $\sim$1.5) at 5 GPa relative to that at 1 bar. Dilution of the metallic phase by a light component produces a concomitant decrease in the mole fractions of the existing elements (in this case, 10 \%). \\

Oxygen fugacities for core-mantle equilibrium, assuming 0.1 $<$ \textit{a}FeO $<$ 0.2, are in the range $\Delta$IW-1.6 to $\Delta$IW-0.7. Knowledge of the FeO content of the lunar mantle (and hence, its activity in the liquid phase) is therefore key for constraining the \textit{f}O$_2$ of core-mantle equilibrium, with lower values giving rise to more reducing conditions, all else being equal. Oxygen fugacities for lunar core formation are never as low as those inferred for core-mantle equilibrium on the Earth \citep[$\Delta$IW-2.2;][]{frost2008} as, even for BSE-like FeO contents, the higher pressures and temperatures under which Earth's core formed resulted in much lower \textit{X}(Ni), and hence lower \textit{f}O$_2$ (Fig. \ref{fig:Ni_fO2_core}b)\footnote{Although this effect may have been partially offset by increased dissolution of Si into Earth's core relative to that in the Moon, see \cite{frost2008,Rubieetal2011}.}. On the other hand, dilution of the metal by a light element decreases \textit{X(M)} while leaving \textit{a}FeO unaffected, thereby increasing the relative \textit{f}O$_2$ by a constant factor (here 0.09 log units). \\

In order to use this thermodynamic model to predict the observed abundances of siderophile elements in the lunar mantle, the mass fraction of the core, \textit{f}$_{core}$ is required. Geophysical inversions indicate that 0.008 $<$ \textit{f}$_{core}$ $<$ 0.015, positively correlated with the Mg\# of the lunar mantle (Table \ref{tab:inv_com}), with the higher value corresponding to Earth's mantle-like cases (Mg\# = 0.89) and the former to FeO-rich compositions (Mg\# = 0.80). This permits the calculation of the abundance of a given metal in the lunar mantle by mass balance:

\begin{equation}
X(M)_{mantle} = \frac{X(M)_{bulk}}{f_{core}D(M)+(1-f_{core})}
\label{eq:mass_bal}
\end{equation}

To do so therefore also requires an estimate of that in the bulk Moon, \textit{X}(M)$_{bulk}$, which is not known \textit{a priori}. Typically, this quantity is taken to be \textit{X}(M)$_{bulk}$ in the bulk silicate Earth \cite[e.g.,][]{Righter2018, steenstra2020}, on the strength of the observation that the Earth and Moon have identical compositions in various mass-independent isotope systems \citep[particularly the elements, Ca, Ti, Cr and O;][]{wiechert2001,mougel2018,simondepaolo2010,zhang_ti2012} that vary widely among other planetary materials. Second, the composition of the bulk Moon, to first order, resembles that of the bulk silicate Earth \citep[][Table \ref{tab:lunar_abu}]{ringwoodkesson1977}. Consequently, these observations are used to justify the assumption that \textit{X}(M)$_{bulk, Moon}$ = \textit{X}(M)$_{bulk, BSE}$, one that is adopted herein. The veracity of this assumption will be discussed in light of the partitioning models. \\

\begin{figure}[!ht]
     \centering
     \includegraphics[width=0.5\textwidth]{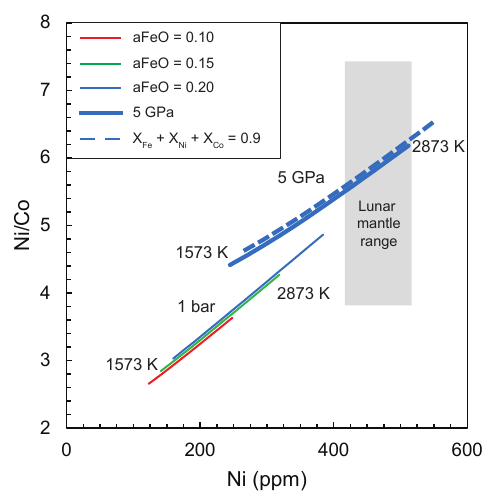}
     \caption{The Ni/Co ratio of the lunar mantle as a function of its Ni abundance as predicted by the thermodynamic model outlined in eq. \ref{eq:met-sil} to \ref{eq:mass_bal}, assuming core mass fraction, \textit{f}$_{core}$ = 0.01 (see Table \ref{tab:inv_com}) and \textit{X}(M)$_{bulk, Moon}$ = \textit{X}(M)$_{bulk, BSE}$. Assuming higher \textit{f}$_{core}$ would shift the curves to the bottom-left, while higher \textit{X}(M)$_{bulk, Moon}$ would shift the curves to the right by the factor \textit{X}(M)$_{bulk, Moon}$/\textit{X}(M)$_{bulk, BSE}$, at constant Ni/Co. Colours and symbols as per Fig. \ref{fig:Ni_fO2_core}.}
     \label{fig:Ni_Co_core}
 \end{figure}

\begin{figure}[!ht]
     \centering
     \includegraphics[width=0.5\textwidth]{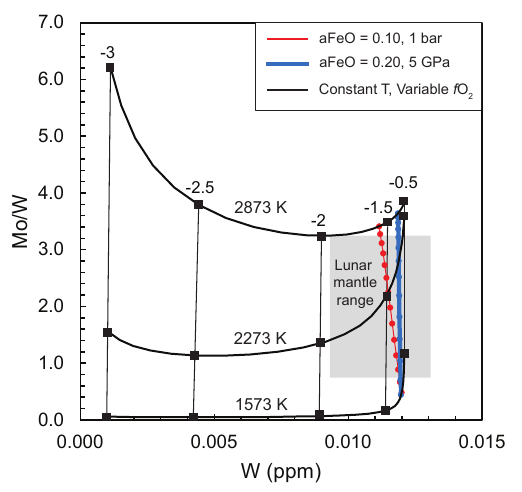}
     \caption{The Mo/W ratio of the lunar mantle as a function of its W abundance as predicted by the thermodynamic model outlined in eq. \ref{eq:met-sil} to \ref{eq:mass_bal}, assuming core mass fraction, \textit{f}$_{core}$ = 0.01 (see Table \ref{tab:inv_com}) and \textit{X}(M)$_{bulk, Moon}$ = \textit{X}(M)$_{bulk, BSE}$. Black curves show models of Mo and W abundances as a function of \textit{f}O$_2$, contoured at 0.5 log unit intervals expressed relative to the iron-wüstite (IW) buffer (legibility permitting), for three different temperatures (1573 K, 2273 K, 2873 K) at 1 bar, using the model of \cite{oneill2008solubility}.}
     \label{fig:Mo_W_core}
 \end{figure}

To constrain the pressure-temperature-\textit{f}O$_2$ conditions of lunar core formation whilst invoking the fewest number of assumptions, the analysis is restricted only to those elements \textit{i)} whose abundances are precisely constrained in the lunar mantle, \textit{ii)} that behave in a non-volatile manner during the formation of the Moon, meaning volatile depletion does not contribute to their depletion, and \textit{iii)} that are only weakly- to moderately siderophile, such that their abundances in the lunar mantle are not readily overprinted by accretion of small fractions of metal (\textit{e.g.}, the late veneer). Applying these criteria leaves but four elements, Ni, Co, Mo and W, with which to assess core formation models (Figs. \ref{fig:Ni_Co_core}, \ref{fig:Mo_W_core}).\\

The Ni/Co ratio of the lunar mantle (5.2) is an excellent arbiter of the conditions of core formation on the Moon, not only because of the high abundances of Ni and Co that lead to a low uncertainty in the Ni/Co ratio \citep[$\pm$1.2,][]{delano1986}, but also because the ratio of their partition coefficients, D(Ni)/D(Co) is independent of oxygen fugacity (that is, \textit{n} = 2 in eq. \ref{eq:met-sil} for both Co and Ni). Although increasing pressure decreases both D(Ni) and D(Co), its effect is more pronounced for Ni, leading to the conclusion that, under the current set of assumptions, only core formation at elevated pressures ($\sim$ 5 GPa) can simultaneously produce the observed Ni depletion and Ni/Co ratio in the lunar mantle (Fig. \ref{fig:Ni_Co_core}). At lower pressures, unrealistically high temperatures are required \citep[$\geq$ 3000 K, even for \textit{a}FeO = 0.2, see also][]{steenstra2020}. The effect of adding a light component is marginal, noting that this does not account for potential changes in $\gamma M$ with the addition of the light element. At the presumed S contents of the lunar core \citep[\textit{X}(S) $\leq$0.15,][]{steenstra2018evidence} $\gamma M$ diverges little relative to the ideal case, as D(Ni) values decrease by $\sim$ 10\% relative per 0.05 \textit{X}(S). \\

As the bulk Moon is depleted in iron, a siderophile element, with respect to the bulk Earth by $\sim$ 3/4 (cf. section \ref{sec:geophys_core}), then Ni, an element even more siderophile than Fe under all plausible conditions, must therefore be depleted by at least this factor in the bulk Moon, setting a strict upper limit of $X$(Ni)$_{bulk}$ = 4300 ppm. The corresponding apparent D(Ni) (eq. \ref{eq:mass_bal}) would increase from $\sim$200 assuming $X$(Ni)$_{bulk}$ = 1860 ppm (BSE-like) to $\sim$450 in order to reach the observed lunar mantle Ni abundance. Yet, as Ni/Co remains unchanged given their similar siderophility, all models in Fig. \ref{fig:Ni_Co_core} shift to the right by a factor 4300/1860, meaning temperatures and/or pressures needed to satisfy the Ni and Co abundances of the lunar mantle are correspondingly lower, or $f_{core}$ correspondingly higher. The 5 GPa cases would require temperatures below the liquidus of Fe-Ni alloy ($<$1573 K), making such high pressures unlikely for $X$(Ni)$_{bulk}$ = 4300 ppm (not shown). Moreover, solutions for the 1 bar cases intersect the lunar mantle range only at X(Ni)bulk > BSE and high temperatures ($>$2300 K; not shown). Even at a reasonable upper limit of $X$(Ni)$_{bulk}$ = 4300 ppm, the \textit{a}FeO = 0.1, 1 bar case would require implausibly high temperatures ($\geq$3000 K) to reach the high Ni/Co ratio of the lunar mantle, independent of core mass fraction and the Ni and Co budgets of the bulk Moon (provided the bulk Moon Ni/Co ratio remains that of the BSE). These results highlight that $X$(Ni, Co)$_{bulk}$ is likely close to that for the BSE, and that conditions near $\Delta$IW-1 remain a robust feature of core formation on the Moon.  \\

Molybdenum and tungsten provide complementary information on the conditions of core formation, owing to the sensitivity of their incorporation into Fe-Ni alloy on \textit{f}O$_2$ \citep[cf.][]{schmitt1989}. Molybdenum is expected to be predominantly tetravalent at $\Delta$IW-1 to -2 \citep{leitzke2017redox,oneill2008solubility,righter2016valence}, while W has been shown to exist predominantly as W$^{6+}$, even under oxygen fugacities relevant to core formation \citep{oneill2008solubility,jennings2021}. Consequently, \textit{n} in eq. \ref{eq:met-sil} is equal to 4 and 6 for Mo and W, respectively. Compared to Ni or Co (\textit{n} = 2), their D$_{met-sil}$ increases a factor of 10$^{(n_{Mo}-n_{Ni,Co})/4}$ = 10$^{0.5}$ for Mo and 10$^{(n_{W}-n_{Ni,Co})/4}$ = 10 for W with decreasing \textit{f}O$_2$ relative to D(Ni,Co)$_{met-sil}$. Figure \ref{fig:Mo_W_core} illustrates the sensitivity of the mantle abundances of Mo and W to temperature, pressure and \textit{f}O$_2$. \\

As the \textit{f}O$_2$ varies only modestly for the equilibrium between the modelled Fe-Ni alloy and silicate phases (between $\Delta$IW-1.6 and $\Delta$IW-0.7; Fig. \ref{fig:Ni_fO2_core}), the variation in Mo/W and W abundances in the lunar mantle are commensurately minor, meaning nearly all models lie within the lunar mantle range (Fig. \ref{fig:Mo_W_core}). That the W abundances in the terrestrial- ($\sim$0.012 $\pm$0.02 ppm) and lunar mantles $\sim$0.011 $\pm$0.02 ppm are indistinguishable (section \ref{sec:geochem_RSEs}, Table \ref{tab:lunar_abu}) indicates that the oxygen fugacity during lunar core formation must have been relatively high, as already implied by Ni-Co systematics (Figs. \ref{fig:Ni_fO2_core}, \ref{fig:Ni_Co_core}). This is in stark contrast to the $\sim$10-fold depletion of W in the terrestrial mantle with respect to chondritic meteorites \citep[0.08 - 0.19 ppm][]{wassonkallemeyn1988}, reflective of the lower \textit{f}O$_2$ and/or higher mass fraction of the terrestrial core. The Mo abundance is comparatively poorly constrained, such that plausible lunar mantle Mo/W ratios (and their abundances) only definitively preclude core formation under conditions more reducing than $\Delta$IW-1.8, as this would have otherwise resulted in resolvable W depletion (Fig. \ref{fig:Mo_W_core}). This constraint can be relaxed slightly presuming X(W)$_{bulk}$ was higher than the BSE value, but, as a refractory element, this seems implausible. 
Even so, oxygen fugacities much below $\Delta$IW-2 remain unfeasible.

\subsection{A terrestrial origin for the Moon?}
\label{sec:physchem_terrestrial}
Assessing one of the prevailing hypotheses that the Moon is derived from the BSE is testable in the context of chemical and isotopic data on lunar samples. A clear distinction between BSE-like source material \citep[cf.][]{Dauphasetal2014} and direct derivation from the BSE \citep[cf.][]{ringwoodkesson1977} should be made. The former postulates that the two bodies originated in the same part of the protoplanetary disk, while the latter suggests the Moon was a physically ejected from the Earth's mantle after it experienced core formation. 
\\

Mass independent isotopes unequivocally indicate identical compositions for Earth and Moon, identified initially for O \citep{wiechert2001} to be equivalent to a level of $\pm$ 30 ppm (0.3 $\permil$). Although \cite{Herwartzetal2014} detected a difference of 12 $\pm$ 3 ppm between the Moon and Earth, more recent measurements showed that the two bodies were homogeneous to a precision of 5 ppm (Young et al. 2016). \cite{Canoetal2020} argue that there may exist internal heterogeneities in lunar lithologies up to $\sim$10 ppm, but the processes leading to these variations are poorly understood. Recent unpublished data indicate O isotopic homogeneity to within 1 ppm \citep{fischer2021}. Regardless, the $<$5 ppm divergence between the Earth and Moon in the context of the $\sim$6 $\permil$ (600 ppm) range of chondritic meteorites \citep{mittlefehldt2008oxygen} implies a close genetic connection. The isotopic similarity in O, as a partially volatile element, lead to the development of models that involved efficient mixing in a gaseous phase between Earth- and Moon-forming material \citep{PahlevanStevenson2007}. However, the discovery that the equivalence in the isotopic composition between the two bodies extends to more refractory elements, notably Ti \citep{trinquier_etal2009,zhang_ti2012} and Cr \citep{qin_etal2010,mougel2018} and more recently also Ca \citep{schiller_etal2018} and Zr \citep{AkramSchoenbaechler2016}, permits two possible scenarios for the formation of the Moon; \textit{i)} that Moon-forming material mixed perfectly with the proto-Earth \citep[see][]{Locketal2018,lock2020geochemical} or \textit{ii)} that the impactor shared (near-)identical mass-independent isotopic compositions with the (proto-)Earth, obviating or reducing the need for physical mixing of Earth- and Moon-forming material \citep[see][]{Dauphasetal2015,MastrobuonoBattistietal2015}. \\

There is a strong precedent for the plausibility of option \textit{ii)}; the enstatite chondrites and Earth have similar mass-independent isotopic compositions for a multitude of elements, including O, Cr and Ti \citep{Dauphas2017}. Mass--independent isotopic concordance arising by formation of material from the same feeding zone \citep[e.g.,][]{MastrobuonoBattistietal2015} is, however, more difficult to reconcile with the observation that Mn, Cr and V are present in the Moon in Earth's mantle-like abundances (or nearly so; Table \ref{tab:lunar_abu})\footnote{Note that, although the Mn contents of mare basalts are \textit{higher} than those of terrestrial basalts by a factor $\sim$1.5, so are their Fe contents, such that the lunar MnO/FeO is actually \textit{lower}, 0.014 (see section \ref{sec:geochem_MnCrV}) than in the BSE (0.017), meaning it cannot be excluded that the Moon is actually marginally depleted in Mn relative to the BSE.}. The significance of this observation was pointed out by \cite{ringwoodkesson1977}, noting that it implies that the Moon formed almost exclusively from the mantle of the proto-Earth. The basis for this conclusion comes from experimental evidence for the lithophile behaviour of Mn, Cr and V at 1 bar \citep[e.g.,][]{rammensee_etal1983,drake_etal1989}, relative to their increasing tendency to partition into the core-forming metallic phase at increasing \textit{T} and \textit{P} conditions, namely, those present on Earth during the formation of its core \citep{ringwood1991partitioning,mann_etal2009,Siebertetal2011,siebert_etal2018}. Moreover, an assessment of the inferred Mn, Cr and V abundances of small telluric parent bodies to basaltic achondrites (Mars, Vesta, Angrite Parent bodies) demonstrated that lunar and terrestrial basalts are unique with respect to their depletions of Mn, Cr and V \citep{drake_etal1989,ruzicka_etal2001,oneillpalme2008}, lending credence to the notion that their depletion must have resulted from processes endemic to the Earth. 
Indeed, V is a refractory element and Cr behaves almost as such during nebular condensation \citep[though it becomes volatile under oxidising conditions,][]{sossietal2019}. Constraints on the extent of depletion of Mn, the most volatile of the trio, can be placed by the thermodynamics of exchange reactions akin to eq. \ref{eq:Li-Na}, which indicate that, for 90 \% condensation of Li and \textit{f}O$_2$ = IW, Mn will be condensed to a level of 99 \% \citep{oneill1991origin}.  Consequently, their depletions in the Moon are unlikely to have been brought about by volatility.  \\

It seems probable, therefore, that the depletions of Mn, Cr and V in the Moon (relative to chondrites) were inherited from those in Earth's mantle \textit{post}-core formation. On the other hand, \cite{chabotagee2003} postulate that, were core formation on the proto-Moon (or a putative impactor) to have occurred at very high temperatures ($>$ 3000 K) and reducing conditions ($\Delta$IW-2 and below), then the depletions of Mn, Cr and V would have been sufficient to account for their abundances in the BSMoon. Even under these conditions, endogenous core formation on the Moon is incapable of modifying their abundances owing to its small mass \citep[0.8 - 1.5 \%, Table \ref{tab:inv_com}, see also][]{steenstra2020}, meaning that, in this scenario, their depletions must have been imputed by an impactor. An isotopically Earth-like impactor cannot satisfy this requirement unless the impactor itself underwent core-formation at elevated \textit{P/T} ($\sim$ 50 GPa) in order to produce liquidus temperatures that were sufficiently elevated to render Mn, Cr and V siderophile. Assuming a similar core mass fraction in the impactor, its mass and internal constitution must have been similar to that of the Earth. Appealing to Earth-like core formation conditions on a precursor impactor is therefore an \textit{ad-hoc} claim that is difficult to verify.   \\

The tungsten isotope composition of the Moon, discussed in section \ref{sec:Hf-W}, is indicative of its time-integrated Hf/W ratio during the lifetime of $^{182}$Hf. As both elements are refractory, variations in Hf/W ratios arise chiefly by metal/silicate fractionation processes, namely core formation or segregation, where W is more siderophile than Hf, leading to high Hf/W ratios in differentiated planetary mantles and correspondingly low values in their cores. Owing to the short half-life of $^{182}$Hf \citep[$t_{1/2}$ = 8.9 Myr][]{Vockenhuber_etal2004}, variations in $^{182}$W/$^{184}$W from $^{182}$Hf decay can develop only in first $\sim$ 60 Myr of Solar System evolution. The lunar mantle, as inferred from measurements of both mare volcanic- and highlands rocks, has an excess, ($^{182}$W/$^{184}$W)$_{\mathrm{Moon}}$/($^{182}$W/$^{184}$W)$_{\mathrm{BSE}}\times10^6$, of 26$\pm$3 ppm \citep{kruijer2017tungsten} relative to the BSE. This additional in-growth of $^{182}W$ is deemed to reflect either the disproportionately large delivery of the late veneer on the Earth relative to the Moon \citep{kruijer2017tungsten,kruijer_etal2021NatGeo} or the early formation of the lunar core \citep[$<$75 Myr after \textit{t$_0$},][] {thiemens2019early,Thiemensetal2021}. While the merits of both scenarios are discussed in section \ref{sec:Hf-W}, both imply that prior to, late veneer delivery on the Earth or core formation on the Moon, respectively, the lunar mantle and the BSE had identical $^{182}$W/$^{184}$W ratios. Because core formation in the early Solar System during planetary accretion fractionates W from Hf, and occurs at different times, a wide range of $^{182}$W/$^{184}$W ratios in the mantles of protoplanets are expected, and indeed observed among achondrites \citep{kleinewalker2017tungsten}. Given this information, \cite{kruijer2017tungsten} calculated that $\leq$5 \% of accretion scenarios would lead to $\sim$15 ppm concordance in the W isotope composition between the Earth and Moon (see section \ref{sec:Hf-W}). 

It should also be noted that, even for an isotopically identical Moon-forming impactor, some additional degree of chemical fractionation is required to account for the iron deficiency of the Moon with respect to the Earth and chondrites. This may be achieved, for example, by core-core merging \citep{Landeauetal2016}, and thereby precludes simple gravitational capture of an otherwise enstatite chondrite-like body for the origin of the Moon (see section \ref{sec:dynamics_history}). An isotopic distinction between an Earth-like Moon and enstatite chondrite-like Moon may also be made on the basis of the isotopic compositions of elements that show differences between the BSE and ECs, for example, Mo and Ni. These elements are particularly important, as their siderophile nature (section \ref{sec:geochem_core}) means their abundances in the BSE are likely to be diagnostic of its unique, time-integrated accretion pathway \cite[cf.][]{Rubieetal2015, Dauphas2017}. Data for these isotopic systems in lunar rocks, however, remain sparse \citep{burkhardt2014evidence,wang_etal2023nickel}. 

Although mass dependent isotopic variations are not as reliable tracers of provenance as mass independent isotope ratios due to the fact that the former can be modified by mass transfer processes, the similarity of the Earth and Moon observed in stable Si isotopes is striking \citep[Fig. \ref{fig:Stables_Tc},][]{armytage2012silicon}. This observation is significant as it distinguishes the Earth-Moon system from enstatite chondrites, whose isotopic composition is markedly ($\sim$ 0.4 $\permil$ in $\delta ^{30}$Si) lighter than that of the Earth and Moon \citep{armytage2011silicon,FitoussiBourdon2012}. Although this difference was initially ascribed entirely to endogenous core formation on the Earth \citep{georg2007silicon}, experimentally determined core-mantle Si isotope fractionation factors at inferred core-mantle equilibration temperatures ($\sim$3000 K) required some $\sim$30 wt.\% metallic Si \citep{shahar2009Sicore} or more \citep{hin2014Sicore} were the Earth to have formed from an enstatite chondrite-like precursor, far in excess of the $\leq$7 wt.\% permitted based on the observed deficit of the core relative to pure Fe-Ni alloy \citep[e.g.,][]{Badroetal2016}. The notion that the Earth-Moon system is the heavy end-member among planetary materials in its Si isotope composition and thus bears special significance was overturned by the measurement of similarly heavy Si enriched values in angrites \citep{Pringleetal2014, Dauphasetal2015}, thereby obviating the need for the Earth's heavy isotope composition to have been generated exclusively via core formation. Therefore, although the isotopic similarity of the Earth and Moon in Si isotopic composition may be fortuitous, the chemical and isotopic evidence taken together implies a genetic relationship between them. 

The W isotopic composition and abundances of Mn, Cr and V elements in the lunar and terrestrial mantles are sufficiently similar so as to suspect the proto-Earth's mantle as the most likely source of the Moon, as envisaged by \cite{ringwoodkesson1977}, rather than from an isotopically Earth-like impactor. Indeed, chemical fractionation from any potential chondritic source to form the Moon is unavoidable owing to its low bulk Fe content, a property that is difficult to explain without appealing to derivation from Earth's mantle (here, it is specified that the Earth's mantle may also denote the end-product of impactor-proto Earth mixing). Whether these chemical characteristics find explanations in dynamical scenarios for the Moon's formation is discussed in section \ref{sec:dynamics}.

\section{Timing of the formation of the Moon}
\label{sec:timing}

\subsection{Sm/Nd}
\label{sec:Sm-Nd}

The Sm-Nd system is of particular importance to determining the timing of lunar formation and differentiation, owing to its two isotopic decay systems, $^{147}$Sm to $^{143}$Nd (\textit{t}$_{1/2}$ = 106 Gyr) and the shorter-lived $^{146}$Sm to $^{142}$Nd system, with \textit{t}$_{1/2}$ = 103 Myr \citep{meissner1987Sm146}\footnote{Although a subsequent study reported a shorter \textit{t}$_{1/2}$ of 68 Myr \citep{kinoshita2012shorter}, recent calibrations against other short-lived chronometers support the longer half-life \citep{carlson_etal2014,fang2022Sm146}}. Consequently, the two systems may be used in tandem to quantify the extent and timing of Sm and Nd fractionation on the Moon, a feature that is likely to have accompanied magmatic differentiation, rather than volatility or core formation, as both Sm and Nd are refractory lithophile elements. \\

Should the Moon have a chondritic Sm/Nd ratio, then the resulting $\varepsilon$Nd (the part-per 10,000 deviation of the $^{143}$Nd/$^{144}$Nd ratio from that of chondrites at a given time) of the oldest lunar rocks, if they were derived from sources representative of the bulk Moon, should also be chondritic (i.e., equal to 0). Latest determinations of Sm-Nd in five FANs and Mg Suite rocks indicate that $\varepsilon ^{143}$Nd ranges between -0.29$\pm$0.09 (FAN sample 60025) to +0.04$\pm$0.11 for Mg Suite sample 67667 with internal isochron ages between 4302$\pm$28 Ma (FAN sample 60016) and 4359$\pm$3 Ma for 60025 \citep[][and references therein]{borg_etal2022}. In the lunar magma ocean paradigm \citep{Woodetal1970}, the early stages of crystallisation (up to 65--75 \%) are typified by olivine and orthopyroxene \citep{snyder1992chemical,charlier_etal2018,schmidtkraettli2022}, both of which incorporate negligible quantities of rare-earth elements, such that the Sm/Nd ratio of the residual liquid at the saturation of plagioclase (which remains on the liquidus until $\sim$ 95 \% crystallisation) is expected to remain chondritic, or marginally subchondritic (i.e., LREE-enriched) owing to the slight preference of orthopyroxene for Sm over Nd \citep[e.g,][]{dygert_etal2020}. Insofar as most whole rock FANs have marginally negative $\varepsilon ^{143}$Nd \citep[$\varepsilon ^{143}$Nd down to -0.3, cf.][]{borg_etal2022}, this expectation is borne out. \\

The picture becomes more complicated when considering that FANs are almost exclusively breccias, many of which are polymict, that is, they contain clasts from a variety of  lithologies \citep[e.g.,][]{WarrenWasson1977}. \cite{torcivia_neal2022} identified several populations of plagioclase crystals in FAN sample 60025. The calculated parental liquids of a subset of these crystals are broadly consistent with an initially chondritic bulk Moon, while others appear to have crystallised from LREE-depleted sources. \cite{ji_dygert2023} came to similar conclusions, after accounting for subsolidus equilibration with coexisting pyroxenes that indicate temperatures around 1323--1423 K, in agreement with those inferred from the granulitic textures and two-pyroxene thermometry of some FANs \citep{mcgee1993lunar,mcleod_etal2016}. The LREE abundances in plagioclase should be more robust than the HREE with respect to subsolidus re-equilibration with pyroxene, owing to the fact that D$_{LREE}^{plag/px}$ $>$ 1, with the ratio tending to unity at Sm \citep{graff_etal2013}. The significance of FAN sample 60025 lies in that it is the only known sample to have been measured that gives concordant ages for the $^{147}$Sm/$^{143}$Nd,
$^{146}$Sm/$^{142}$Nd, and Pb-Pb systems, yielding 4360$\pm$3 Ma \citep{borg_etal2011}. Yet, both negative \citep[-0.24$\pm$0.09;][]{borg_etal2011} and positive \citep[+0.9$\pm$0.5;][]{CarlsonLugmair1988} $\varepsilon ^{143}$Nd values have been reported for this sample, perhaps reflective of the distinct provenance of the plagioclase crystals contained therein. It should be noted that other studies derive older ages for some FANs \citep[e.g., 4430$\pm$30 Myr;][]{nyquist_etal2006}, which, in view of the heterogeneity among plagioclase clasts in 60025, leave open the possibility for a more ancient age for Moon formation. \\  

\begin{figure}[!ht]
     \centering
     \includegraphics[width=0.5\textwidth]{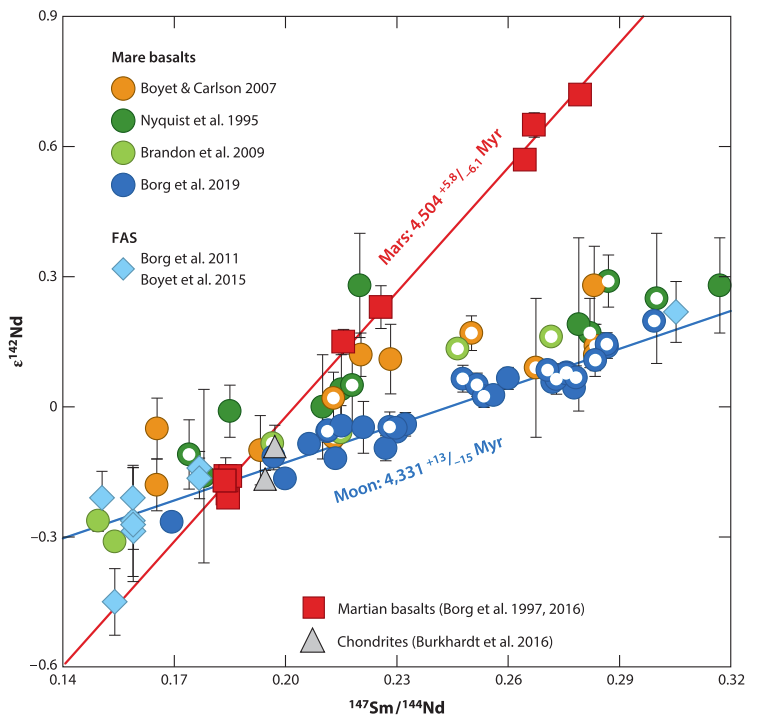}
     \caption{Combined time-integrated $^{146}$Sm/$^{142}$Nd (as shown by $\varepsilon ^{142}$Nd) and $^{147}$Sm/$^{144}$Nd systematics of a wide range of whole-rock lunar lithologies, compared with those of Mars. After \cite{borg_carlson2023}.}
     \label{fig:Sm-Nd_Moon}
 \end{figure}

\cite{borg_carlson2023} compiled $^{147}$Sm/$^{143}$Nd and $^{146}$Sm/$^{142}$Nd data to illustrate that variation among $^{142}$Nd/$^{144}$Nd ratios in lunar rocks, across mare basalts, FANs and Mg Suite samples, is relatively restricted (of the order of 60 ppm, or 0.6 $\varepsilon$), despite the large variation in $^{147}$Sm/$^{144}$Nd ratios (0.15 to 0.32; Fig. \ref{fig:Sm-Nd_Moon}). The slope of the line in Fig. \ref{fig:Sm-Nd_Moon} can be used to derive a `planetary isochron' age \citep[cf.][]{carlson_etal2014} that is independent of the initial $^{142}$Nd/$^{144}$Nd of the Moon. Although this approach groups together lithologies with a wide range of Sm/Nd ratios that also formed at different times, that the majority of the samples plot along the same line that defines an age of 4331$^{+13}_{-15}$ Ma \citep[Fig. \ref{fig:Sm-Nd_Moon},][]{borg_carlson2023} suggests that the Sm/Nd ratios of the sources of FANs, Mg Suite rocks and mare basalts all differentiated at this time. Given \textit{t}$_{1/2}$ of 103 Myr, this occurred at a mature stage in the lifetime of $^{146}$Sm (2.5 half-lives). This `late' lunar differentiation relative to other planetary bodies is evident from the larger spread of $\varepsilon ^{142}$Nd values for Martian shergottites ($\sim$ 100 ppm or 1 $\varepsilon$), despite their more restricted range of $^{147}$Sm/$^{143}$Nd ratios (0.18--0.28). If such an event were consistent with crystallisation of a lunar magma ocean, then the uncertainty on the ages indicate that it likely occurred over a period not longer than $\sim$30 Myr. \\

Isotope ratios in KREEP-rich rocks \citep[e.g.,][]{edmunson2009mgsuite,borg2017chronologic} also show slightly negative $\varepsilon ^{143}$Nd values, indicative of an LREE-enriched source, consistent with, but not proof of, enrichment of Nd over Sm in a crystallising lunar magma ocean. Mare basalts have positive $\varepsilon ^{143}$Nd values that, although too scattered to define an accurate model age, tend to become more positive with younger crystallisation ages \citep{snyder_etal2000chronology}. In the sense that mare basalts have strong negative Eu anomalies\footnote{defined as Eu/Eu* = Eu$_N$/$\sqrt{{Gd_N}\times{Sm_N}}$, where $N$ indicates normalisation to some reference reservoir, typically chondrites.)}, their sources are supposed to represent the geochemical complement to FANs \citep[e.g.,][]{NealTaylor1992}, that is, their sources have had a plagioclase-rich component subtracted from them, leaving them with superchondritic Sm/Nd ratios. Were this complementarity to hold, then the the $\varepsilon ^{143}$Nd values of FANs would be expected to decrease over time, owing to their derivation from LREE-enriched (slightly subchondritic) sources \cite[e.g.,][]{snyder1992chemical}. However, the opposite is observed, and their $\varepsilon ^{143}$Nd increase as their ages become younger, following the trends defined by mare basalts \citep{borg_etal2011}. This may indicate either a disturbance in the Sm-Nd systematics of some FANs, or that the paradigm for their crystallisation from a magma ocean no longer holds. Therefore, it is unclear as to whether the bulk Moon is LREE-depleted, or chondritic, though both options remain plausible. \\

\subsection{Hf/W}
\label{sec:Hf-W}

The utility of the short-lived \citep[$t_{1/2}$ = 8.9 Myr;][]{Vockenhuber_etal2004} $^{182}$Hf-$^{182}$W chronometer lies in the fact that the lithophile geochemical behaviour of the parent isotope contrasts with that of the moderately siderophile daughter. Consequently, core formation on small differentiated bodies in the early Solar System, as recorded by iron meteorites \citep[e.g.,][]{kruijer2014protracted} leads to high Hf/W ratios in their silicate-rich mantles and complementarily low Hf/W ratios in their cores. Because both elements are cosmochemically refractory, these variations are unlikely to have arisen via volatility, though the somewhat more compatible behaviour of Hf relative to W in ferromagnesian silicates can lead to small Hf/W variations by partial melting \citep[e.g.,][]{konig2011earth}. Like other radiogenic isotopic systems, the measured $^{182}$W/$^{184}$W is the result of time-integrated ingrowth of $^{182}$W, such that only average Hf/W ratios over the lifetime of $^{182}$Hf can be derived from a given sample. In the simplest case, core formation on the Earth can be modelled to have occurred in a single event. Assuming the bulk Earth had, initially, a chondritic $^{182}$W/$^{184}$W ratio, then the Hf/W of the BSE \citep[25$\pm$5;][]{munker2010high,Dauphasetal2014,kleinewalker2017tungsten} combined with the measured $\sim$190 ppm excesses in $^{182}$W in the BSE relative to chondritic meteorites, imply the core separated from the mantle at $\sim$34$\pm$3 Myr after \textit{t$_0$} \citep{Kleinetal2002,Yinetal2002,kleinewalker2017tungsten}. However, as core formation on Earth likely proceeded in a more periodic manner, a myriad of other scenarios, in which the core grew more quickly or more slowly, remain plausible \citep[e.g.,][]{rudge_etal2010,FischerNimmo2018,Fischeretal2021}. \\

The first high precision measurements of the W isotopic composition of lunar samples revealed systematic variations among lunar lithologies, leading to the idea that their sources differentiated within the lifetime of $^{182}$Hf \citep[less than about 60 Myr;][]{Leeetal1997}. However, subsequent work showed that this variability was, in large part, due to nuclear reactions induced by galactic cosmic rays, namely, $^{181}$Ta(\textit{n,$\gamma$})$^{182}$Ta(\textit{$\beta ^-$})$^{182}$W and $^{182}$W(\textit{$n$})$^{183}$W \citep{leya_etal2000}. In order to circumvent this issue, \cite{Toubouletal2015} determined $^{182}$W/$^{184}$W in metallic phases, which contain negligible Ta relative to W, separated from KREEP basalts, detecting an excess of 20.6$\pm$5.1 ppm relative to the BSE. Through the use of $^{180}$Ta as a neutron capture dosimeter to correct for the exposure times of whole rock KREEP samples to cosmic rays, \cite{Kruijeretal2015} independently reached the same conclusion, identifying an excess of 27$\pm$4 ppm relative to the BSE. This value was later refined, through studies of not only KREEP, but low-Ti, high-Ti and Mg Suite rocks, to 26$\pm$3 ppm \citep{kruijer2017tungsten}. \\

Importantly, the Hf/W ratios of KREEP-rich basalts ($\sim$15--20) low-Ti mare basalts ($\sim$30--50), and high-Ti mare basalts ($\sim$50--150) differ markedly \citep{munker2010high,fonseca_etal2014,kruijer2017tungsten}. Should this variation translate to that in their magmatic source regions, then the lack of $\mu ^{182}$W variability (within 3 ppm) indicates that their mantle sources must have differentiated from one other \textit{after} $\sim$60 Myr, so as not to engender differential radiogenic ingrowth of $^{182}$W in accordance with their Hf/W ratios \citep{kruijer2017tungsten}. The tendency of Hf/W in lunar basalts towards super-BSE values likely reflects the contribution of Hf-rich Fe-Ti oxides to the petrogenesis of high-Ti basalts, whereas the sub-BSE Hf/W of KREEP is reconciled with their derivation from evolved liquids that experienced oxide- and silicate crystallisation and removal \citep{fonseca_etal2014}. Experimental work detected a mild increase in W compatibility at low \textit{f}O$_2$s below the IW buffer \citep{fonseca_etal2014}, inferred to be due to small amounts of W$^{4+}$ \cite[e.g.,][]{oneill2008solubility}. However, the magnitude of the effect influences mantle Hf/W ratios by $\sim$20 \% relative, comparable to the precision to which the ratio can be determined in the Moon, and a factor $\sim$20 smaller than the range observed in lunar basalts. The range of Hf/W ratios in lunar rocks therefore hampers the definition of a bulk Moon Hf/W ratio, with two scenarios being explored below. \\

\begin{figure*}[!ht]
     \centering
     \includegraphics[width=1\textwidth]{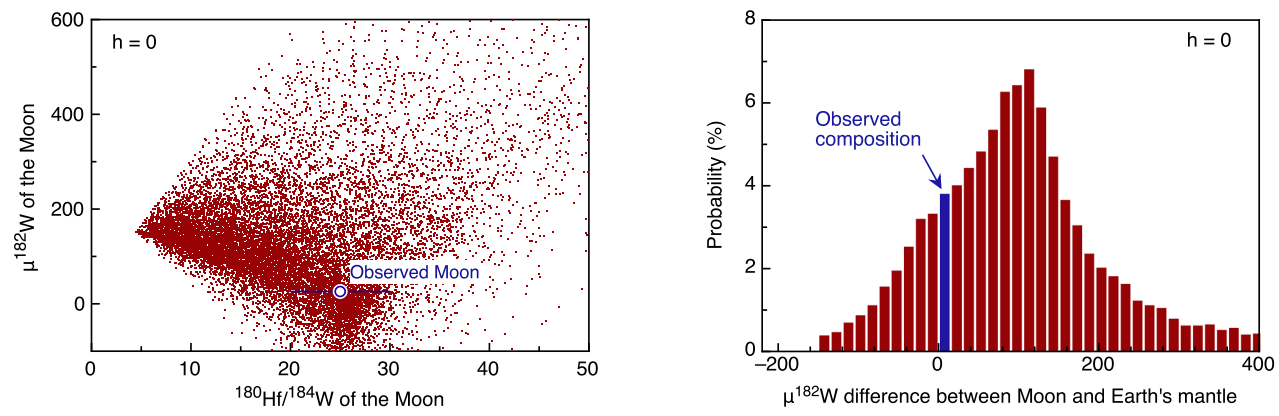}
     \caption{Results of Monte-Carlo simulations showing the likelihood of the pre-late veneer Earth and Moon having the same composition in a giant impact scenario in which the Moon is derived entirely from the proto-Earth's mantle. After \cite{kruijer2017tungsten}.}
     \label{fig:Hf-W_Moon}
 \end{figure*}

In the event that Hf/W ratios between the BSE and the BSMoon are indistinguishable, as adopted by \cite{kruijer2017tungsten}, then the $^{182}$W/$^{184}$W difference between the BSE and BSMoon could not have arisen by \textit{in-situ} radiogenic growth of $^{182}$W. Instead, \cite{kruijer2017tungsten} suggest that the present-day excess of $^{182}$W in the Moon relative to the BSE is due to differential accretion of the late veneer on the Earth \citep[$\sim$0.3--0.8 Earth masses;][]{walker_etal2015lateveneer} relative to that received by the Moon \citep[at most $\sim$0.023 $\pm$ 0.002 lunar masses;][]{Dayetal2016}. The rationale for this explanation lies in the fact that chondritic material, as is thought to have characterised the late veneer, has a  $\mu^{182}$W = -190 ppm, and is rich in W \citep[100--200 ppb;][]{wassonkallemeyn1988} compared to the silicate mantles of the Earth and Moon ($\sim$ 10 ppb), meaning small quantities are able to shift their W isotope compositions, provided the late veneer is well-mixed. \cite{Dauphasetal2014}, in the context of a canonical Mars-sized giant impact, modelled scenarios in which the Moon is comprised of proto-Earth mantle, impactor mantle, impactor core and 0.2 wt. \% late veneer, and attempted to find the fractions of the aforementioned components, together with their Hf/W and $^{182}$W/$^{184}$W compositions, that would satisfy the present-day $^{182}$W/$^{184}$W of the Earth and Moon (i.e., $\Delta \mu ^{182}$W$_{Moon-BSE}$ = 26$\pm$3 ppm). Although \cite{Dauphasetal2014} found numerous solutions to the present-day W isotope compositions of the two bodies, they do not readily account for their identical pre-late veneer composition. \cite{kruijer2017tungsten} illustrated that such an accord is a very rare outcome of reasonable giant impact scenarios; less than 5 \% of all cases examined yield a pre-late veneer $\mu ^{182}$W$_{Moon-BSE}$ $<$ 15 ppm (Fig. \ref{fig:Hf-W_Moon}). The likelihood of achieving the same result drops to $<$ 1 \%, should the Moon have been made from 80 \% impactor material. The rarity of such outcomes arises from the fact that the $\mu ^{182}$W composition of a differentiated body (such as a putative Moon-forming impactor) reflects its time-integrated Hf-W evolution. Consequently, models in which Moon was largely derived from impactor material imply that the impactor fortuitously underwent the same \textit{P-T-t} evolution as the Earth's mantle with respect to core formation. \\

The alternate scenario, which states that the Hf/W ratio of the lunar mantle is \textit{higher} than that of the BSE, as deduced from the relatively higher compatibility of W over Hf under reducing conditions and the higher Hf/W of low-Ti basalts (30--50) compared with that inferred for Earth's mantle ($\sim$25), was explored by \cite{thiemens2019early}. In this scheme, the observed lunar 26$\pm$3 ppm excess in $\mu ^{182}$W$_{Moon-BSE}$ can arise through endogenous core formation on the Moon while $^{182}$Hf was still extant, which, in turn, requires that the Moon formed within $\sim$60 Myr of \textit{t$_0$}. In order to substantiate that the BSMoon, as sensed by low-Ti basalts, has a higher Hf/W ratio higher than that of the BSE, \cite{thiemens2019early} employed partition coefficients in which ferromagnesian silicates have preference for W over Hf, which is found experimentally only in melt compositions with $\geq$10 wt. \% TiO$_2$ and at low \textit{f}O$_2$, $\leq$$\Delta$IW-1.7 \citep{leitzke2016HfW}. As the TiO$_2$ content of silicate liquids is far lower than 10 wt. \% during \textit{i)} the petrogenesis of low-Ti basalts (which contain 3--5 wt. \%) and \textit{ii)} crystallisation of a lunar magma ocean \citep[initially containing 0.2--0.3 wt. \%,][Table \ref{tab:lunar_abu}]{Dauphasetal2014}, it is likely Hf is similarly- to more compatible in ferromagnesian silicates than is W during the formation of the low-Ti basalts (or their source regions). Hence, an elevated Hf/W (30--50) in the BSMoon with respect to the BSE, as argued by \cite{thiemens2019early}, could have resulted from a BSE-like Hf/W ratio for the bulk Moon. Empirically, the near-constant U/W and Th/W ratios in mare basalts, despite a range of W abundances (Fig. \ref{fig:W-U}, Table \ref{tab:lunar_abu}), and the likely absence of residual metal in their sources \citep{munker2010high,daypaquet2021}, suggest that these three elements behave in a lithophile manner on the Moon. Indeed, such correlations were used to define bulk lunar Hf/W ratios of $\sim$25 \citep{munker2010high}. Instead, elevated Hf/W ratios in low-Ti mare basalts likely stem from the modest addition of ilmenite \citep[in which Hf is moderately compatible, with \textit{D$_{ilm/melt}$} near unity;][]{klemme2006partitioning} to their source regions, whereas KREEP-rich rocks have slightly sub-BSE-like Hf/W. \\

In either model, the (pre-late veneer) BSE and Moon are assumed to have had the same $\mu ^{182}$W ratio, a feature difficult to explain by any giant impact model, unless the two bodies perfectly equilibrated \citep[cf.][]{kruijer2017tungsten,Locketal2018,lock2020geochemical}. As for Sm-Nd ages, the lack of $\mu ^{182}$W variability among any lunar lithology (once corrected for spallation reactions), indicates a relatively young age for the formation of their source regions, after 4.5 Ga \citep{kruijer2017tungsten}. Evidence for core formation on the Moon having taken place prior to this time is permitted based on the 26$\pm$3 ppm excess in $\mu ^{182}$W over that in Earth's mantle \citep{thiemens2019early}, however, the factor $\sim$20 lower fraction (proportional to the size of the body) of the late veneer received by the Moon with respect to the Earth is an equally viable explanation for this difference \citep{kruijer2017tungsten}. No unequivocal evidence for a lunar formation age older than 4.5 Ga is proffered by Hf-W systematics.

\subsection{Rb/Sr}
\label{sec:Rb-Sr}

The isotope $^{87}$Rb decays into $^{87}$Sr with a half life of 49.61 $\pm$ 0.16 Myr \citep{nebel_etal2011,villa_etal2015iupac}. Rubidium is moderately volatile in both the solar nebula \citep{woodetal2019condensation} and during evaporation from silicate melts \citep{sossietal2019}, whereas Sr is refractory under all conditions \citep[e.g.,][]{fegley_etal2020}. Because Rb and Sr are both lithophile, their fractionation during early Solar System processes arose largely owing to their differential volatility. The locus at which this fractionation occurred, however, is less certain. Many chondritic meteorites, with formation ages $<$5 Myr after the formation of CAIs, show evidence for Rb/Sr fractionation, as do the terrestrial planets, despite their more protracted formation timescales \citep[e.g.,][]{oneillpalme2008}. Indeed, the $^{87}$Rb/$^{86}$Sr ratio of the Moon, 0.019$\pm$0.001 (Table \ref{tab:lunar_abu}) is significantly lower than that of the Earth \citep[0.080$\pm$0.001][]{palmeoneill2014}, implying that specific processes acted on the Moon to lower its Rb/Sr ratio relative to the Earth. \\

The deduction of the integrated Rb/Sr ratio of the material from which lunar rocks formed is tractable, provided i) the $^{87}$Sr/$^{86}$Sr from which the material separated is known ii) the age of the rocks are known independently, and iii) the present-day $^{87}$Sr/$^{86}$Sr can be determined. The FANs and Mg Suite rocks, with crystallisation ages of $\sim$ 4300 to 4360 Ma \citep{borg_carlson2023} fulfil these criteria. They are particularly amenable to Rb-Sr dating, owing to the fact that Rb is less readily incorporated into anorthositic plagioclase than is Sr. As a result, the correction for in-situ growth of $^{87}$Sr by the decay of $^{87}$Rb since the formation of FANs is minimal, thereby maximising the precision of $^{87}$Sr/$^{86}$Sr$_{i}$ (where \textit{i} = initial) obtained through Rb-Sr measurements. Compared to the Basaltic Achondrite Best Initial (BABI) value of 0.6989707$\pm$0.000028, taken as the initial value for the Solar System, FANs and Mg Suite rocks show only slight radiogenic ingrowth, with most recent determinations ranging from 0.699050$\pm$0.000010 to 0.699116$\pm$0.000022 \citep[][and references therein]{borg_etal2022}. \\

\begin{figure}[!ht]
     \centering
     \includegraphics[width=0.5\textwidth]{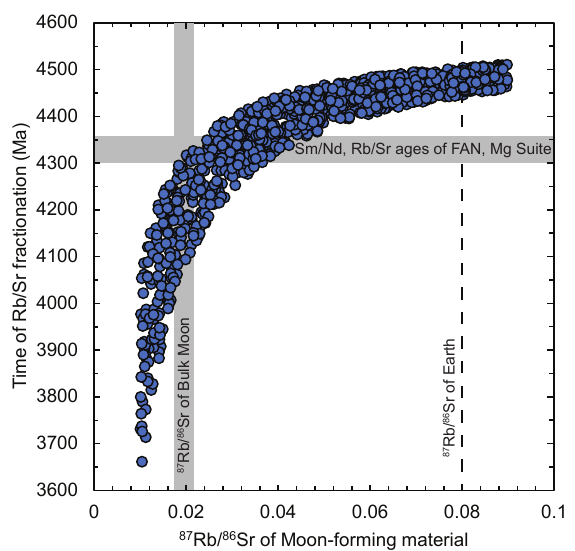}
     \caption{The timing of chemical fractionation of Rb from Sr in source regions of Ferroan Anorthosite (FAN) and Mg Suite rocks as a function of the time-integrated $^{87}$Rb/$^{86}$Sr ratio of their sources (taken to be that of the bulk Moon) as deduced by Monte Carlo simulation (\textit{n} = 1000) taking into account uncertainties in measured $^{87}$Sr/$^{86}$Sr$_{i}$ values in FAN and Mg Suite as determined by \cite{borg_etal2022}, combined with a Basaltic Achondrite Best Initial (BABI) of 0.69898 \citep{hans_etal2013rb} and a decay constant for $^{87}$Rb of 1.3972$\times$10$^{-11}$ yr \citep{nebel_etal2011,villa_etal2015iupac}. The array passes to the right of present-day lunar Rb/Sr value \citep[0.019 $\pm$ 0.001; Table \ref{tab:lunar_abu};][]{borg_etal2022} at the independently determined ages of FAN and Mg Suite rocks, implying a higher time-integrated Rb/Sr ratio (i.e., more volatile-rich) in Moon-forming material relative to its present value. }
     \label{fig:Rb-Sr_Moon}
 \end{figure}

The $^{87}$Sr/$^{86}$Sr$_{i}$ of BABI shows no resolvable difference with the $^{87}$Sr/$^{86}$Sr$_{i}$ of ADoR (Angra dos Reis) or with CAIs, which yield 0.6989787$\pm$0.000004 and 0.6989757$\pm$0.000008, respectively \citep{hans_etal2013rb}. On this basis, it is reasonable to assume an initially well-mixed solar nebula with respect to $^{87}$Sr/$^{86}$Sr, the reservoir from which all planetary bodies inherited their $^{87}$Sr/$^{86}$Sr$_{i}$. Considering the uncertainties in both the ages and $^{87}$Sr/$^{86}$Sr$_{i}$ of measured FANs and Mg Suite rocks, a Monte Carlo simulation is performed to describe how the apparent time of Rb/Sr fraction depends on the Rb/Sr ratio of the precursor material (Fig. \ref{fig:Rb-Sr_Moon}), by solving the equation:

\begin{equation}
t (Myr) =  \frac{\ln \left( \frac{ \frac{ ^{87}Sr }{ ^{86}Sr  }_{i,smp} - \frac{ ^{87}Sr }{ ^{86}Sr  }_{BABI} } { \frac{ ^{87}Rb }{ ^{86}Sr  } } \right)}{\lambda} \times 10^{-6}
\label{eq:rb-sr_decay}
\end{equation}

where $\lambda$ is the decay constant for $^{87}$Rb of 1.3972$\times$10$^{-11}$ yr \citep{nebel_etal2011,villa_etal2015iupac} and in which it is assumed $^{87}$Sr/$^{86}$Sr of the Earth and Moon are equivalent and equal to $^{87}$Sr/$^{86}$Sr$_{i}$ of FANs and Mg Suite rocks at $t$ of Rb/Sr fractionation. \\

The required range of $^{87}$Rb/$^{86}$Sr under these assumptions fall in between $\sim$0.02 to $\sim$0.08 (Fig. \ref{fig:Rb-Sr_Moon}). These values are, as found by \cite{borg_etal2022}, strongly volatile-depleted with respect to CI chondrites (0.86) and even the present-day BSE (0.081$\pm$0.009). If it is assumed that the Moon separated from the BSE and hence Moon-forming material evolved with the present-day Rb/Sr ratio of the BSE from $t_0$ to the time of Rb/Sr fractionation ($y$-axis on Fig. \ref{fig:Rb-Sr_Moon}), then this fractionation must have occurred at 4478$\pm$18 Ma according to eq. \ref{eq:rb-sr_decay}. If the event is supposed coincident with the age of the Moon (i.e., 89$\pm$18 Myr after $t_0$), then it is in agreement with that determined by \cite{Halliday2008}, 87$\pm$13 Myr, \cite{KleineNimmo2024}, 75$\pm$27 Myr and 51--79 Myr \citep{yobregat2024} under the same assumption. It also overlaps with the 61$\pm$15 Myr age quoted by \cite{mezger_etal2021}, assuming the proto-Moon (i.e., pre-volatile loss) and the present-day Earth are formed from a mixture of 15~\% CI-chondritic material and 85~\% proto-Earth material devoid of Rb. As highlighted by \cite{halliday_porcelli2001}, it is critical to recognise that the Rb/Sr of Moon-forming material is a time-integrated value from t$_0$ (4.567 Ga) to t$_{crys}$. Any inference as to the nature of the Rb/Sr ratio of the precursor body (or bodies) is therefore necessarily model-dependent. The ages quoted above are maxima (i.e., the Moon could be older), because the calculation is made on the assumption that the BSE $^{87}$Rb/$^{86}$Sr ratio was set at $t_0$. \\

These low $^{87}$Rb/$^{86}$Sr ratios are in contrast to the higher, Mars-like $^{87}$Rb/$^{86}$Sr ($\sim$0.2) inferred for the precursor material of the Moon by \cite{halliday_porcelli2001}, which comes from their use of the 4515 Ma Pb-Pb age for FAN sample 60025 \citep{CarlsonLugmair1988,hanan1987_60025}. More modern Sm-Nd and Lu-Hf measurements for a wider suite of FANs, including 60025, yield 4300--4360 Ma for their formation \citep{borg_etal2022}. Adopting these younger ages shows that the required time-integrated $^{87}$Rb/$^{86}$Sr ratio could not have been lower than the present-day ratio ($\sim$0.02), lest the Rb/Sr fractionation age is younger than the formation of FANs and Mg Suite rocks themselves (Fig. \ref{fig:Rb-Sr_Moon}). These estimates coincide with the observation that Mg Suite and FAN whole rocks plot along a single Sr isotope evolution line requiring $^{87}$Rb/$^{86}$Sr = 0.035 \citep{carlson_etal2014}. The key result is that the time-integrated Rb/Sr ratio of Moon-forming precursor material must have been, in the first $\sim$ 200 Myr of the Solar System, higher than that inferred for the present-day BSMoon (Table \ref{tab:lunar_abu}). \\ 

\subsection{U-Pb}
\label{sec:U-Pb}

The other most commonly examined radiogenic isotopic systems in lunar rocks are U-Pb and Lu-Hf, cases in which both the parent and daughter are incompatible elements, and, with the exception of Pb, are cosmochemically refractory. Lead is shown in experiments to be moderately siderophile / chalcophile with D$_{met/sil}$ between $\sim$ 1--10 for S-poor metal compositions thought to characterise the lunar core \citep[see][]{wood_etal2014sulfur,ballhaus2017great}, and, together with the small mass of the lunar core (0.8--1.5 \% of the Moon; Table \ref{tab:inv_com}), is not likely to have been sequestered in large quantities into metal. Despite the expected lack of Pb in the lunar core, the $^{238}$U/$^{204}$Pb, or \textit{$\mu$} ratio, has long been known to be elevated \citep[\textit{$\mu$} $\sim$300,][]{tatsumoto1970age,Teraetal1974,oneill1991origin}, not only with respect to the CI-chondritic ratio (0.19), but also relative to the BSE \citep[$\sim$9,][]{palmeoneill2014}. These considerations have been taken as evidence to suggest that much of the Pb loss relative to U occurred during the formation of the Moon itself \citep{ConnellyBizzarro2016lead}. \\

The low Pb contents of lunar mare basalts and the propensity for their contamination with common Pb hindered the precise estimation of systematic differences in the \textit{$\mu$} ratios of their source regions. Recently, however, \textit{in-situ} determination of U-Pb systematics in constituent phases via SIMS has spurred a renaissance of Pb isotope geochemistry in lunar samples \citep{Snapeetal2016,snapeetal2019,connellyetal2022}. This 
has permitted the identification of clear, systematic differences in the $^{238}$U/$^{204}$Pb ratios of products of mare volcanism. Assuming the bulk Moon evolved with an initial $\mu$ ratio of 460, prior to a differentiation event at 4376$\pm$18 Ma, producing $\mu$ values ranging from 2600--3700 (KREEP), 360--390 (high-Ti) and 410--650 (low-Ti) to explain the coupled $^{207}$Pb/$^{206}$Pb--$^{204}$Pb/$^{206}$Pb variations \citep{snapeetal2019}. The relative differences between KREEP and low-Ti volcanism are not readily explained by protracted fractional crystallisation in the magma ocean paradigm, as the residual liquids do not reach the extremely elevated $^{238}$U/$^{204}$Pb in the KREEP source \citep{snape_etal2022experimental}, an issue that remains unresolved. \\

Nevertheless, the absolute $\mu$ values inferred are model-dependent in the sense that the observed $^{207}$Pb/$^{206}$Pb--$^{204}$Pb/$^{206}$Pb values can be created via a trade-off between $\mu$ and the age of the differentiation event, with younger ages requiring higher $\mu$ than for older ages of U/Pb fractionation, all else being equal. \cite{connellyetal2022} explored scenarios in order to quantify the effect of the timing of a fractionation event in the source regions of mare basalts (taken by these authors to be equivalent to the giant impact) at $\sim$4.35 Ga and $\sim$4.5 Ga on their Pb isotope systematics. For the younger age, $\mu$(KREEP) is 4600 and $\mu$(low-Ti) is 340, while these values decrease assuming U/Pb fractionation at 4.5 Ga, to 2650 and 255, respectively \citep{connellyetal2022}. As noted by \cite{snapeetal2019} and \cite{connellyetal2022}, the degeneracy between fractionation times and $\mu$ values means that no discrete age for the formation of the Moon can be constrained by Pb isotopes alone. However, it is clear that the Moon, even on the basis of Low-Ti basalts, possesses very high $^{238}$U/$^{204}$Pb likely set by the same process(es) that led to its volatile depletion (section \ref{sec:physchem_volatile}) and Rb/Sr fractionation (section \ref{sec:Rb-Sr}), and that a young age concordant with Sm-Nd constraints for this U/Pb fractionation is permissible. \\

Other, more sporadic Pb-Pb determinations of lunar samples return a larger spread of ages. Five zircons analysed in highlands breccia 72215 yield ages of 4417$\pm$6 Ma \citep{nemchin2009timing}, yet the remaining 36 analyses of zircons from the same sample result in somewhat younger ages ($\sim$4380 Ma). In Apollo 14 breccias, zircon U-Pb ages based on SIMS \citep{Tayloretal2009} and TIMS \citep{Barbonietal2017} measurements yield ages for their crystallisation up to $\sim$4350 Ma, most of which lie on the concordia in Pb isotope space. \cite{Barbonietal2017} further determined, on the basis of Hf isotope model ages, 4510 $\pm$ 10 Ma for the age of the source of the zircons, assuming an Lu/Hf = 0, representing an infinite enrichment of Hf over Lu (values $>$ 0 would give older ages). This approach is associated with considerable uncertainty given the \textit{i)} uncertainty of the correction applied to account for cosmic ray exposure-induced neutron capture and \textit{ii)} this age is based on the oldest four zircons to define the age, whereas the majority of the zircon population analysed by \cite{Barbonietal2017} cluster around ages of 4400 - 4450 Ma. These ages also overlap with those recently determined for concordant regions of zircon grains in impact melt breccia sample 72255, 4460$\pm$31 Ma \citep{greer2023}.

\subsection{Age of the Moon}
\label{sec:age_moon}

Much of the significance of the ages and time-integrated geochemical properties of samples from the lunar highlands and mare volcanism stems from their implied origin from crystallisation of a lunar magma ocean. This concept, itself, rests upon the idea that the Moon originated in a giant impact, which, in the canonical scenario, occurred between the proto-Earth and a Mars-sized impactor \citep[][section \ref{sec:giant_impact}]{CameronWard1976,Canup2004}. This event is thought to have provided the energy required to engender wholesale melting of the accreting Moon \citep[e.g.,][]{ThompsonStevenson1988,Melosh1990}, and its subsequent crystallisation lead to the chemical (and, over time, isotopic) fractionation of the systems described in section \ref{sec:timing} \citep[e.g.,][]{snyder_etal2000chronology,charlier_etal2018}. \\

Chronological support of a global-scale event is found in the recurrence of ages around 4350 Ma for a wide range of lunar lithologies; namely, the Mg Suite, the FANs, and urKREEP, as recently reviewed by \cite{borg_carlson2023}. This event, if ascribed to the differentiation and solidification of the magma ocean, would provide an age constraint for the Moon only insofar as the timescales over which this chemical variability was imparted were brief. The spread of these ages ($\pm$30 Myr, Fig. \ref{fig:Sm-Nd_Moon}) provide observational evidence that the thermal event was short-lived, and such heterogeneity, in the LMO model, would therefore be near-concordant with the age of the Moon itself. An extended timescale for LMO crystallisation has been postulated by \cite{Mauriceetal2020} \citep[see also][]{michautneufeld2022} on the grounds that the Nd- and Hf-based model ages for the formation of the urKREEP reservoir would yield younger ages ($\sim$4270--4190 Ma) when considering the evolution of Lu/Hf and Sm/Nd ratios to subchondritic values during progressive fractional crystallisation of the LMO. These ages are $\sim$100--150 Myr younger than those computed from the assumption that the Lu/Hf and Sm/Nd ratios of the urKREEP source were chondritic. \cite{borg_carlson2023} point out, however, that these younger ages would conflict with the $\sim$4350 Ma crystallisation ages of Mg Suite samples, which harbour a significant KREEP component \citep{shearer2015MgSuite}, implying urKREEP must have formed prior to 4350 Ma. In either case, both interpretations are non-unique, as the model ages depend on the pre-4350 Ma histories of the Lu/Hf and Sm/Nd ratios of the urKREEP reservoir, which are unknown. \\ 

More problematic for relating the frequency of the $\sim$4350 Ma ages to the formation of the Moon is the complex nature of the FANs themselves. Although the formation of FANs from an LMO, either directly or indirectly, still remains widely accepted, evidence against this model has been mounting \citep{Haskin_etal1982,floss1998lunar}. Much of the skepticism for a single, FAN-forming event from the LMO has come from the realisation that trace element ratios in plagioclase imply that they were once in equilibrium with silicate liquids that had a wide range of incompatible trace element ratios
\citep[e.g., 0.001 $<$ Th/Sm $<$ 0.186 and 0.13 $<$ Zr/Sm $<$ 79.7;][]{pernetfisher2019}. These authors interpret trace element variations as being due to the primary magmatic conditions at which plagioclase and pyroxene in FANs crystallised, citing the $\sim$ 10$^9$ yr timescales over which solid-state diffusion would be required to reached the observed homogeneous composition of plagioclase grains of the sizes found in FANs. On the other hand, \cite{ji_dygert2023} find that timescales for subsolidus equilibration to reach completion are $\sim$1--100 Myr, provided that temperatures exceeded 1273 K and grains were no larger than 3 mm. On this basis, \cite{ji_dygert2023} provided a scenario in which FANs could be produced by differing degrees of KREEP-rich intercumulus melt percolation and re-equilibration, as suggested prior on the basis of the lack of correlation between An\# of plagioclase and the Mg\# in coexisting pyroxenes \citep[e.g.,][]{RaedekeMccallum1980,mcgee1993lunar,JolliffHaskin1995}. \\

The fact that the FANs, and, in particular, Mg-Suite rocks have granulitic textures indicative of recyrstallisation lead to the proposition that the isotopic ages record cooling in the crust rather than primary igneous crystallisation \citep[cf.][]{McCallumOBrien1996,MccallumSchwartz2001}. Lending credence to the importance of secondary `resetting' events over the primacy of LMO-based processes for the ages of highlands rocks is the geological association of Mg-Suite rocks that intrude FANs 
, yet record indistinguishable ages \citep[$\sim$4350 Ma,][]{borg_etal2011,carlson_etal2014}. Indeed, the FANs are not monomict rocks, and the archetypal sample, 60025, known to yield concordant ages in different isotopic chronometers \citep{borg_etal2011} is composed of a diverse collection of plagioclase grains of different heritage \citep{torcivia_neal2022}. The vast majority of the anorthosites, including 60025, come from the Descartes formation of the lunar nearside and hence sample a limited range around the Apollo landing sites, now known to be anomalously rich in the KREEP component relative to the lunar farside \citep[e.g.,][]{WieczorekPhillips2000}. Similarly, the observation that anorthosites from the lunar meteorite collection typically possess less marked positive Eu anomalies and have higher Mg\#s than returned samples \citep{korotev2005lunar, Gross_etal2014} permits large-scale heterogeneity in FAN progenitors. \\

The well-defined peak in ages at 4350 Ma for a wide range of lithologies therefore suggests a Moon-wide process that resulted in a thermal maximum, with the consequence being resetting of chronometric ages. While the petrological complexity of highlands rocks renders the canonical LMO model, in which FANs crystallise directly from an evolving, single batch of magma, difficult to test, the `serial magmatism' model has gained traction in recent years \citep{GrossJoy2016}. This model postulates that FANs formed by intrusion of broadly basaltic magmas into a now-unsampled primary crust, with plagioclase flotation occurring \textit{in-situ}. Recent models ascribe this to a cumulate overturn event \citep{floss1998lunar, li_etal2019overturn, pernetfisher2019} or progressive growth of a stagnant lid via compaction of a slushy interior \citep{michautneufeld2022}, both of which posit an LMO in order to produce a density inversion in the cumulate architecture in the first place \citep[cf.][]{ElkinsTanton2011}. Alternative sources of heat to induce melting include an impact that led to the dichotomy between the lunar far- and nearside, tentatively associated with the Procellarum Basin \citep{Tartese_etal2019}, the South Pole-Aitken basin \citep{barboni2024} or a less catastrophic, protracted interval of impact-induced melting on the lunar surface \citep{Haskin_etal1982}. Nevertheless, there is no mechanism, as yet, that is as well-developed as the flotation hypothesis set out by \cite{Woodetal1970} capable of explaining the widespread distribution of FAN in the lunar crust, together with the concordant whole-rock isochron ages \citep{borg_carlson2023} and chemical properties complementary to those expressed in mare volcanics \citep{ringwoodkesson1977}. Consequently, the significance of this event for the formation of the Moon remains unclear. However, the chronometric model ages (most significantly that of Rb-Sr at $\sim$4490 Ma) and zircon ages back to 4460 Ma, suggest the Moon likely formed close to $\sim$4500 Ma.

\section{Dynamical scenarios for the formation of the Moon}
\label{sec:dynamics}

\subsection{Historical hypotheses}
\label{sec:dynamics_history}
While today the giant impact is widely accepted as the leading hypothesis for the origin of the Moon, historically, this has not always been the case. In particular, the Apollo and Luna missions, in addition to more recent remote- and lander missions, have provided a host of data, both geochemical and geophysical, by which models for the formation of the Moon can be tested.  Consequently, nowadays, a successful model should account for, at least (1) the angular momentum of the Earth-Moon system, (2) the mass, density, and moment of inertia of the Moon (section \ref{sec:geophys_inv}), (3) the small mass of the lunar core (section \ref{sec:geophys_inv_core}) (4) the depletion of elements more volatile than Li and the degree of mass-dependent isotopic fractionation observed in the Moon relative to the Earth (section \ref{sec:physchem_volatile}) and (5) the similarity in the mass-independent isotopic composition of the Earth and the Moon (section \ref{sec:physchem_terrestrial}). Below, we outline several pre-eminent Moon formation models, and highlight their strengths and weaknesses, prior to  exploring the giant impact hypothesis in more detail 
(see also reviews by \citealt{Stevenson1987}).

  \begin{figure*}[!ht]
     \centering
     \includegraphics[width=0.85\textwidth]{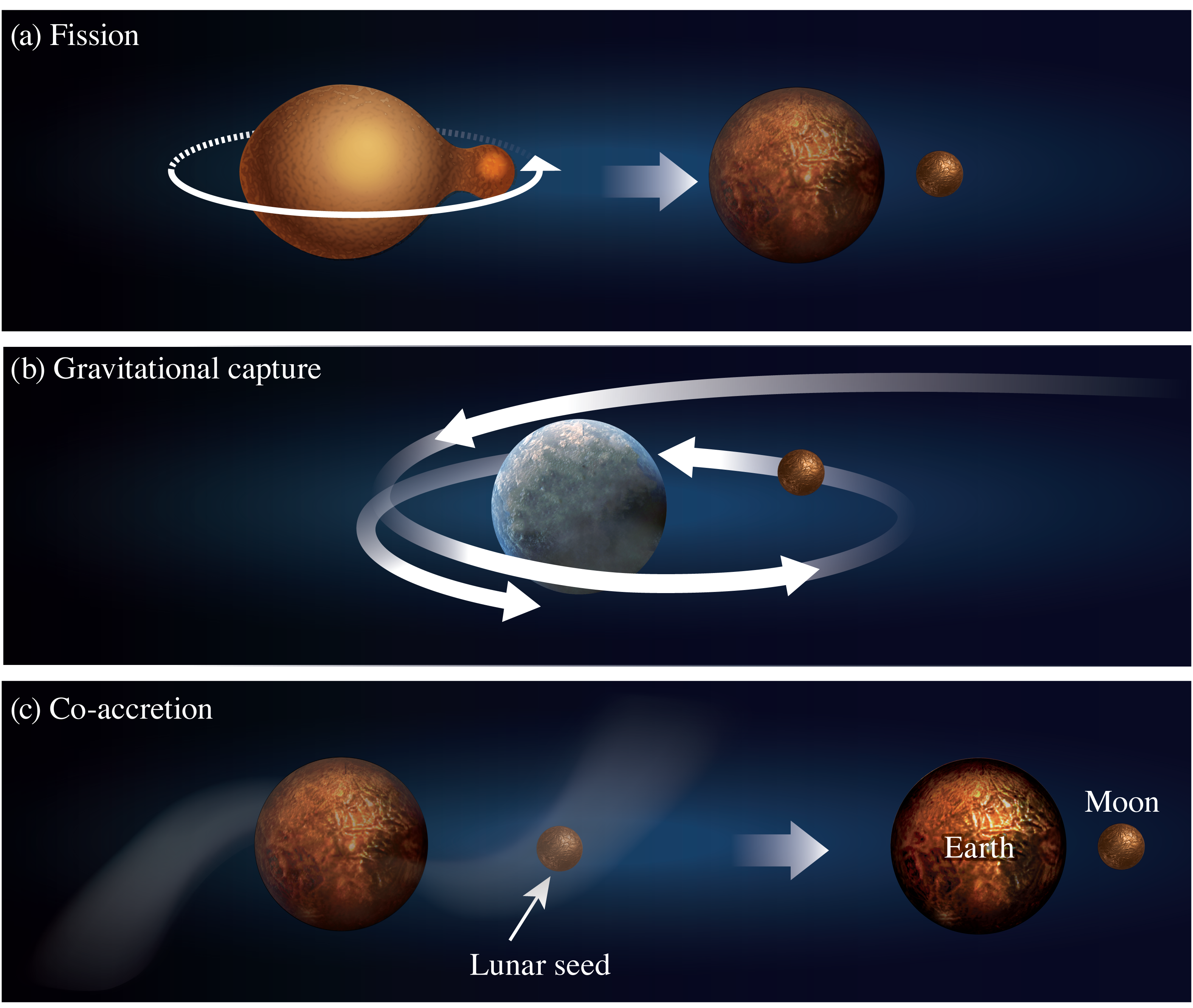}
     \caption{Historical models of the origin of the Moon. (a) Fission: the Moon spun out of the rapidly rotating Earth. (b) Gravitational capture: an asteroid (the Moon) was captured in the Earth's gravity. (c) Co-accretion model: Earth and the Moon formed simultaneously in the protoplanetary disk. Modified from \cite{Hull2024}. }
     \label{fig:Moon_old_models}
 \end{figure*}

\textit{Fission}. \cite{darwin1880, Darwin1962} demonstrated that, were the angular momentum of the Earth-Moon system to have been conserved over time, tidal evolution results in an initial rotational period of the proto-Earth (Earth + Moon) of approximately 4 hours. This means that the period of diurnal solar tides acting on the proto-Earth would have been every two hours, a frequency similar to the slowest free oscillation of a fluid body with the average density of the proto-Earth. Such resonance would have caused an enormous tidal bulge, which culminated in the break up of the proto-Earth, with the resulting debris forming the Moon (Figure \ref{fig:Moon_old_models}a). This hypothesis, however, has several major caveats, including the assumption that the rotating body had a very low viscosity (friction) and ignoring the existence of the Earth's core \citep{Jeffreys1930}. Moreover, the spin period of 4 hours is insufficient for a classical break up scenario, which would require a rotation period of $\sim 2$ hours
\citep{KokuboGenda2010}. \cite{Ringwood1972} suggests that the proto-Earth may have had higher angular momentum than today and this excess was removed by atmospheric loss, but this would have been dynamically challenging.
While the fission hypothesis necessarily results in mass-independent isotopic similarity between the Earth and the Moon, the presence of a lunar core \citep[e.g.,][see section \ref{sec:geophys_core}]{weber_etal2011seismic} is harder to account for, given the difficulty of ejecting metallic iron into orbit around the Earth. Therefore, in such a scenario, the Moon's core would need to have formed endogenously (\textit{e.g.,} through preferential loss of oxygen from the Moon). 
As a result, this hypothesis is not considered as a promising model, though it is successful in accounting for the chemical and isotopic similarity between the Moon and the Earth. Nevertheless, a derivative of the fission hypothesis was, in combination with a giant impact \citep{CukStewart2012}, revisited, as discussed in Section \ref{dynamics_giantimpact} in detail.

\textit{Capture}. 
Another lapsed hypothesis is that the Moon was gravitationally captured by Earth (Figure \ref{fig:Moon_old_models}b). Irregular moons around gas giants, such as Triton \citep{ AgnorHamilton2006}, are thought to have formed by this process. In general, capture requires significant dissipation of kinetic energy in order to convert the heliocentric orbit of a planetary object into a circumplanetary orbit. One way to remove the excess kinetic energy is via strong tidal interactions during a close encounter between a planet and an embryo. However, if the encounter is too close, the embryo can be destroyed, depending on the viscosity of the object \citep{MizunoBoss1985}. 
Earth's atmospheric drag has been proposed as an energy dissipation mechanism by \cite{Nakazawaetal1983}, but the model requires a very small encounter velocity \citep{Stevenson1987}. In addition to dynamical challenges, this model would have difficulties in explaining the mass-independent (O, Cr, Ti) and particularly the W isotope similarity of the Earth and Moon (sections \ref{sec:physchem_terrestrial} and \ref{sec:Hf-W}), as well as the dearth of iron in the bulk Moon \ref{sec:geophys_inv_core}, considering that the bulk compositions of captured objects would be anticipated to be broadly chondritic.

\textit{Co-accretion (binary formation)}. 
\cite{Ruskol1960} argue that debris formed by planetesimal collisions could orbit the Earth, with the debris eventually accreting beyond the Roche Limit to form the Moon (Figure \ref{fig:Moon_old_models}c). The major issue with this hypothesis is that the resulting Moon would be expected to have the same composition as the bulk Earth (and not its mantle). Consequently, it would overestimate the mass of the lunar core, unless there were a mechanism to preferentially strip metallic iron from the Moon-forming region. Moreover, such planetesimal impacts are expected to be random, and therefore achieving a high angular momentum would be challenging. 
One of the more modern binary formation processes would be by streaming instability, which is a rapid way to form objects up to $\sim 100$ km in size by spontaneous concentration of dust particles (pebbles) followed by gravitational collapse. This hypothesis was originally proposed for planetesimal formation \citep{YoudinGoodman2005}, but would later be used to predict the size distribution of the Kuiper Belt Objects (KBOs) and their population of binaries \citep{Nesvornyetal2010, Nesvornyetal2019}. These planetesimal binaries may subsequently evolve by pebble accretion, potentially forming a system akin to that of Pluto-Charon \citep{Konjinetal2023}. However, this was unlikely to have been the case for the Earth-Moon system, given that the solar nebular gas would have already dissipated at the time of Moon formation ($\gg$ 5 Myr, section \ref{sec:age_moon}) and pebble accretion would have ceased to be efficient. 

Thus, while some of the hypotheses above were actively discussed in the past, especially that of fission, all have difficulties in accounting for present-day observables of the Earth-Moon system. Subsequently, the community now perceives the giant impact hypothesis as the most promising model, despite its caveats, as discussed below. \\

\subsection{Review of the giant impact hypothesis}
\label{dynamics_giantimpact}
The various giant impact hypotheses have been recently reviewed by \cite{Canupetal2021}. Here, we provide further reviews as well as discuss recent developments in the field.

  \begin{figure*}[!ht]
     \centering
     \includegraphics[width=0.75\textwidth]{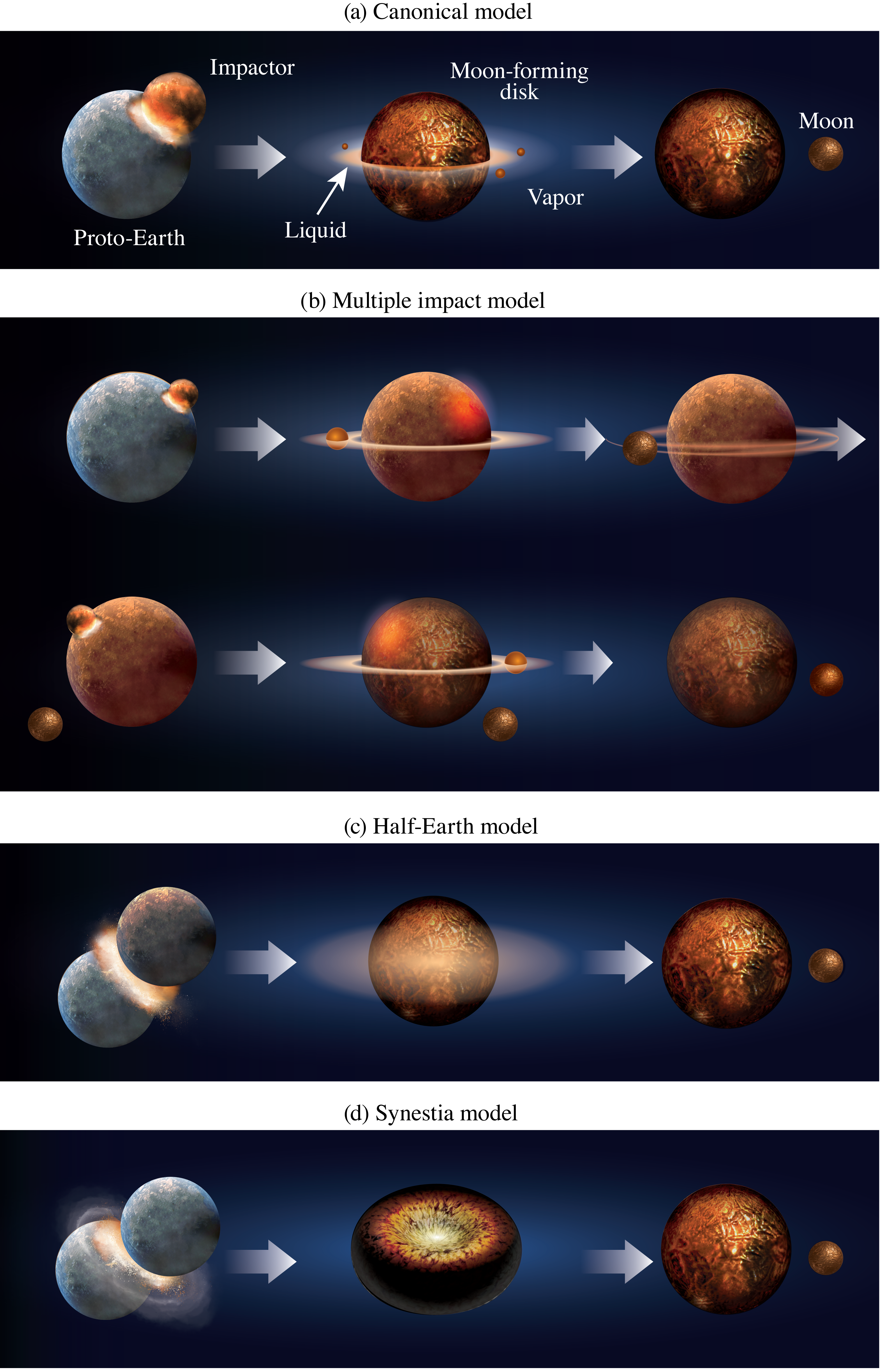}
     \caption{Giant impact hypotheses for the origin of the Moon. (a) Canonical model: a Mars-sized impactor (Theia) hit the proto-Earth. (b) Multiple impact model: the Moon formed from multiple impacts. (c) Half-Earth model: The Moon formed from a collision between two half-Earth sized planetary objects. (d) Synestia model: An energetic impact (similar to the half-Earth impact) formed a donut-shape silicate vapour disk, from which the Moon formed. Modified from \cite{Hull2024}.}
     \label{fig:Moon_impact_models}
 \end{figure*}

\textit{Canonical model.}
The idea that the Moon formed as a result of a large collision was first proposed by \cite{HartmannDavis1975} and \cite{CameronWard1976}. \cite{HartmannDavis1975} argued that such a planetary collision would be sufficiently energetic so as to eject material from the proto-Earth's mantle into orbit. The resulting material, which would be enriched in rocky (as opposed to metallic) components, would then accrete to form the Moon. The high temperatures \citep[$\sim$6000 K,][]{Melosh1990} associated with such a process would have also engendered vaporisation of a substantial fraction of Moon-forming matter, potentially accounting for the Moon's iron- and volatile-depleted nature, as well as its isotopic similarity to Earth. The study of \cite{CameronWard1976} provided support for the collision hypothesis by pointing out that the angular momentum of the Earth-Moon system is compatible with a collision of the sort imagined by \cite{HartmannDavis1975}.

The adoption of the impact hypothesis by the community was initially sluggish; other hypotheses were still actively debated. In 1984, the Origin of the Moon conference in Kona, Hawaii, USA, was instrumental in promoting the giant impact hypothesis as the leading theory in Moon formation. It motivated numerical simulations of the impact itself, though computational resources were still rather limited compared to what is available today. \cite{KippMelosh1986} conducted impact simulations to illustrate that a large fraction of the Earth's mantle materials are injected into orbit around the Earth. 
However, these simulations were conducted in 2D, assuming that two infinitely long cylinders collided without self-gravity, and were therefore of limited relevance to two-body collisions. In contrast, \cite{Benzetal1986, Benzetal1987, Benzetal1989} conducted 3D impact simulations with self-gravity using a method called smoothed particle hydrodynamics (SPH). With the standard version of this method, the fluid mass is represented by Lagrangian particles, whose effective radius (smoothing length) $h_i$ changes depending on how many particles exist nearby (compressible fluid). The density of an isolated particle is approximately proportional to $m_i/h_i^3$, where $m_i$ is the mass of each SPH particle. When a particle is surrounded by a large number of particles (neighbours), $h_i$ becomes small, and the particle has a high density. The density of a given location is determined by summations of densities of $\sim 30$ neighbouring particles \citep{Hopkins2015}. SPH is a very efficient method to investigate impact processes, because it does not need to resolve vacuum space and provides high resolution in a dense region. Early SPH simulations with $10^3$ particles showed that a large impact with energy dictated by mutual escape velocities injects a mass of material into orbit around the Earth that is sufficient to form the Moon (the impactor to proto-Earth mass ratio is 0.1--0.25, \citealt{Benzetal1986}). Moreover, the material that is injected is enriched in silicates relative to metallic iron, consistent with the observation of a small lunar iron core (section \ref{sec:geophys_core}). On the other hand, these early SPH simulations failed to explain the angular momentum of the Earth-Moon system \citep{Benzetal1986, Benzetal1987}. Further SPH simulations showed that an impact between the proto-Earth and a Mars-sized impactor travelling at mutual escape velocities can reproduce the angular momentum of the current Earth-Moon system, assuming that the angular momentum has been conserved since formation \citep{CanupAsphaug2001}. Here, we call this model as the canonical model (Fig. \ref{fig:Moon_impact_models}a). 

The canonical model has had considerable success in explaining many of the observed features as stated above, such as the angular momentum of the Earth-Moon system, the small lunar core, and the total mass of the Moon, but its major drawback is that the isotopic similarity between the Earth and the Moon is difficult to achieve in such a model (Section \ref{sec:intro}). According to SPH simulations \citep{CanupAsphaug2001, Canup2004}, most of the disk particles originate from the impactor, which likely had different isotopic ratios from those of Earth (more discussion on this in Sections \ref{sec:physchem_terrestrial} and \ref{sec:Hf-W}). Given that the Moon originates from the Moon-forming disk, the chemical and isotopic compositions of the Moon should reflect that of the impactor. This result, at odds with observations, is therefore the major problem befalling the canonical model, referred to as the so-called ``isotopic crisis'' \citep[cf.][]{Melosh2017}. 

Several potential solutions to this problem have been proposed. One mechanism that may promote homogenisation between the proto-Earth and the impactor is convective mixing in the vapour portion of the disk and the Earth's atmosphere during the disk stage \citep{PahlevanStevenson2007}. At the time this idea was proposed, oxygen, a partially volatile element, was the only element shown to have an identical isotopic composition between the Moon and the Earth \citep{wiechert2001}. For this isotopic similarity to have arisen, O would have had to have been nearly quantitatively vaporised during the disk phase, consistent with the temperatures extant in the aftermath of a canonical impact. However, the subsequent discovery of isotopic overlap for refractory elements, such as Cr and Ti in the Earth and Moon (cf. section \ref{sec:physchem_terrestrial}) rendered this mechanism unsatisfactory. Alternative solutions to this issue proposed that, because of extensive radial mixing in the inner solar system, the impactor may have happened to have the same isotopic ratios to Earth \citep{Dauphasetal2014}. Though fortuitous, such a scenario would inherently be able to account for the O, Cr and Ti isotopic similarity of the Earth and Moon, as these ratios track only the provenance of material within the protoplanetary disk. However, the Mn, Cr and V abundances, as well as tungsten isotopic similarities of the two bodies identified thereafter (section \ref{sec:Hf-W}), require an identical core formation history, both in time and extent \citep{KruijerKleine2018, Fischeretal2021}, thereby compromising the suitability of any such 'Earth-like impactor' models. \\

\textit{Hit-and-run models.}
In hit-and-run scenarios, the impactor collides with the proto-Earth but does not form a Moon-forming disk after the first encounter. Instead, it orbits the Sun together with the proto-Earth and eventually hits the proto-Earth again. Such multiple hit-and run collisions can launch more Earth-like materials into the orbit \citep{Reuferetal2012, Asphaugetal2021} promoting mixing between the proto-Earth and the Moon. Despite the enhanced efficiency of mixing, the remarkably similar isotopic ratios between the Earth and the Moon remain out of reach because the protolunar disk composition is still dominated by the impactor.

\textit{Instantaneous Moon formation model.}
\cite{Kegerreisetal2022} conducted SPH simulations with much higher resolution than conventional ones ($N=10^8$ where $N$ is the number of SPH particles). The work finds that the impactor divides into two fragments in higher resolution studies ($N\geq 10^6$), whereas it remains whole when examined in lower resolution ($N\leq 10^6$) a few hours after the initial impact. This trend has also been confirmed in other studies \citep{Hulletal2023}. In high resolution cases, under certain conditions, one of the fragments starts orbiting the proto-Earth. Interestingly, instantaneous clump formation has been previously observed in low resolution simulations (e.g., $N\leq 10^{4-5}$, \citealt{Benzetal1986, Canup2004}). The clumps consist of a mixture of the proto-Earth- (40-50 wt\%) and impactor-derived materials (50-60 wt\%), but the upper 10 \% in radius of the clump can be more proto-Earth rich ($>60$ wt \%). Assuming that the Moon eventually forms from this fraction of the clump, the isotopic similarity between the Earth and the surface of the Moon could be explained, if the Moon's interior is compositionally and isotopically different from the surface. This lunar interior heterogeneity has been previously proposed (e.g. \citealt{SalmonCanup2012}) and there is some potential isotopic observations (e.g. oxygen isotopes, \citealt{Canoetal2020}), but there is no direct evidence that the lunar interior is significantly different from more surficial regions (see Section \ref{sec:geophys_inv}). Moreover, since this model does not experience a disk phase, volatile loss due to the impact or from the magma ocean is required to explain the lunar volatile depletion.

\textit{Magma ocean model.}
\cite{MeloshSonett1986} proposed that a jet induced by the Moon-forming impact may launch mixtures of the target and impactor materials into orbit before eventually forming the Moon. \cite{Karato2014} proposes a scenario in which the surface of the proto-Earth was covered by a magma ocean at the time of the Moon-forming impact. Since the bulk modulus of magma (silicate liquid) is smaller than that of solid, magma is more compressible and shock heated by impact than a solid mantle would have been.  Their analytical model suggests that the impact could form a jet originating from the proto-Earth's mantle, which then becomes the Moon. \cite{Hosonoetal2019, Hosonoetal2022} conducted giant impact simulations to test the feasibility of this hypothesis. In their calculations, a hard-sphere model was used for the equation of state (EOS) of the SPH particles that were initially located near the surface of the proto-Earth (representing a silicate liquid). They used the standard SPH as well as the density independent SPH (DISPH) method. In the standard SPH, spatially smooth density is required by design, but compromises accuracy in reproducing density discontinuities (e.g. core-mantle boundary). In contrast, DISPH requires smooth pressure, which appears to describe material boundaries and instability growth more faithfully \citep{Hosonoetal2016}. Improvements in boundary descriptions have been also investigated elsewhere (e.g. \citealt{Ruiz-Bonillaetal2022}).
While impact simulations with DISPH are able to produce a disk in which material primarily originates from the proto-Earth, those based on the standard SPH implementation do not. The origin of this phenomenon is still unknown, partly because whether this can be seen in other SPH simulations with improved boundary conditions has not been reported (e.g. \citealt{Ruiz-Bonillaetal2022}). Moreover, the SPH simulations conducted by \cite{Hosonoetal2019, Hosonoetal2022} may not be fully equivalent to the original jet formation model because SPH is not able to handle two phases separately (i.e. vapour jets and melt) given that SPH assumes that these two phases move with the same velocity. Thus, SPH is not able to fully capture jet dynamics. 
Direct comparisons among different SPH methods as well as different impact codes would be helpful to assess how such modifications affect the final outcome and origin of lunar material (e.g. \citealt{Wadaetal2006, Canup2013, Dengetal2019, Ruiz-Bonillaetal2022}). 

\textit{Multiple impact model.}
\cite{Rufuetal2017} propose that the Moon formed as a result of multiple small impacts (Figure \ref{fig:Moon_impact_models}b). A relatively small impactor whose mass is a few wt\% of the Earth mass with a high impact velocity (2-4 $v_{\rm esc}$, where $v_{\rm esc}$ is the escape velocity) can strip off Earth's mantle materials and form a disk originating from Earth. Each impact produces a relatively small disk mass and moonlet mass, and therefore this process has to occur multiple times. If these moons merge \citep{Canupetal1999}, this model can produce the Moon, which is isotopically similar to Earth.  However, such high velocity impacts are less common towards the end of the planet accretion stage (see Section \ref{sec:giant_impact}).

\textit{Half-Earth model.}
\cite{Canup2012} proposed a model that involves a collision between two half-Earth sized planetary objects. Such an energetic impact effectively mixes the proto-Earth and the impactor, naturally explaining the isotopic similarity (Figure \ref{fig:Moon_impact_models}c). At the end of the simulation, the angular momentum of the Earth-Moon system is $\sim 2.5$ times as large as the current value. Some of the excess can be removed via resonances, but its removal efficiency is under active scrutiny (see Section \ref{sec:angular_momentum}). Additionally, collisions between two protoplanets with subequal masses at the end of the planetary accretion phase are typically rare, but the statistics depend on the planetary growth model (see Section \ref{sec:dynamics} and \citealt{Canupetal2021}). 

\textit{High spin and high angular momentum model.}
\cite{CukStewart2012} proposed a model in which a small impactor hits the rapidly rotating proto-Earth, which has a large tidal bulge due to its rapid rotation period. The impactor hits the bulge and launches a portion of the Earth's mantle into the orbit, which eventually becomes the Moon. Since the Moon would form largely from material ejected from the Earth, this model could naturally explain the similar mass-independent, and, importantly, W isotopic ratios relative to the Earth's mantle. They investigate a scenario in which the proto-Earth's spin angular momentum and the orbital angular momentum of the impactor are aligned, which is not very likely. \cite{LockStewart2017} and \cite{Locketal2018} generalised this model by expanding it to impacts that have high-angular momenta and high-energies. The impact is so energetic that it transforms the disk, in part, into a supercritical fluid that behaves in a contiguous manner. \cite{LockStewart2017} and \cite{Locketal2018} dubbed this structure a synestia (Figure \ref{fig:Moon_impact_models}d). The synestia would convect vigorously, thereby erasing any pre-existing heterogeneity. 
\cite{Locketal2018} illustrated that silicate droplets condense as the synestia cools and coalesce to form moonlets, and, eventually, the Moon itself. However, the growth of the Moon to its present-day mass may face a challenge; these silicate droplets and moonlets would experience a strong pressure gradient (or headwind) in the disk, limiting the efficiency of their accretion (see Section \ref{sec:gas_drag}). \\

\subsection{Dynamics and predictions of the giant impact hypothesis}
\label{sec:dynamical_problems}
\subsubsection{Angular momentum evolution in the Earth-Moon system}
\label{sec:angular_momentum}
Historically, the angular momentum of the Earth-Moon system was thought to have been conserved for the $\sim$4500 Myr period between the time of the putative impact and the present-day. \cite{ToumaWisdom1998} find that the Moon would encounter an evection resonance at $4.6 R_\oplus$ (where $\oplus$ denotes the Earth), which occurs when the period of the lunar perigee precession is the same as the Earth's orbital period (1 year). This resonance can transfer the angular momentum of the Earth-Moon system to the Sun-Earth system. This effect is limited for the canonical model assuming that the Earth-Moon system's initial angular momentum has been conserved throughout its history (the Earth's spin period would have been $\sim5$ hours \citep{Canup2004} at the time of the Moon formation) because the Moon would have escaped the evection resonance quickly \citep{ToumaWisdom1998}. \cite{CukStewart2012}, in contrast, find that in the fast spinning Earth scenario, a significant amount of the Earth-Moon angular momentum would have been lost in the evection resonance because the Earth's high spin and oblateness moves the evection resonance location at $\sim 6.8 R_\oplus$ and the Moon could stay in the resonance for nearly 70,000 years. Approximately, the angular momentum of $\sim 1 L_{\rm EM}$, where $L_{\rm EM}$ is the current Earth-Moon angular momentum, could be removed by this process. 

Evection resonance appears to be an important process in the evolution of the angular momentum of the Earth-Moon system, yet, how much angular momentum can be drained from the system remains unclear. \cite{CukStewart2012} employ a simplified tidal evolution model (tidal Q), whereas \cite{WisdomTian2015, Tianetal2017} find that a more conventional tidal Q model does not remove sufficient angular momentum from the Earth-Moon system. The work finds an alternative mechanism called a limited cycle, which is a cycle associated with the evection resonance. \cite{Tianetal2017} argue that the evection resonance capture is not likely once the thermal interior evolution of the Moon is considered. The key parameter here is the so-called tidal parameter $A$, which is described as
\begin{equation}
A = \frac{k_{2M}}{k_{2E}}\frac{Q_{E}}{Q_{M}}\frac{m^2_{E}}{m^2_{M}}\frac{R^5_M}{R^5_E},
\label{eq:tidalA}
\end{equation}
where $k_2$ is the Love number, $Q$ is the dissipation factor, $m$ is the mass, and $R$ is the radius, and the subscripts $E$ and $M$ represent the Earth and the Moon. $k_2$ and especially $Q$ are expected to change during magma ocean crystallisation. In the original \cite{CukStewart2012} model, the large angular momentum loss occurs at a limited $A$ range ($A=0.8-1.7$, i.e. $0.5<Q_E/Q_M<1.2$), but \cite{Tianetal2017} find that the $A$ can change by orders of magnitude when the interior evolution of the Moon is considered, which makes the angular momentum removal by evection resonance less promising. The limit cycle can remove the excess angular momentum, but an acceptable range of $A$ depends on factors such as $Q_E$. In addition to these numerical studies, analytical calculations conducted by \cite{Wardetal2020, RufuCanup2020} find that angular momentum loss by evection resonance appears to be inefficient. Another configuration, starting from a highly inclined lunar orbit, has also been discussed \citep{Cuketal2016, Cuketal2021}, while \cite{TianWisdom2020} argue that this configuration may not reproduce the component of the current Earth-Moon angular momentum perpendicular to the ecliptic plane. Thus, further research is needed to investigate the evolution of the angular momentum of the Earth-Moon system.

\subsubsection{Dynamical constraints on the giant impact}
\label{sec:giant_impact}
The dynamical requirements for the impact models, as well as upon the planet formation scenario(s) that underpin them, may differ significantly. 
In the classical planet formation scenario (\textit{Planetesimal collision model}), planetesimals collide with each other until they reach the Mars-sized protoplanets \citep{Safronov1966,wetherill1986}. After the dissipation of the nebular gas, the orbits of protoplanets become unstable and they collide with one another, pertaining to the so-called giant impact stage \cite[e.g.][]{Weidenschilling1977, KokuboIda1998}. An alternative end-member scenario is the \textit{pebble accretion model}, where small, mm- to cm-sized particles (pebbles) efficiently accrete onto growing protoplanets due to gas drag, which requires completing most of the planet accretion before the protoplanetary disk has dissipated \cite[e.g.][]{JohansenLambrechts2017, Johansenetal2021, Onyettetal2023}. The pebble accretion model was originally proposed to explain the rapid formation timescale of Jupiter \citep{LambrechtsJohansen2012}, but has been recently applied to terrestrial planet formation models. If pebble accretion is efficient in the inner solar system, planets do not necessarily have to experience giant impacts to reach their final planetary masses. \cite{Johansenetal2021} argue that the only giant impact in the Solar System may be that which formed Earth's Moon. This may not be fully consistent with other giant impacts proposed in the inner solar system in order to explain, for example, Mercury's large iron core \citealt{Benzetal2007}, Martian moon formation \citealt{Craddock2011} or the Martian crustal dichotomy \citealt{Marinovaetal2008}, but it should be noted that alternative models exist to account for these observations. Moreover, pebble accretion requires planet formation within a few million years, but this appears to be inconsistent with the Earth's tungsten isotopic ratios indicate $\sim30$ Myr formation timescale (unless a giant impact promoted equilibration between Earth's mantle and core) and the late formation age of the Moon (see Section \ref{sec:age_moon}). Pebble accretion may not have been efficient in the inner solar system, and therefore planetesimal collisions and giant impacts played a key role to shape the planetary system \citep{Burkhardtetal2021, Lichtenbergetal2021, BatyginMorbidelli2022}.

\begin{figure*}[!ht]
     \centering
     \includegraphics[width=\textwidth]{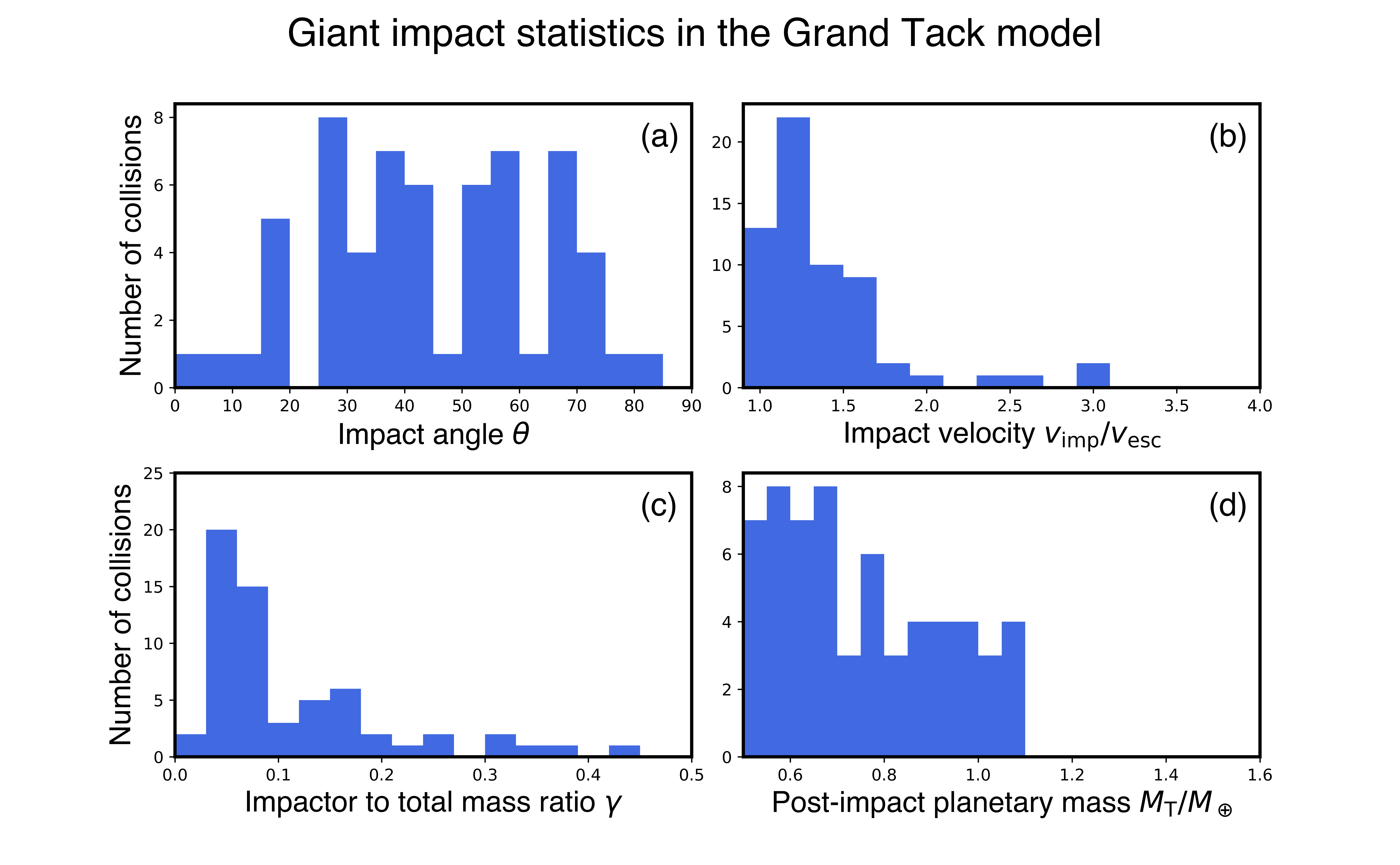}
     \caption{Giant impact statistics of the Grand Tack simulations in \citealt{Rubieetal2015}. These are histograms of (a) the impact angle $\theta$, (b) the normalized impact velocity $v_{\rm imp}/v_{\rm imp}$, (c) the impactor-to-total mass ratio $\gamma$, and (d) the normalised post-impact planetary mass $M_{\rm T}/M_\oplus$.}
     \label{fig:Rubie}
 \end{figure*}

Since the pebble accretion model does not provide constraints as to the nature and frequency of giant impacts, we focus on the planetesimal collision model and investigate the likelihood of various impact hypotheses (see also \citealt{Canupetal2021}).
Figure \ref{fig:Rubie} shows the statistics of giant impacts from the orbital dynamics simulations \citep{Rubieetal2015}. Here, we define giant impacts as the impacts whose total masses exceed 0.5 $M_\oplus$, where $M_\oplus$ is the Earth's mass. These simulations are based on the Grand Tack model \citep{Walshetal2011}, where the planetesimal disk is truncated at 1 AU due to Jupiter's inward migration.
Figure \ref{fig:Rubie}a-d show the impact angle ($\theta$), the impact velocity $v_{\rm imp}/v_{\rm esc
}$, where $v_{\rm esc}$ is the escape velocity, the impactor-to-total mass ratio ($\gamma$), and the total mass ($M_{\rm T}/M_\oplus$) distribution in the orbital dynamics calculations. In the Grand Tack model, the configuration required for the canonical impact ($\theta \sim45^\circ$, $v_{\rm imp}\sim v_{\rm esc}$, and $\gamma\sim0.1$) is relatively common. The required $\gamma$ is slightly larger than the most probable $\gamma$ ($ \leq  0.05$) in these specific simulations, but more common in the conventional orbital dynamic calculations (a few - 10\%, \citealt{Canupetal2021} and references therein). In contrast, the half-Earth model requires $\gamma=0.4-0.5$, $v_{\rm imp}\sim v_{\rm esc}$, $\theta \sim 45^\circ$. By the end of the accretion stage, high values of $\gamma=0.4-0.5$ are not very likely (Fig. \ref{fig:Rubie}, see also \citealt{Canupetal2021}). While a synestia may form by various impact conditions, impacts similar to the half-Earth model are preferable because such impacts can easily provide sufficient energy and angular momentum (a collection of small impacts tend to cancel out the total angular momentum). Thus, such a scenario is dynamically less favoured in the conventional planet formation and Grant Tack models. 

\begin{figure*}[!ht]
     \centering
     \includegraphics[width=\textwidth]{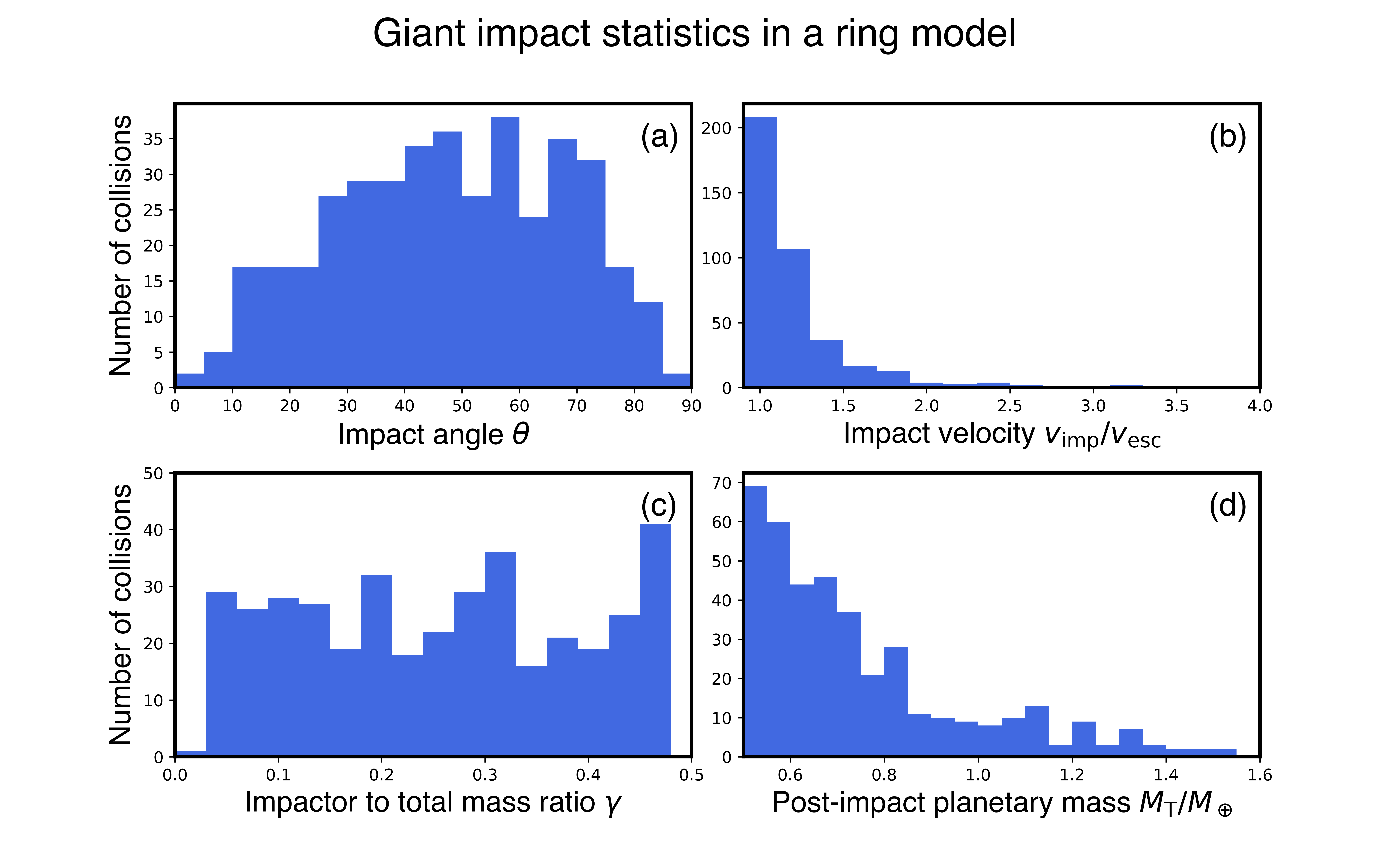}
     \caption{Impact statistics of the ring model  in \citealt{Wooetal2024}. The parameters shown in this figure are the same as Figure \ref{fig:Rubie}.}
     \label{fig:Woo}
 \end{figure*}

However, a new study indicates that nature and frequency of giant impact events depend upon the type of planet formation model invoked. Figure \ref{fig:Woo} shows the giant impact statistics in a recent model \citep{Wooetal2024}, where terrestrial planets grow from a `ring' of planetesimals at $\sim 1$ AU, where the streaming instability is an efficient way to form planetesimals due to an enhanced dust-to-gas ratio \citep{Morbidellietal2022}. The overall statistics are similar to those in the Grand Tack model, but there are several notable differences. First, the impact velocities are typically smaller in the ring model than in the Grand Tack model, as shown in Figure \ref{fig:Woo}b.  Secondly, the impactor-to-total mass ratios $\gamma$ are much larger in the ring formation model, which is because planetesimals grow relatively uniformly to similarly sized embryos. These high $\gamma$ values make half-Earth and synestia-forming impacts more dynamically probable.

As such, the canonical impact model is dynamically highly probable across the conventional, Grant Tack, and the ring models. Impact styles that are required or preferred for the half-Earth and synestia models are relatively common in the ring model, whereas they are rare in the other dynamical models. The multiple impact scenario requires high velocity impacts (2.5-4$v_{\rm esc}$) in order to strip material from the Earth's mantle. Such high velocity impacts are possible (Figures \ref{fig:Rubie}c and \ref{fig:Woo}c) but not as commonplace as low velocity impacts. 


\subsubsection{Problems with gas drag and the streaming instability}
\label{sec:gas_drag}
The coalescence of the Moon from a protolunar disk around the Earth depends upon not only the amount of material injected into orbit, but the facility with which such material can settle to the midplane. \cite{Nakajimaetal2022} proposed that strong gas drag in an initially vapour-rich disk onto silicate particles may prevent moonlets from growing larger than a few km. Gas drag originates from the velocity difference between particle (Keplerian) and gas (sub-Keplerian due to the pressure gradient in the disk). This issue was also pointed out by \cite{Canupetal2021}, but the implications were not discussed in detail. \cite{Nakajimaetal2022} calculated the structure and vapour mass fraction (VMF) of the disk based on standard SPH simulations. The disk VMF critically depends on the model; the canonical VMF is $0.1-0.3$ \citep{NakajimaStevenson2014}, the VMF of the half-Earths and high-energy high-angular momentum scenarios are 0.7-1.0 \citep{NakajimaStevenson2014}, and VMF for the multiple Moon model is 0.1-0.5 \citep{Rufuetal2017}. If the initial disk is completely vaporised, as is the case for the energetic models, the radial pressure gradient of the disk is so large that growing moonlets fall onto the Earth on short timescales. This timescale critically depends on the particle size; the residence time for a 2 km-sized particle is short ($\sim1$ day), while much smaller particles are completely coupled with the gas and larger particles are entirely decoupled from the gas, and therefore these particles are less affected by the gas drag. Mass loss due to gas drag continues to occur until the disk cools sufficiently so as to have caused it to have largely condensed. Once VMF becomes small enough, the effect of gas drag weakens and the accretion of the Moon gathers pace, but by this time a significant fraction of the mass of the disk ($>\sim 80 \%$) has been lost \citep{Idaetal2020, Nakajimaetal2022}. Thus, even if the remaining condensed material can accrete, the resulting mass of the Moon may be too small relative to that observed. This problem applies only to initially vapour-rich disks, whereas particles in vapour-poor disks experience less gas drag and are able to remain in orbit. 

The impediment of gas drag on the accretion of a growing object in a disk was originally recognised for planet formation in the protoplanetary disk, where growing $\sim 1$ m-sized planetesimal fall onto the Sun in $\sim 80 $ years \citep{Adachietal1976, Weidenschilling1977}, which is much shorter timescale than the planet formation timescale (a few to 10s of Myr, according to $N$-body simulations). A likely solution comes in the form of the streaming instability, a process that permits rapid formation of much larger planetary bodies by concentrating particles that then gravitationally collapse \citep[e.g.][]{YoudinGoodman2005}. By this process, accreting planets can avoid the problematic ``1 meter barrier''. This same mechanism, however, does not appear to have been relevant for the Moon-forming disk. Simulations of streaming instability in the Moon-forming disk show that the maximum size of the streaming instability is $\sim 100$ km \citep{Nakajimaetal2023}, too small to overcome gas drag. Thus, energetic models, in which much of the disk is initially vaporised, may still encounter difficulties in reaching the mass of the Moon. 


\subsection{Lunar volatile loss}
\label{sec:lunarvolatileloss}
As discussed in detail in Section \ref{sec:physchem_volatile}, how and when volatiles were lost from the Moon can place important constraints on dynamical models. Chemical and isotopic evidence are consistent with low temperature ($\sim$ 1500 K) volatile loss from the Moon during gas-condensed phase equilibrium, though higher temperatures ($\sim$ 3000--3500 K) are possible, should the mechanism depart marginally from equilibrium. Typically, two key sites for vapour loss are mooted; (1) the Moon-forming disk, and (2) the lunar magma ocean (post-accretion). 
We discuss the proposed dynamics of the models as below.

\subsubsection{Mechanisms for volatile loss from the Moon}

Following the Moon-forming impact, a partially vaporised disk forms around the Earth and the Moon accretes in 10s--100s of years, depending on the disk evolution model \citep{ThompsonStevenson1988, SalmonCanup2012, Locketal2018}.  
The initial condition of the disk is sensitive to the impact model. As discussed in Section \ref{sec:gas_drag}, the canonical model tends to produce a disk with a (silicate) vapor mass fraction of 0.1--0.3 (i.e. 70--90 wt \% silicate liquid) and a midplane temperature of 4000--5000 K, while more energetic models, such as the high angular momentum and high energy impact, can produce disk vapor mass fractions of 0.7--1.0 with midplane temperatures of 6000--7000 K (e.g. \citealt{NakajimaStevenson2014}). The initial disk temperature can affect the extent of volatile loss and the gaseous species involved, as discussed in detail below.

\textit{Atmospheric escape from the Moon}
 \cite{GendaAbe2003b} calculate a one-dimensional hydrostatic disk structure and estimate the condition of the breakdown of the hydrostatic assumption for the gas phase, which then leads to the onset of hydrodynamic escape of the atmosphere. \cite{DeschTaylor2013} propose that hydrodynamic escape would occur from the Moon if the Jeans parameter;
 
 \begin{equation}
 \lambda = \frac{GM_\oplus m}{RTr}, 
 \label{eq:esc_param}
 \end{equation}
 falls below 2 \citep{Parker1963}, where $G$ is the gravitational constant, $M_\oplus$ is the Earth's mass, $m$ is the mean molar mass of the atmosphere, $R$ is the gas constant, $T$ is the disk temperature, $r$ is the distance from Earth. This condition may be met when the disk is predominantly composed of hydrogen, H$_2$ ($m$ = 2), but H can also speciate as H$_2$O (H$_2$O=H$_2$+$\frac{1}{2}{\rm O}_2$). As O$_2$ is a minor gas species ($<$10$^{-10}$ bar at 1500 K), the relative proportion of H$_2$ to H$_2$O controls the value of $m$ of the atmosphere. The \textit{f}H$_2$/\textit{f}H$_2$O depends on the \textit{f}O$_2$ of the lunar atmosphere at the time of its formation, which remains uncertain, with the implication that reducing atmospheres promote escape relative to their oxidising counterparts ($m_{\rm H_2}$ $\ll$ $m_{\rm H_2O}$), all else being equal. 
 Escaping H$_2$ could drag other, heavier elements, predicting loss of water and other volatile elements present in the gas. Extensive loss of water may or may not be consistent with the estimated water abundance of the Moon, which is uncertain (a few- to 100s of ppm, \citealt{Boyceetal2010, Haurietal2011, McCubbinetal2010, Huietal2013, Saaletal2013, McCubbinBarnes2019}). \cite{DeschTaylor2013} assume that all the volatiles would escape simultaneously, whereas \cite{NakajimaStevenson2018} find that volatile loss from the disk would be diffusion-limited because light elements and molecules (e.g. H$_2$) need to first diffuse out from the disk atmosphere which is dominated by heavy elements and molecules (such as SiO$_2$ at $T_{\rm mid}>2500-2800$ K and H$_2$O at $T_{\rm mid}<2500-2800$ K, where $T_{\rm mid}$ is the midplane disk temperature). \cite{NakajimaStevenson2018} conclude that volatiles are not readily lost from the the Moon-forming disk to space, owing to the difficulty of escaping the gravitational potential well of the Earth. It should be noted, however, that this conclusion is contingent upon H$_2$O being a major component of the gas at $T_{\rm mid}<2500-2800$ K, whereas H$_2$ could 
 have prevailed were \textit{f}O$_2$ to have been sufficiently low. 

\cite{charnoz2021tidal} pointed out that the energy needed for a gaseous particle to escape from the Moon and be captured by the Earth is significantly lower than that for escape to space (eq. \ref{eq:esc_param}), particularly when the Moon forms, close to the Earth's Roche Limit at $\sim$3 $R_\oplus$. \cite{tangyoung2020} suggested that, even under this energetically favourable configuration, atmospheric escape from the Moon is limited, owing to the Myr timescales required. However, at high temperatures ($>$ 1500 K), the lunar atmosphere may have been hydrostatically unstable down to the surface of a lunar magma ocean, and mass in the atmosphere would have been, in part, transported advectively (i.e., by hydrodynamic escape). However, advective transport would engender isotopic fractionation according to the differential velocity of the advecting gas and the thermal velocity given by a Maxwell-Boltzmann distribution \citep{charnoz2021tidal}. The magnitude of the isotopic fractionation engendered by this process is too large relative to observations (Table \ref{tab:stables}) for hydrodynamic escape to have played a prevailing role for volatile loss on the Moon \citep{tangyoung2020,charnoz2021tidal,Dauphasetal2022_MVE}. 

\cite{canup2015} conducted a numerical study of the evolution of the Moon-forming disk. The region of the disk exterior to the Roche radius, which should reflect the initial bulk composition of the Moon-forming disk, is expected to have accreted quickly and to have formed moonlets that represent the interior part of the Moon \citep{SalmonCanup2012}. By contrast, the part of the disk inside the Roche radius is not able to accrete at first because the gravity of the Earth continuously disrupts self-gravitating moonlets. However, the inner disk slowly expands outwards viscously over time. Material originating from the inner disk would be initially volatile-depleted, because volatiles would be still in the vapour phase in the hotter, inner disk. Once this volatile-depleted portion of the disk has crossed the Roche radius, it can accrete onto the pre-existing and volatile-rich moonlets. The inner disk would become more volatile-rich over time (closer to the initial bulk composition) as the disk cools and more material condenses, but these more volatile-rich materials may not efficiently accrete onto the moonlets outside the Roche radius because they tend to be gravitationally scattered by the growing large moonlets. As a result, this model would predict that, initially, or in the absence of a magma ocean, the interior of the Moon, which formed from the Roche-interior disk, would be volatile-rich while the exterior, which formed from the Roche-exterior disk, is more volatile-depleted. Whether such signatures would be preserved despite mantle convection of the Moon or observable remain unknown, though it appears that at least the pyroclastic glasses, which come from deeper regions of the lunar interior \citep{elkins2003experimental} may be more volatile-rich than the shallower source regions of Low- and High-Ti mare basalts (see Section \ref{sec:geochem_mantle}.  \cite{Locketal2018} propose that the Moon formed from condensates within a synestia under high pressure and temperature conditions ($\sim3500-4000$ K, $\sim10-100$ bars) could explain the chemical depletion pattern. However, this model may be difficult to explain the enrichment of heavy isotopes of the Moon (such as K, \citealt{wangjacobsen2016}) because equilibrium condensation at this high tempearatures would produce negligible isotopic fractionation, whereas kinetic isotopic fractionation would predict isotopically light condensates (see section \ref{sec:physchem_volatile}). 

Several studies suggest that Earth's magnetic field may have played an important role for lunar volatile depletion. \cite{CharnozMichaut2015} propose that vapour in the Moon-forming disk may have been turbulent due to MRI (magneto-rotational instability). Due to the high temperature of the disk, vapour could have been partially ionised \citep{VisscherFegley2013_MVE}. A terrestrial or stellar magnetic field coupled with ions in the atmosphere can trigger MRI in the disk, which can significantly affect its viscosity. Typically, the viscosity of vapour is described as $\nu=\alpha c_s H$, where $\alpha$ is a dimensionless parameter that describes the extent of turbulence, $c_s$ is the sound speed, and $H$ is the scale height. This formulation was originally developed for the protoplanetary disk, but has been used to describe general disk viscosities. The value of $\alpha$ is uncertain and it can change temporary and spatially by orders of magnitude \citep{Carballidoetal2016, MullenGammie2020}. Assuming $\alpha=10^{-4}$, \cite{CharnozMichaut2015} argue that the vapour portion of the disk spreads more quickly than liquid and would accrete onto Earth. If the Moon accreted from the liquid portion of the disk, the Moon would become volatile depleted. \cite{Carballidoetal2016} follow up on this work and propose that extensive mixing and isotopic homogenisation could occur, even though the long term average of $\alpha$ is small ($\alpha\sim 7 \times 10^{-6}$). 

\begin{figure}[!ht]
     \centering
     \includegraphics[width=0.5\textwidth]{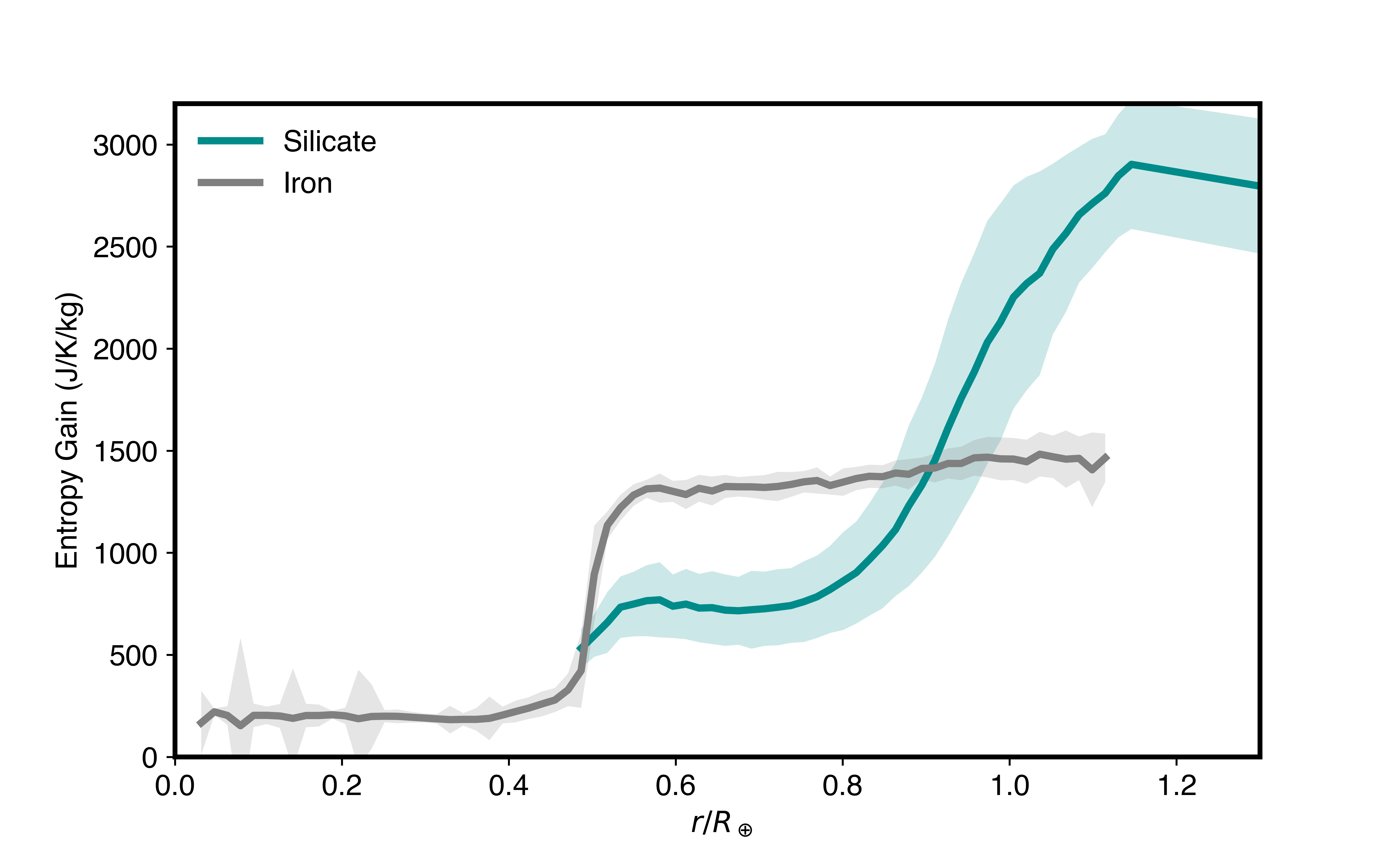}
     \caption{Entropy gain of the Earth's interior after the canonical impact based on an SPH simulation \citep{NakajimaStevenson2015}. The dark green line represents the silicate mantle and grey line represents the iron core. The entropy is averaged as a function of the radius with a 100 km interval. Regions that contain less than 10 SPH particles are omitted. The standard deviations are shown by the shaded region. The impactor's iron core is entrained in the Earth's mantle, and therefore the iron is distributed all the way to the surface. Both the silicate mantle and iron core are thermally stratified, which prohibits core convection. }
     \label{fig:entropy}
 \end{figure}

Here, we suggest that the influence of the magnetic field of Earth after the impact, however, is likely to have been suppressed. Shortly after the impact, the Earth's mantle and core are thermally stratified (Figure \ref{fig:entropy}). The top of the mantle would have begun convecting owing to radiative cooling from the surface, but it would take at least $\sim 1000$ years for the mantle to reach close an adiabatic state \citep{NakajimaStevenson2015}. The core itself is also thermally stratified (Figure \ref{fig:entropy}), which would delay the onset of core convection. Given that the timescale for formation of the Moon is 10s--100s years, it is not likely that Earth would have had a strong magnetic field, though whether the stellar magnetic field could trigger an MRI needs further investigation. 

Alternatively, \cite{MullenGammie2020} propose that MRI could have occurred had the proto-Earth and Theia possessed strong magnetic fields prior to the collision. These fields may be exponentially amplified in the early disk phase. The vapour portion of the disk would have been turbulent and spread out very rapidly ($\sim 600$ hours, \citealt{Gammieetal2016}) to large radii ($\sim 10 R_\oplus$). Simultaneously, some vapour would have fallen onto Earth, potentially making the Moon volatile-depleted. This model may require a strong pre-impact magnetic field ($\sim 1 $kG, while the current Earth's magnetic field strength on the surface is $10^{-1}$ G), and whether a protoplanet can have such a strong magnetic field has not been demonstrated. Thus, MRI has the potential to affect the disk dynamics, mixing, and distribution of chemical elements and isotopes, but whether MRI can develop in the disk is uncertain and this requires further research.

\subsection{Interim grades for the proposed dynamical models}
Thus, all of the proposed models have strengths and weaknesses, whereas their weights are different. The transcript of the proposed models in our opinion is shown in Table \ref{tb:transcript}. The models considered here are (a) the canonical model, which also includes the hit-and-run (HR) model, the magma ocean model, (b) the multiple impact model, (c) the half-Earth model, and (d) the Synestia model.

The mass-independent isotopic similarity for the various models have been discussed in Sections \ref{sec:physchem_terrestrial} (isotopes) and \ref{dynamics_giantimpact} (models). The observed chemical depletion pattern and mass-dependent isotopic fractionations and model predictions are listed in Sections \ref {sec:physchem_volatile} and \ref{sec:lunarvolatileloss}. The angular momentum, dynamical likelihood of the impact conditions, and lunar mass are discussed in Sections  \ref{sec:angular_momentum}, \ref{sec:giant_impact}, and \ref{sec:gas_drag}, respectively.

\begin{table*}[!h]
\begin{center}
\begin{tabular}{ c c c c c } 
\hline
& (a) Canonical  & (b) Multiple & (c) Half-Earth & (d) Synestia \\
\hline
 Isotopic similarities & C & C & A & A \\ 
 Volatile depletion (chemistry) &  B & B & B & B \\ 
  Volatile depletion (isotopes) &  B & B & C & C \\ 
 Angular momentum & A & B & C & C \\ 
Dynamical likelihood & A & B & B & B \\

Lunar mass & A & B & C &C  \\ 
\hline
\end{tabular}
\caption{Qualitative assessment of the success of various dynamical models in explaining physical and chemical properties of the Moon. A is considered the best grade, followed by B and finally C. }
\label{tb:transcript}
\end{center}
\end{table*}

\section{Summary and Outlook}
\label{sec:outlook}

The story of the Moon is intimately tied to that of the Earth. The two bodies are separated by just 384,400 km, implying a genetic connection between them. The low mean density of the Moon, 3340 kgm$^{-3}$ though similar to the Earth's upper mantle, already attests to its peculiar constitution. Indeed, the lunar core is shown by new joint inversions to comprise only 0.8--1.5 \% of its mass (radius of 300$\pm$20 km); a lower fraction than in any other terrestrial body in the Solar System. The inversions show that, if the bulk composition of the silicate portion of the Moon is akin to that of the Earth's upper mantle (8.1 wt. \%), then the lunar core has a density of $\sim$7800 kgm$^{-3}$, and is required to be comprised almost entirely of Fe-Ni alloy. On the other hand, should the bulk silicate Moon have higher FeO contents ($\sim$13 wt. \%) than in the Earth's mantle, then only considerably lower core densities ($\sim$6100 $\pm$ 800 kgm$^{-3}$ can satisfy the mass and moment of inertia of the Moon, and therefore imply considerable quantities of S and/or non-metallic components in the lunar core. At present, the unambiguous resolution of the nature and composition of the lunar core is limited by the attenuation of seismic waves below $\sim$1100 km in the lunar interior and the insensitivity of remote geophysical techniques to core properties. The corollary is that, were future seismic surveys, such as the upcoming Farside Seismic Suite \citep{panning_etal2022} and potential Lunar Geophysical Network in 2030 \citep{haviland2022LGN}, to detect core-traversing phases, distinctions between genetic models for the formation of the Moon would be possible. \\

The identification of indistinguishable O isotope compositions between lunar and terrestrial rocks \citep{wiechert2001} is consistent with a scenario in which the Moon and Earth were derived from material with the same nucleosynthetic signature. However, the dependence of this interpretation on a single isotope system allowed considerable flexibility as to erstwhile models of Moon formation, with a leading explanation suggesting that the Earth and Moon formed in the same region of the protoplanetary disk \citep{Dauphasetal2014}. Indeed, the emergence of the paradigm that the inner- and outer solar system were populated by material of distinct nucleosynthetic heritage \citep{warren2011} in the isotopes of Cr and Ti permits the use of these isotopes as tracers of the provenance of material from which the planets were built. That the Earth and Moon also possess identical Cr, Ti, Ca and Zr mass-independent isotopic compositions \citep{zhang_ti2012, AkramSchoenbaechler2016, mougel2018, schiller_etal2018} paints an increasingly clear picture of their derivation from a common precursor. This view is reinforced not only by the similarity in the abundances of Mn, Cr and V in the mantles of the Earth and Moon \citep{ringwoodseifert1986}, but also in the overlap of their inferred initial $^{182}$W/$^{184}$W isotope compositions \citep{kruijer2017tungsten}. The latter constraint is particularly exacting because the evolved $^{182}$W/$^{184}$W ratio depends not only on the extent of Hf/W fractionation, but on how and when it occurred. Nevertheless, there are considerable uncertainties associated with both the Hf/W of the lunar mantle and the timing of core formation on the Moon \citep{thiemens2019early}. Sampling of the Moon away from the Procellarum KREEP terrane, upon which the Apollo missions exclusively landed, and particularly discovery of samples of the lunar mantle (e.g., near the South Pole-Aitken basin) would be invaluable in determining the bulk composition of the Moon, both chemically and isotopically. Sample return missions from the far side, including the Chang'e 6 and Endurance, which is a proposed mission, would bring new and key insights to this issue.  \\

Present estimates of lunar composition are based on the products of mare magmatism, notably the pyroclastic green glasses and low-Ti basalts, which are in equilibrium with mantle sources that have vastly different Mg\#s ($\sim$ 0.85 and 0.75, respectively). Despite our inability to determine the major element composition of the mantle precisely with this approach \citep{Ringwoodetal1987}, these rocks reveal a stark depletion of moderately- and highly volatile elements by a factor $\sim$5 and 100--500, respectively, relative to the Earth's mantle \citep{wolfanders1980,gleissneretal2022}. Chemical fractionation patterns among alkali metals favour volatile loss at $\sim$1400 K \citep[cf. also][]{oneill1991origin}. Mass-dependent stable isotope fractionation among a wide range of lunar lithologies attest to widespread volatile depletion of the Moon during its formation, with the observed isotopic differences between the Earth and Moon consistent with (near-) equilibrium fractionation \citep{Tartese2021,Dauphasetal2022_MVE}. Unique definition of the pressures and temperatures of volatile depletion is limited by the lack of experimental data on vaporisation reactions between condensed phases and gas, and the isotopic fractionation associated with them. \\

The physical setting in which such volatile depletion occurred is frequently discussed in the context of a giant impact event(s). The canonical giant impact \citep{HartmannDavis1975,CameronWard1976, CanupAsphaug2001} describes the collision between a Mars-sized body ($\sim$10 \% the mass of the Earth) and the proto-Earth in aid of accounting for the present-day angular momentum of the Earth-Moon system. However, SPH simulations showed that such collisions would predict that the Moon formed predominantly- or partially from impactor material, violating the chemical- and isotopic kinship of the Earth and Moon attested to by more recent geochemical evidence. Even if O isotope homogeneity were to have been achieved by mixing in the vapour phase of a protolunar disk \citep{PahlevanStevenson2007}, this mechanism would be inadequate to homogenise more refractory elements, notably Cr and Ti. Consequently, higher energy scenarios in which the proto-Earth was in retrograde motion, or collisions between larger impactors were invoked, are found to give rise to a peculiar extended contiguous disk structure, or synestia that facilitate homogenisation \citep{CukStewart2012,Locketal2018}. Although this event would likely give rise to the required level of isotopic homogeneity, the physical and transport properties of supercritical fluids \citep[forming above $\sim$6000 K,][]{caracas_stewart2023} remain poorly understood, and have yet to be incorporated into numerical impact simulations. Moreover, these energetic impacts, which are more common for half Earth-sized collisions, are not dynamically favoured at the late stage of planet formation in the conventional dynamical planet formation- and Grand Tack models, but are more common in a recent ring planet formation model \citep{Wooetal2024}. Whether sufficient mass can coalesce in the Moon-forming regions in such structures is under scrutiny, while equilibrium condensation at very high temperatures ($\sim$3000 K) would not engender the required mass-dependent isotopic fractionation. \\

From a chemical and isotopic perspective, there is no unambiguous evidence for the contribution of an impactor to the Moon. This implies that either \textit{i)} the material that made the Earth and Moon mixed perfectly during an impact event, before the Moon preferentially sampled a silicate-rich part of the disk or \textit{ii)} there was no impact at all. 
Whether these options remain plausible depends upon our ability to resolve the bulk composition of the Moon from its geophysical response and geochemical record. Should the Moon have \textit{exactly} the same composition as the Earth's mantle with respect to major element abundances, then its origin could be reconciled with a `devolatilised' fragment of the Earth's mantle, after core-formation, but \textit{before} the accretion of the late veneer. In the absence of a more plausible physical scenario, the giant impact is currently the worst model we have for the origin of the Moon, aside from all the others that have been tried from time to time.


\appendix
\section{Acknowledgements}
\label{sec:acknowledgements}
We are indebted to an anonymous reviewer and to the editor, Sujoy Mukhopadhyay, for numerous perceptive comments on the manuscript. PAS is grateful for the enlightening discussions over the years with Hugh O'Neill, most recently in the tranquil surrounds of the Melbourne Botanic Gardens. PAS also thanks Jim Connelly, Jeff Taylor, Daniele Antonangeli, Max Schmidt, Josh Snape, Maria Schönbächler, Christian Liebske, Thomas Kruijer, Thorsten Kleine, Francis Nimmo, Raúl Fonseca, Andreas Pack and Ian Jackson for insights on various aspects of lunar physics and chemistry. M.N. thanks David Rubie and Alessandro Morbidelli for providing orbital dynamics data. PAS thanks the Swiss National Science Foundation (SNSF) via an Eccellenza Professorship (203668) and the Swiss State Secretariat for Education, Research and Innovation (SERI) under contract number MB22.00033, a SERI-funded ERC Starting Grant `2ATMO'.   M.N. was supported in part by the National Aeronautics and Space Administration (NASA) grant numbers 80NSSC19K0514, 80NSSC21K1184, and 80NSSC22K0107. Partial funding for M.N. was also provided by NSF EAR-2237730 as well as the Center for Matter at Atomic Pressures (CMAP), an NSF Physics Frontier Center, under Award PHY-2020249. Any opinions, findings, conclusions or recommendations expressed in this material are those of the authors and do not necessarily reflect those of the National Science Foundation. M.N. was also supported in part by the Alfred P. Sloan Foundation under grant G202114194. 


\bibliographystyle{elsarticle-harv} 
\bibliography{cas-refs, Miki-refs, miki-long-refs}





\end{document}